\documentclass[longauth]{aa}  
\usepackage[english]{babel} 
\usepackage{natbib}
\usepackage{amssymb}
\usepackage{amsmath}
\usepackage{url}
\usepackage{txfonts}
\usepackage{rotating} % for sidewaystable
\usepackage{xspace}
\usepackage{color}
\usepackage[switch]{lineno} % use for line numbers
\usepackage{txfonts}
\usepackage{caption}
\usepackage{graphicx}
\usepackage{hyperref}
\newcommand{\fermi}{\textsl{Fermi}\xspace}
\newcommand{\lat}{\textsl{Fermi}/LAT\xspace}
\newcommand{\X}{8.4\,GHz\xspace}

\begin{document} 

   \title{TANAMI: Tracking Active Galactic Nuclei with Austral
     Milliarcsecond Interferometry\footnote{The cleaned VLBI images
       displayed in Figs. 1 to 5 (FITS files) are only available at
       the CDS via anonymous ftp to \url{cdsarc.u-strasbg.fr} (130.79.128.5)
or via \url{http://cdsweb.u-strasbg.fr/cgi-bin/qcat?J/A+A/}.} }

   \subtitle{II. Additional sources}

\titlerunning{TANAMI: II. Additional Sources}
\authorrunning{C.~M\"uller et al.}

   \author{C.~M\"uller\inst{\ref{affil:ru},\ref{affil:wuerzburg},\ref{affil:remeis},\ref{affil:mpifr}}
     \and M.~Kadler\inst{\ref{affil:wuerzburg}}
     \and R.~Ojha\inst{\ref{affil:nasa_gsfc},\ref{affil:umbc},\ref{affil:cua}}
     \and R.~Schulz\inst{\ref{affil:astron},\ref{affil:wuerzburg},\ref{affil:remeis}}
     \and J.~Tr\"ustedt\inst{\ref{affil:wuerzburg}}
      \and P.~G.~Edwards\inst{\ref{affil:csiro}}
       \and E.~Ros\inst{\ref{affil:mpifr},\ref{affil:obsvalencia},\ref{affil:valencia}} 
     \and B.~Carpenter\inst{\ref{affil:nasa_gsfc},\ref{affil:umbc},\ref{affil:cua}}
     \and R.~Angioni\inst{\ref{affil:mpifr}}
   \and J.~Blanchard\inst{\ref{affil:jive}}
   \and M.~B\"ock\inst{\ref{affil:remeis},\ref{affil:mpifr}}
     \and P.~R.~Burd\inst{\ref{affil:wuerzburg}}
     \and M.~D\"orr\inst{\ref{affil:wuerzburg}}
     \and M.~S.~Dutka\inst{\ref{affil:nasa_gsfc},\ref{affil:wyle}}
     \and T. Eberl\inst{\ref{affil:ecap}}
     \and S.~Gulyaev\inst{\ref{affil:warkworth}}
     \and H.~Hase\inst{\ref{affil:bkg}}        
     \and S.~Horiuchi\inst{\ref{affil:csiro_canberra}}
     \and U.~Katz\inst{\ref{affil:ecap}}
     \and F.~Krau\ss \inst{\ref{affil:uva}}
     \and J.~E.~J.~Lovell\inst{\ref{affil:tasmania}}
     \and T.~Natusch\inst{\ref{affil:warkworth}}
     \and R.~Nesci\inst{\ref{affil:inaf}}
     \and C.~Phillips\inst{\ref{affil:csiro}}
     \and C.~Pl\"otz\inst{\ref{affil:bkg}}
     \and T.~Pursimo\inst{\ref{affil:not}}
     \and J.~F.~H.~Quick\inst{\ref{affil:hartrao}} 
     \and J.~Stevens\inst{\ref{affil:csiro}}
  \and D.~J.~Thompson\inst{\ref{affil:nasa_gsfc}}
     \and S.~J.~Tingay\inst{\ref{affil:curtin},\ref{affil:inaf_bologna}}
     \and A.K.~Tzioumis\inst{\ref{affil:csiro}}
     \and S.~Weston\inst{\ref{affil:warkworth}, \ref{affil:csiro}}              
      \and J.~Wilms\inst{\ref{affil:remeis}}
      \and J.~A.~Zensus\inst{\ref{affil:mpifr}} 
     }
     
\institute{
Department of Astrophysics/IMAPP, Radboud University Nijmegen, PO
Box 9010, 6500 GL, Nijmegen, The Netherlands\\
\email{c.mueller@astro.ru.nl}\label{affil:ru}
\and Institut f\"ur
Theoretische Physik und Astrophysik, Universit\"at W\"urzburg, Am
Hubland, 97074 W\"urzburg, Germany \label{affil:wuerzburg} 
\and
Dr.\ Remeis-Sternwarte \& ECAP, Universit\"at Erlangen-N\"urnberg,
Sternwartstra\ss e 7, 96049 Bamberg, Germany \label{affil:remeis}
\and
Max-Planck-Institut f\"ur Radioastronomie, Auf dem H\"ugel 69, 53121
Bonn, Germany \label{affil:mpifr}  
\and NASA, Goddard Space Flight Center, Astrophysics Science
Division, Code 661, Greenbelt, MD 20771, USA \label{affil:nasa_gsfc}
\and CRESST/University of Maryland Baltimore
County, Baltimore, MD 21250, USA \label{affil:umbc} 
\and Catholic University of America, Washington, DC 20064,
USA \label{affil:cua} 
\and ASTRON, the Netherlands Institute for Radio Astronomy, Postbus 2,
7990 AA, Dwingeloo, The Netherlands\label{affil:astron}
\and
Joint Institute for VLBI ERIC (JIVE),
Postbus 2, 7990 AA Dwingeloo, The Netherlands \label{affil:jive}
\and Wyle Science, Technology and Engineering Group, Greenbelt, 
MD 20771, USA \label{affil:wyle} 
\and ECAP, Universit\"at Erlangen-N\"urnberg, Erwin-Rommel-Str. 1, 91058
Erlangen, Germany \label{affil:ecap} 
\and CSIRO Astronomy and Space Science, ATNF,
PO Box 76 Epping, NSW 1710, Australia \label{affil:csiro}
\and 
Observatori Astronòmic, Universitat de València, 46980 Paterna,
València, Spain \label{affil:obsvalencia} 
\and Dept.\ d'Astronomia i Astrof\'isica, Universitat de Val\`encia,
46100 Burjassot, Val\`encia, Spain \label{affil:valencia} 
\and 
Institute for Radio Astronomy \& Space Research, AUT University, Auckland, New Zealand
\label{affil:warkworth}
\and Bundesamt f\"ur Kartographie und Geod\"asie, 93444 Bad K\"otzting,
Germany \label{affil:bkg}
\and CSIRO Astronomy and Space Science,
Canberra Deep Space Communications Complex, P.O. Box 1035,
Tuggeranong, ACT 2901, Australia \label{affil:csiro_canberra} 
\and
GRAPPA \& Anton Pannekoek Institute for Astronomy, University of
Amsterdam, Science Park 904, 1098 XH Amsterdam, The Netherlands
 \label{affil:uva}
\and
School of Mathematics \& Physics, University of Tasmania, Private Bag
37, Hobart, Tasmania 7001, Australia \label{affil:tasmania} 
\and INAF/IAPS, via Fosso del Cavaliere 100, 00133 Roma, Italy \label{affil:inaf}
\and Nordic Optical Telescope Apartado
474, 38700 Santa Cruz de La Palma Santa Cruz de Tenerife,
Spain \label{affil:not} 
\and Hartebeesthoek Radio Astronomy
Observatory, PO Box 443, Krugersdorp 1740, South
Africa \label{affil:hartrao}
\and International Centre for Radio Astronomy Research, Curtin
University, Bentley, WA 6102, Australia \label{affil:curtin}
\and INAF, Istituto di Radio Astronomia, Via Piero Gobetti, 41029 Bologna,
Italy\label{affil:inaf_bologna}
             }
   \date{\today}

% \abstract{}{}{}{}{} 
% 5 {} token are mandatory
 
  \abstract
  % context heading (optional)
  % {} leave it empty if necessary  
   {TANAMI is a multiwavelength program monitoring active galactic nuclei (AGN)
   south of $-30^\circ$ declination including high-resolution very long baseline interferometry
(VLBI) imaging, radio, optical/UV, X-ray, and $\gamma$ ray studies. 
   We have previously
published first-epoch 8.4\,GHz VLBI images of the parsec-scale
structure of the initial sample. In
this paper, we present images of 39 additional sources.
The full sample comprises most of the radio- and $\gamma$-ray brightest AGN in the
southern quarter of the sky, overlapping with the region from which
high-energy ($> 100$\,TeV) neutrino events have been found.
    }
  % aims heading (mandatory)
   { 
We characterize the parsec-scale radio properties of
the jets and compare them with the quasi-simultaneous \lat
$\gamma$ ray data. Furthermore, we study the
jet properties of sources which are in positional coincidence with high-energy 
neutrino events  compared to the full sample. We test the positional
agreement of high-energy neutrino events with
various AGN samples. 

     }
  % methods heading (mandatory)
   {TANAMI VLBI observations at 8.4\,GHz are made with southern hemisphere radio
     telescopes located in Australia, Antarctica, Chile, New Zealand, and
     South Africa.
     }
  % results heading (mandatory)
   {Our observations 
yield the first images of many jets below $-30^\circ$ declination  at milliarcsecond
resolution. We find
that $\gamma$-ray loud TANAMI sources tend to be more compact on
parsec-scales and 
have higher core brightness temperatures than $\gamma$-ray faint jets, indicating higher
Doppler factors.  No significant structural difference is found
between sources in positional coincidence with high-energy neutrino
events and other TANAMI jets.  The 22 $\gamma$-ray brightest AGN
in the TANAMI sky show only a
weak positional agreement with high-energy neutrinos
demonstrating that the $> 100$\,TeV IceCube signal is not simply
dominated by a small number of the $\gamma$-ray brightest blazars. 
Instead, a larger number of sources have to contribute to the signal
with each individual 
source having only a small Poisson probability for producing an event
in multi-year integrations of current neutrino detectors.
}
  % conclusions heading (optional), leave it empty if necessary 
 {}


\maketitle

%________________________________________________________________
\section{Introduction}

Multiwavelength observations of active galactic nuclei (AGN) and their
jets provide an intimate link between the key science categories of
high-resolution radio astronomy, $\gamma$-ray astronomy, and
astroparticle physics.  Most $\gamma$-ray loud AGN detected by \textsl{Fermi}/LAT are blazars, i.e., extragalactic jets 
oriented at a small angle to the line of sight from the observer 
\citep{1fgl,Nolan2012,3fgl}.  The broadband
spectral energy distribution (SED) of flat-spectrum radio quasars
(FSRQs) and BL\,Lac objects, both subtypes of blazars, is dominated by
the strongly Doppler-boosted emission from the approaching relativistic
jet.  Aside from blazars, a small number of ``misaligned'' AGN jets have
also been detected at $\gamma$ rays \citep{Abdo2010_misaligned}. Observations of AGN jets
have revealed a strong connection between their parsec-scale radio
structure and dynamics, as probed by observations using very long baseline interferometry
(VLBI), and their variable $\gamma$-ray emission
\citep[e.g.,][]{Jorstad2001a,Lister2009b,Lister2011,Fuhrmann2014}.
However, a significant fraction of bright and powerful AGN jets
remain undetected at $\gamma$ rays \citep[see
also][]{Lister2015}. Only VLBI observations can resolve the parsec-scale
structure of jets, allowing the investigation of potential differences
between these samples.

\textit{Tracking Active Galactic Nuclei with Milliarcsecond
Interferometry} (TANAMI)  is a multiwavelength program monitoring a sample of
the radio and $\gamma$-ray brightest extragalactic jets of the
southern hemisphere \citep{Ojha2010a,KadlerOjha2015}. The central element of TANAMI is the VLBI
monitoring program at 8.4\,GHz and 22\,GHz \citep[see
Sect.~\ref{sec:observations} and][hereafter Paper~I]{Ojha2010a}.  The initial TANAMI sample consisted of
43~sources, whose first-epoch 8.4\,GHz VLBI images are discussed in
Paper~I.  Here, we present first-epoch VLBI images at 8.4\,GHz of 39
additional sources and discuss their parsec-scale properties. This
study covers many sources that are relevant in the context of recent
TANAMI high-energy astronomy and astroparticle physics studies, for
which no previous VLBI images have been available, namely 8 sources
from the dynamic-SED catalog of southern blazars \citep{Krauss2016},
28 sources whose $\gamma$-ray properties were studied by
\citet{Boeck2016}, and 3 sources  discussed by \citet{Dutka2013}
and by \citet{Nesci2011ATel0447,Nesci2011ATel0402}.

A new link between VLBI observations and high-energy astronomy and
astroparticle astrophysics is emerging in the context of the search
for high-energy neutrino point sources. Based on TANAMI VLBI and
multiwavelength data, we were able to  show \citep{Krauss2014a} that the
population of FSRQs as a whole can explain the neutrino signal at
energies above $\sim$100\,TeV as found in the IceCube High-Energy
Starting-Event (HESE) analysis \citep{Ice14}. Moreover, it was
possible to associate the single neutrino event at 2\,PeV energy
(HESE-35) with the exceptional outburst observed in the FSRQ
PKS\,1424$-$418 \citep{Kadler2016}, provided that the observed
high-energy emission is due to hadronic emission.
A second case of a blazar outburst coinciding with a high-energy
muon-neutrino event was recently published by \citet{Kun17}. In this case,
the blazar PKS\,0723$-$008 is found to be located inside the small
($<1\fdg2$) field of a muon-neutrino event (HESE-5). Similar to
PKS\,1424$-$418, the blazar is found to show a massive radio outburst
over several months around the time of the neutrino detection and a
possible associated jet-component ejection.  However, it is difficult to 
correctly assess the a posteriori likelihoods of these coincidences
given the low number statistics and the long variability timescales
of the blazar outbursts considered. This situation will somewhat improve
as longer data sets of large neutrino telescopes become available.

High-fluence FSRQs are the favored neutrino producers in the scenario
advocated by \citet{Kadler2016} because of the availability of dense UV photon fields
as seeds for photomeson production, while BL Lac objects are
disfavored. However, in stratified-jet scenarios, the neutrino output
can be much higher than expected in simple one-zone models \citep{Tavecchio2014}. 
This implies that  BL\,Lac objects are also possible
neutrino point sources. In fact, \citet{Padovani2016} found
evidence for a spatial correlation between the reconstructed arrival
direction of neutrinos and TeV-emitting BL\,Lacs, while they did not
find a comparably strong correlation with FSRQs or other low-peaked
blazars.
\citet{Ice17} find that
a large number of sources is needed in order to explain the observed
sky distribution of TeV-PeV neutrino events with blazars if there is a
universal neutrino to $\gamma$-ray emission ratio and if steep neutrino
spectral indices are considered. We note that this result is not in
contradiction to the findings of \citet{Kadler2016} that blazars may
dominate the IceCube signal above 100 TeV because of their PeV-peaked
neutrino spectra. In fact, the results of \citet{Ice17} and
\citet{Kadler2016} both imply that a substantial probability for the
detection of a >100\,TeV neutrino from an individual source can only be
expected in unusually bright outbursts, while the majority of events
are expected to be contributed by fainter sources, which are much more
numerous than the brightest ones.  
It is, however, possible that only a subset of blazar jets are
bright neutrino sources. These neutrino-bright jets might reveal
themselves via characteristic parsec-scale radio properties like
limb-brightened jet morphologies \citep[e.g.,][]{Giroletti2004b},
distinguished core brightness temperatures, or multiwavelength properties.

We  investigate and characterize the parsec-scale jet properties
revealed by high-resolution VLBI observations of the radio and
$\gamma$-ray brightest TANAMI AGN in the southern sky where
  $>$100\,TeV neutrinos have been observed by IceCube.  The
remainder of this paper is structured as follows. In Sect.~\ref{sec:observations} we
discuss the data reduction of the VLBI observations using the TANAMI
array. The resulting new images are presented in
Sect.~\ref{sec:results}. Section~\ref{sec:individual} presents notable
parsec-scale and multiwavelength properties of individual sources. We
discuss the VLBI properties of a subsample of sources with respect to
their $\gamma$-ray properties in Sect.~\ref{sec:Boeck}, which is complementary
to \citet{Boeck2016}.  In Sect.~\ref{sec:neutrinos} we investigate
the characteristics of sources in positional agreement with
high-energy neutrino events, and we present our summary and conclusions
in Sect.~\ref{sec:conclusion}.

%__________________________________________________________________
\section{Sample, observations, and data reduction}\label{sec:observations}

The original TANAMI sample was set up as follows (defined in detail
in Paper~I): it consisted of a radio-selected subsample, 
which is a complete flux-density-limited sample of the radio-brightest
compact AGN jets south of $-30^\circ$ and a sample of $\gamma$ ray
associated sources based on EGRET observations.  Since the start of
\lat monitoring in 2008, we added further radio-loud targets due to
their association with significantly detected $\gamma$ ray
sources\footnote{There are only a few definite
identifications of \lat detected sources with radio counterparts,
e.g., based on contemporaneous multiwavelength variability. Most
other sources are associations based on statistical tests
\citep{1fgl,Nolan2012,3fgl}. However, in the following we  refer
to them as ``detected by \lat''.} and a correlated VLBI flux density exceeding $\sim 100$\,mJy.
The full TANAMI sample now includes\footnote{The only source from this sample not covered in this paper is PKS 2326$-$502, which has only recently been added to the VLBI monitoring program.} the 22 $\gamma$-ray brightest AGN from the third catalog of AGN detected by the \lat \citep{3lac} at declinations south of $-30^\circ$. In the following, we will refer to this as \emph{the $\gamma$ ray sample}.
Sources from both the radio and the $\gamma$ ray sample
have been observed for several years as part of the TANAMI VLBI program.
\citet{Boeck2016} discuss a subsample of TANAMI sources for which
quasi-simultaneous radio/VLBI and $\gamma$ ray data were available
from the first year of \textsl{Fermi} data.

The full TANAMI source sample includes 88 AGN jets (see
Table~\ref{table:all}): 46 quasars, 16
BL\,Lacs, 17 radio galaxies, and 9 unclassified AGN (i.e., those without an 
optical identification). Thirty-three TANAMI
sources are located within the median-positional-uncertainty regions
of high-energy ($> 100$\,TeV) neutrino events from the IceCube HESE
analysis. 

Table~\ref{table:all} summarizes the status
of the TANAMI source sample and its relevant subsamples. The optical
classifications (B: BL\,Lac, Q: Quasar, G: Galaxy, U: unclassified)
are based on the catalog by \citet{Veron2006} and the new optical
observations of $\gamma$-ray blazars by \citet{Shaw2012,Shaw2013a}.

\begin{figure*}
\centering
\includegraphics[width=0.45\textwidth]{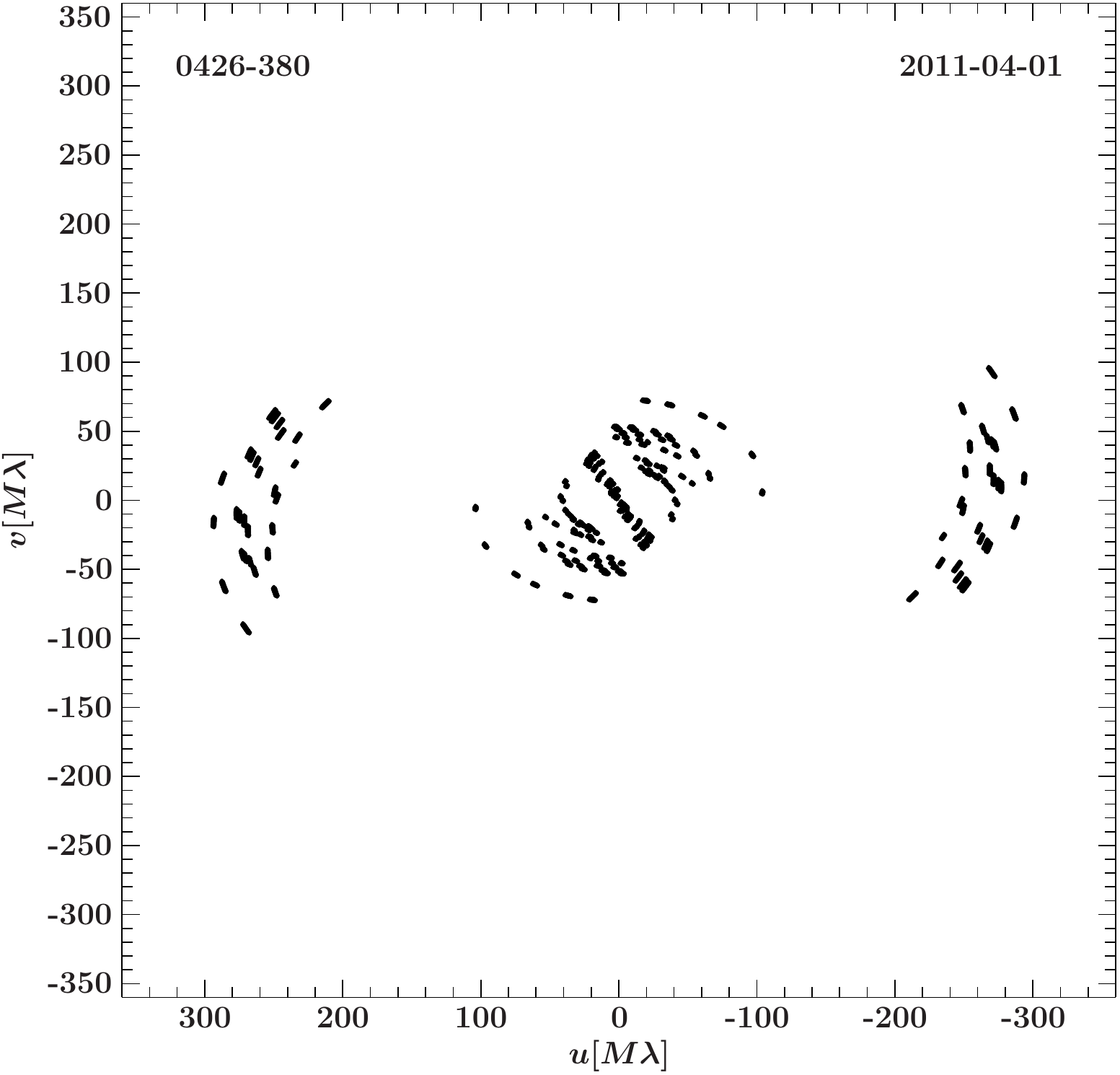}
\includegraphics[width=0.45\textwidth]{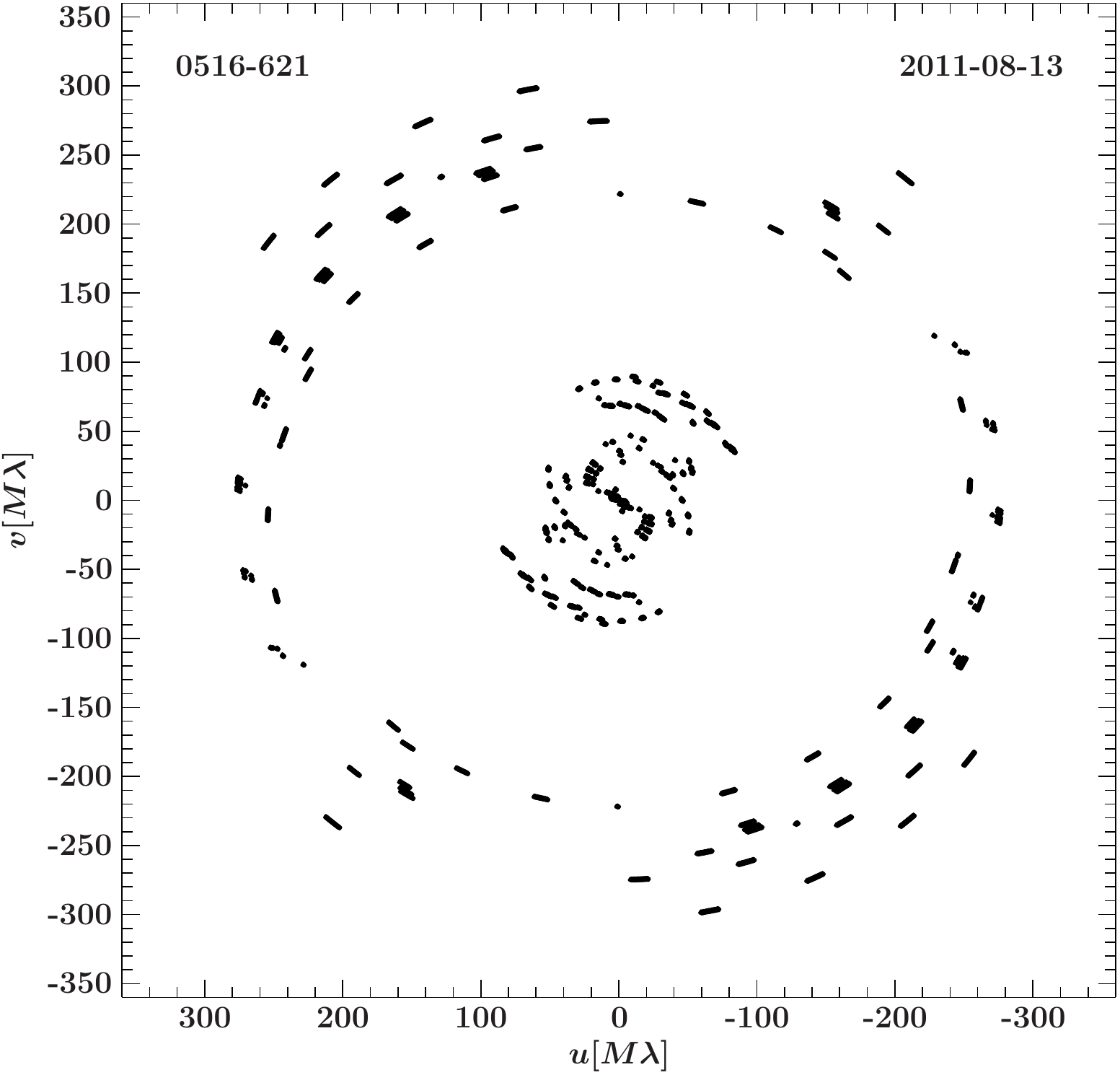}
\caption{$(u,v)$-plane of two representative TANAMI VLBI observations at 8.4\,GHz showing the
  improved coverage at $-38^\circ$ (left) and $-62^\circ$ (right) declination
  due to including new telescopes since 2011 (compare to Fig.~12 in Paper~I). The
  intermediate $(u,v)$-range is covered due to baselines to Warkworth
  (\textit{left}) and Katherine and Yarragadee  (\textit{right}). The
  long baselines are provided by TIGO and Hartebeesthoek.}
\label{fig:uvplots}
\end{figure*}

TANAMI VLBI observations are made at 8.4\,GHz and 22\,GHz using antennas 
that constitute the Australian Long Baseline Array \citep{Edwards2015}, augmented by those at 
Hartebeesthoek (South Africa), TIGO (Chile\footnote{The antenna was moved to La Plata, 
Argentina, and renamed the Argentinean-German Geodetic Observatory
(AGCO) in 2015.}), 
O'Higgins (Antarctica), NASA Deep Space Network antennas at Tidbinbilla (Australia), 
as well as the Auscope telescopes at Katherine and Yarragadee \citep[both
Australia;][]{Lovell2013}, 
and Warkworth \citep[New Zealand;][]{Weston2013}. Telescope information is listed in
Table~\ref{table:TANAMIarray}.  
A typical array configuration provides sub-mas resolution and high image fidelity (see
Fig.~\ref{fig:uvplots} and Paper~I for more details) unprecedented for
the southern hemisphere. The individual
configuration for each observing epoch presented in this paper is
summarized in Table~\ref{table:config}.  

The data presented here were correlated with the DiFX software
correlator \citep{Deller2007,Deller2011} and later calibrated and imaged using AIPS
and DIFMAP as described in Paper~I. TANAMI monitoring provides regular
observations of all sources with a cadence determined by the radio-variability 
timescale of each object,  up to two or three times per year.

%_________________________
\section{Results}\label{sec:results}

\begin{figure*}
\centering
\includegraphics[width=0.32\textwidth]{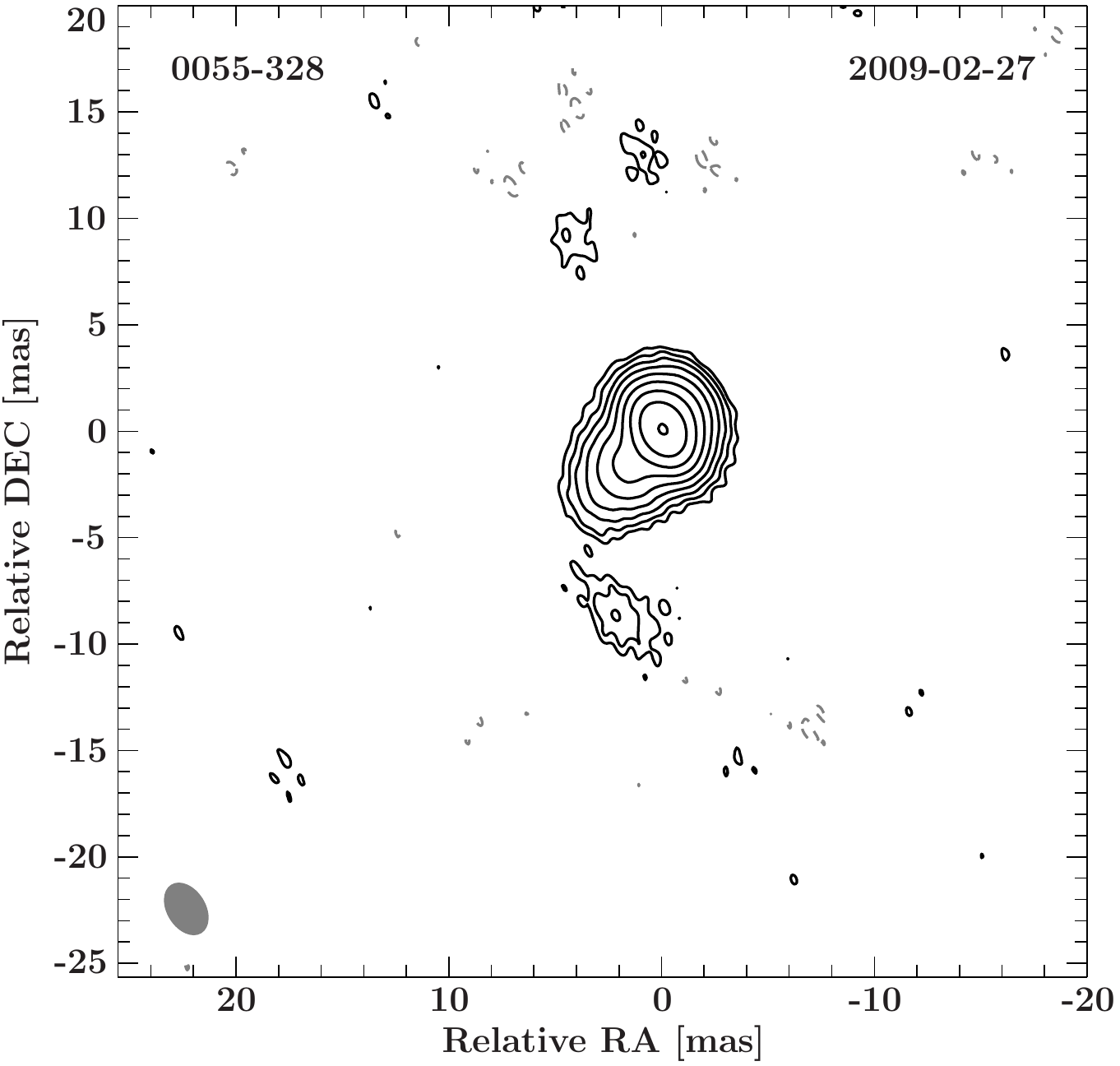} 
\includegraphics[width=0.32\textwidth]{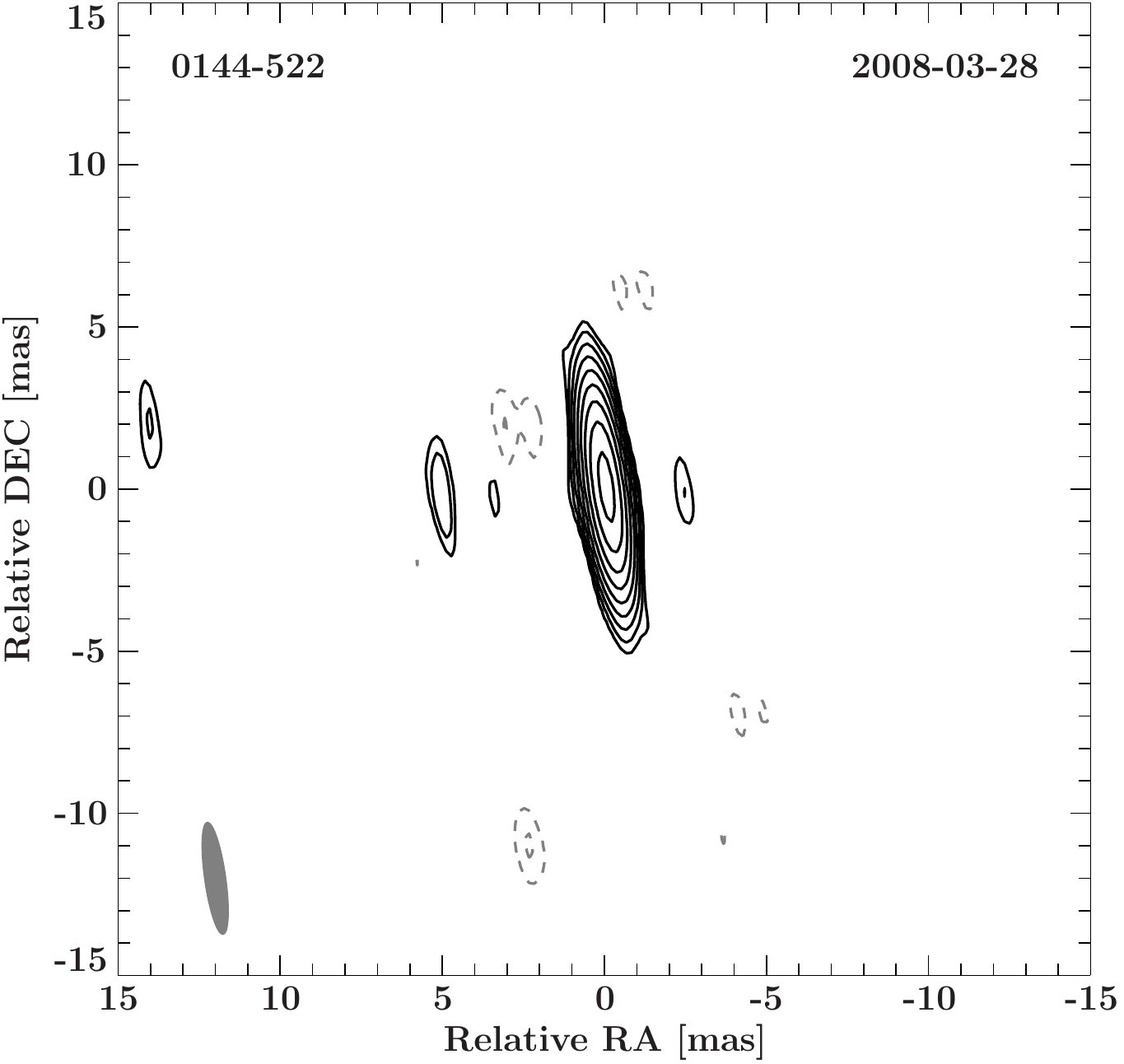} 
\includegraphics[width=0.32\textwidth]{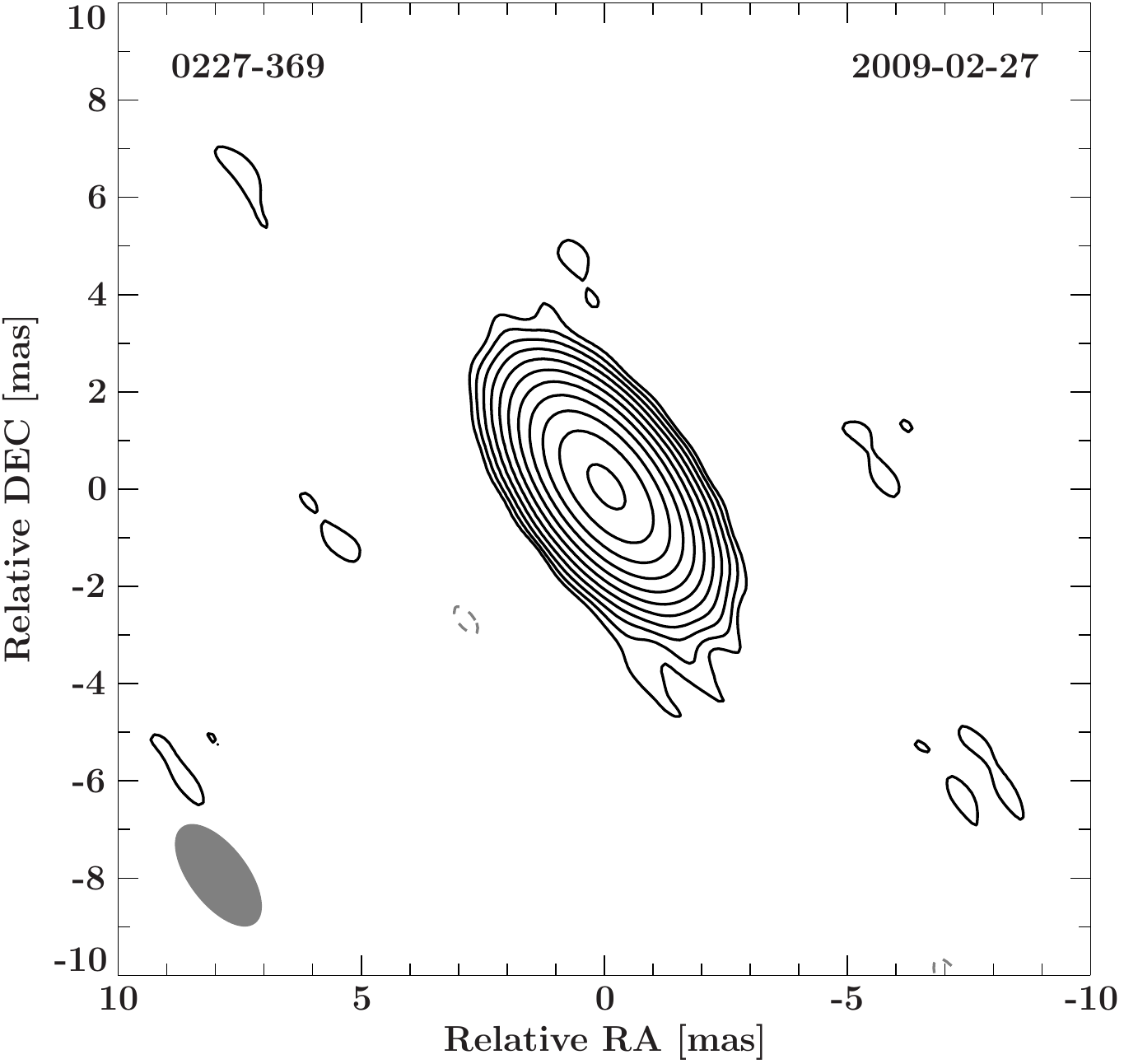}
 
\includegraphics[width=0.32\textwidth]{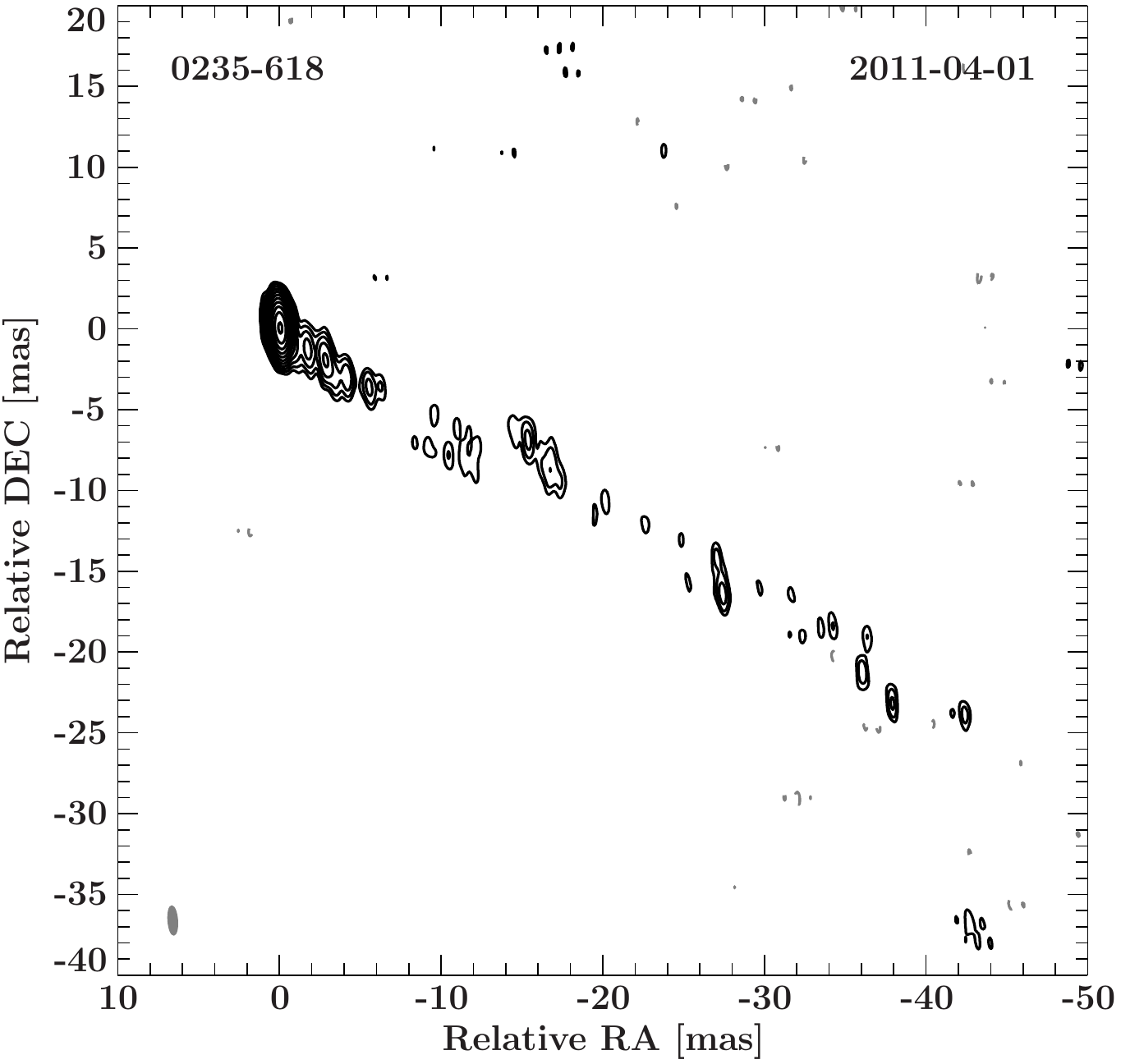}
\includegraphics[width=0.32\textwidth]{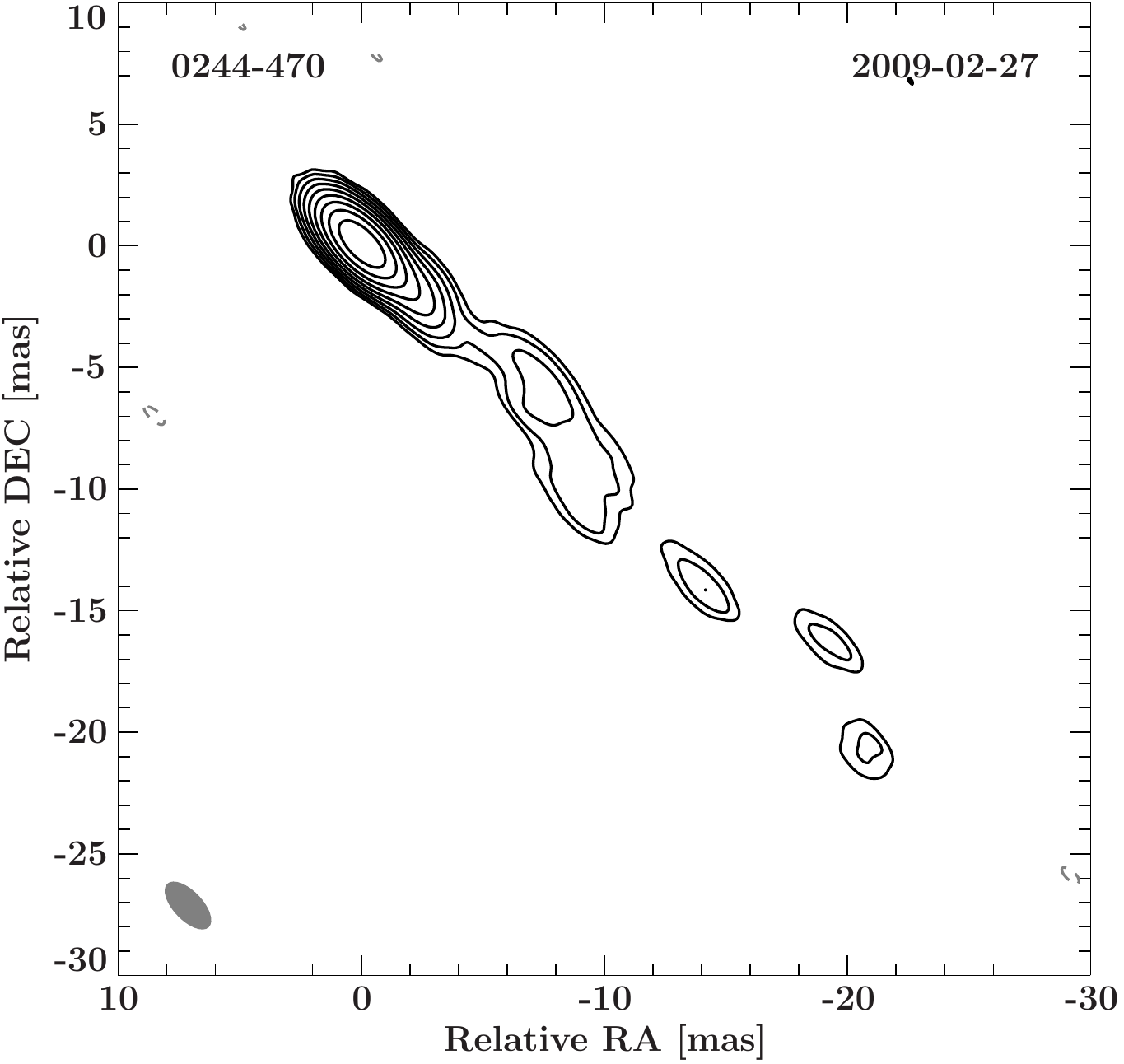} 
\includegraphics[width=0.32\textwidth]{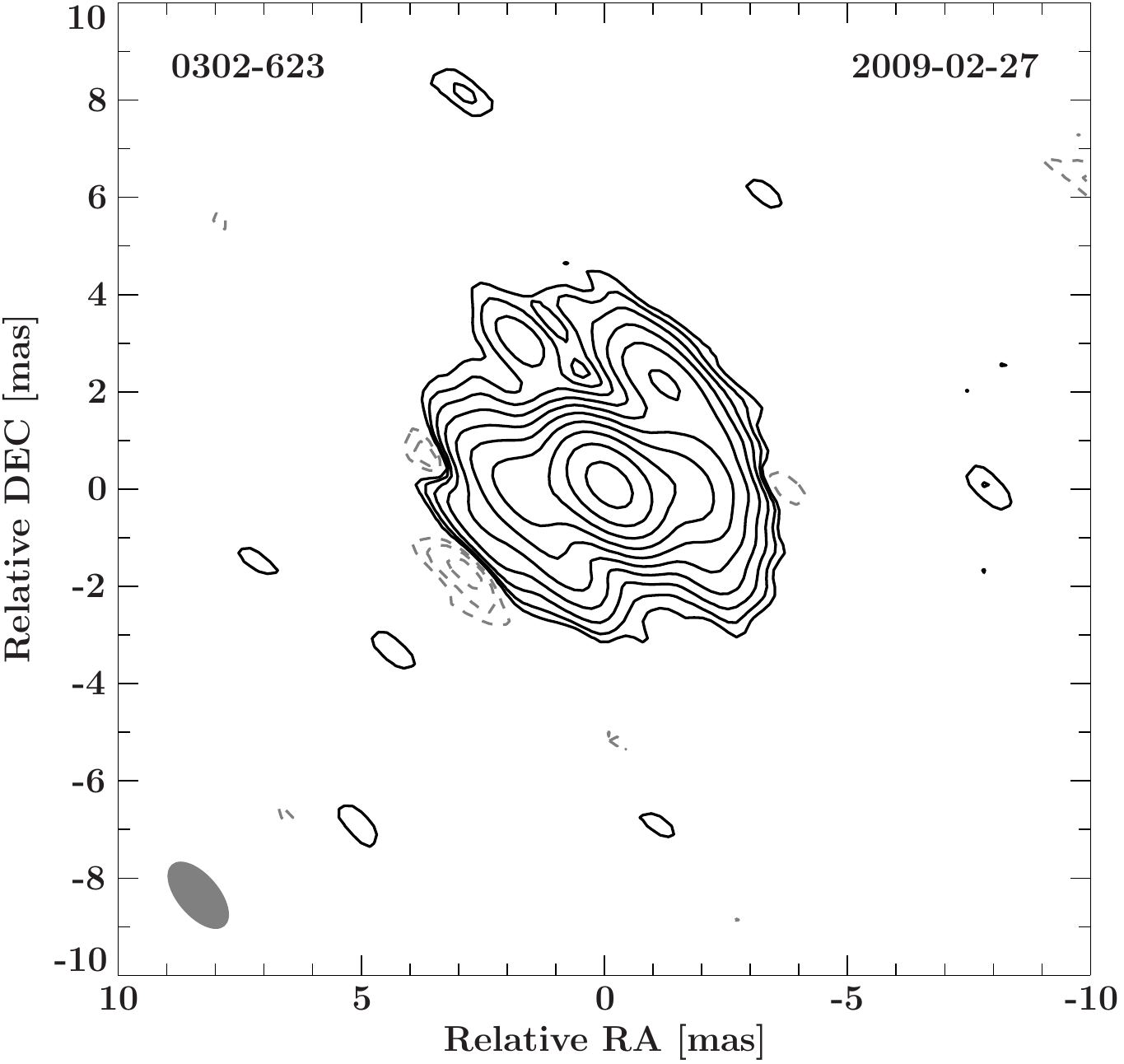}

\includegraphics[width=0.32\textwidth]{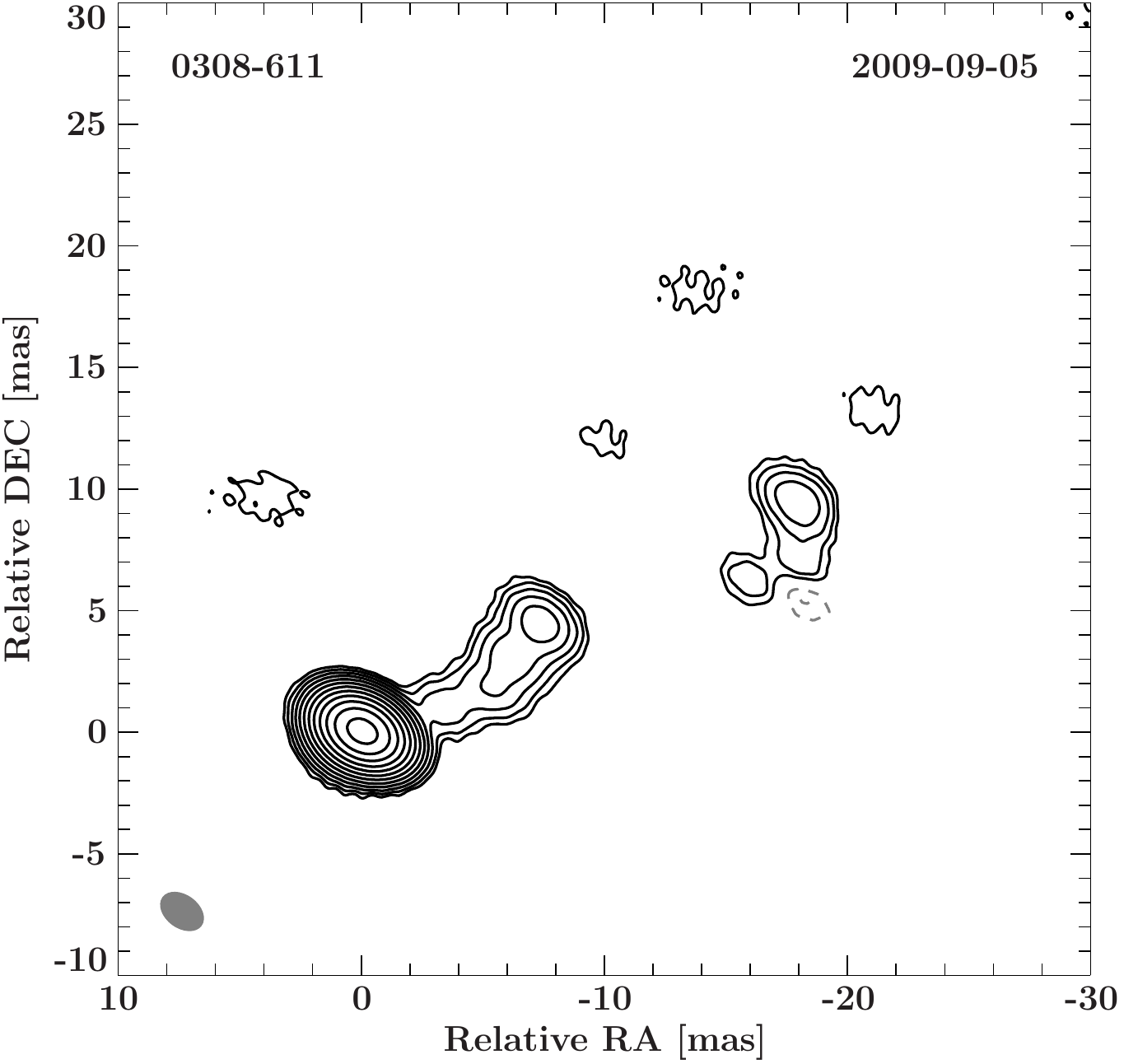}
\includegraphics[width=0.32\textwidth]{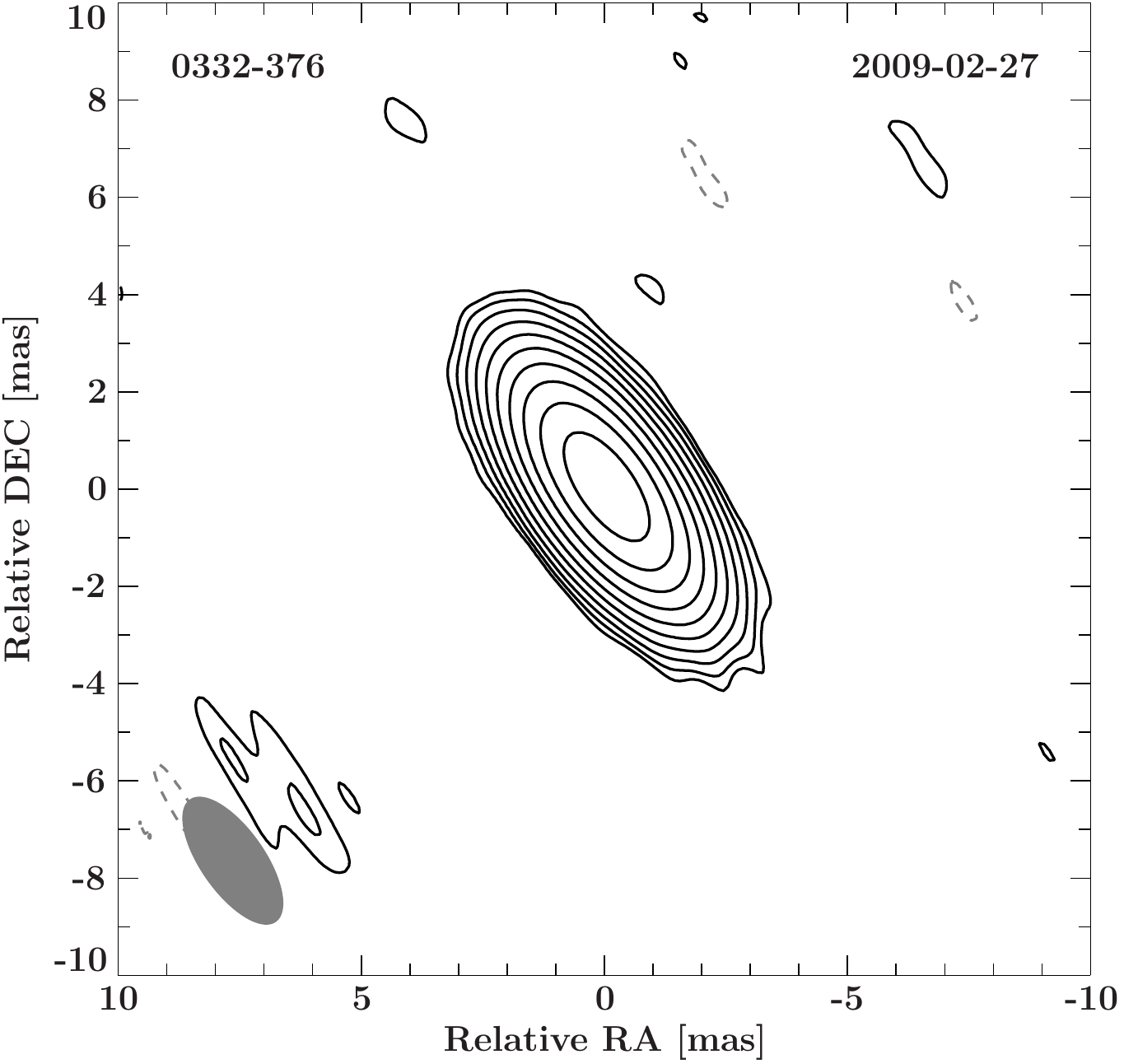}
\includegraphics[width=0.32\textwidth]{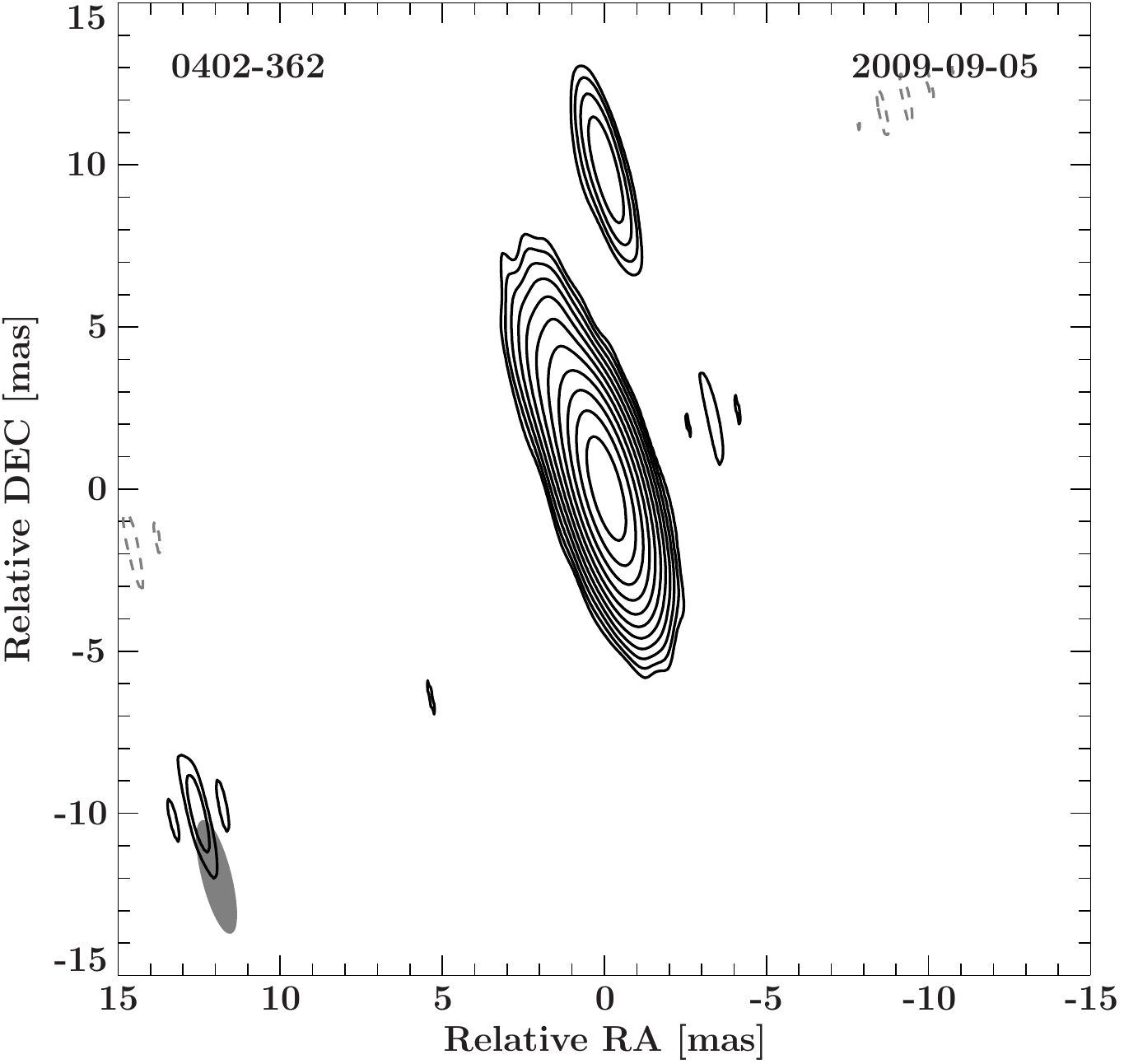}

\includegraphics[width=0.32\textwidth]{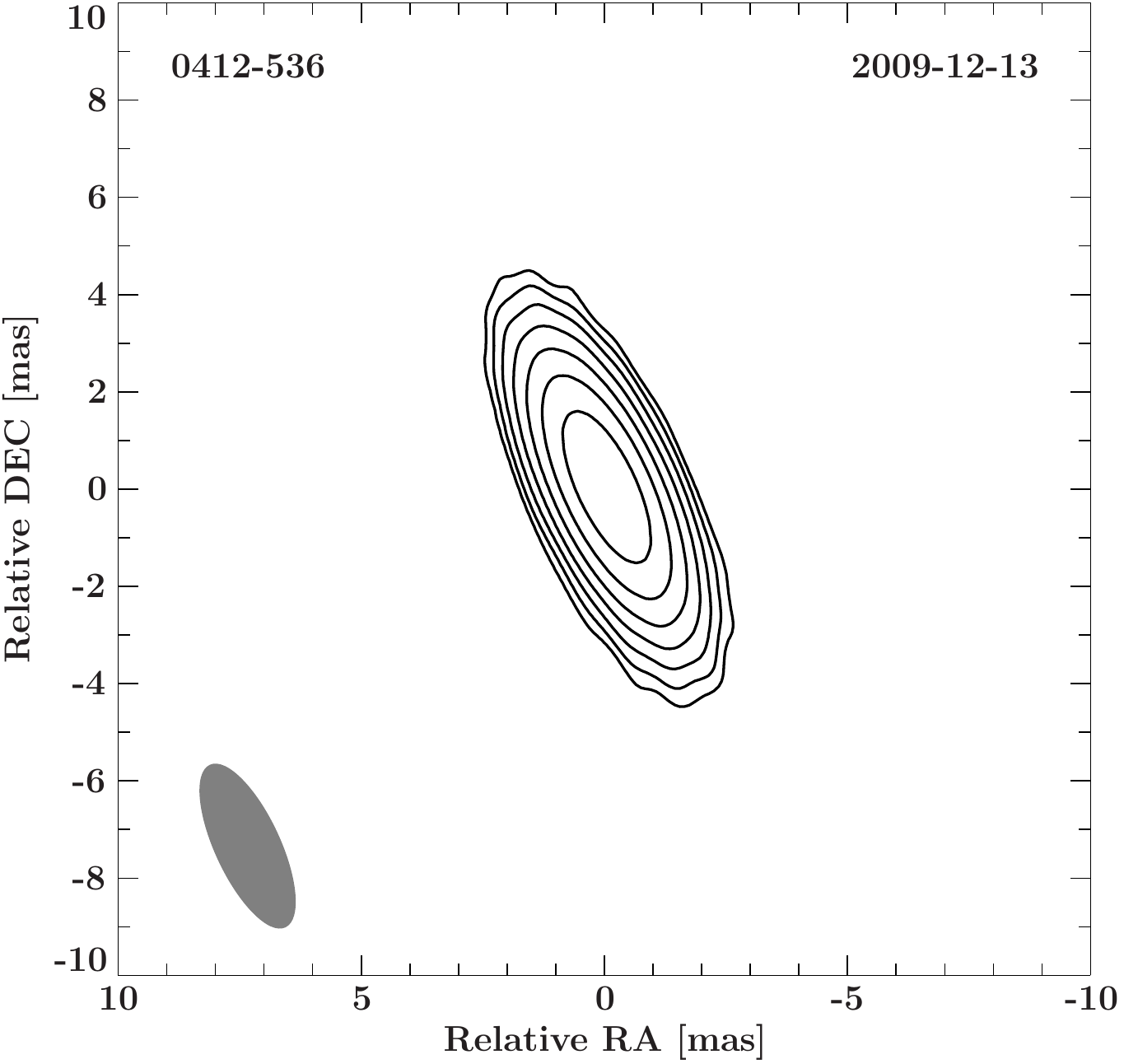}
\includegraphics[width=0.32\textwidth]{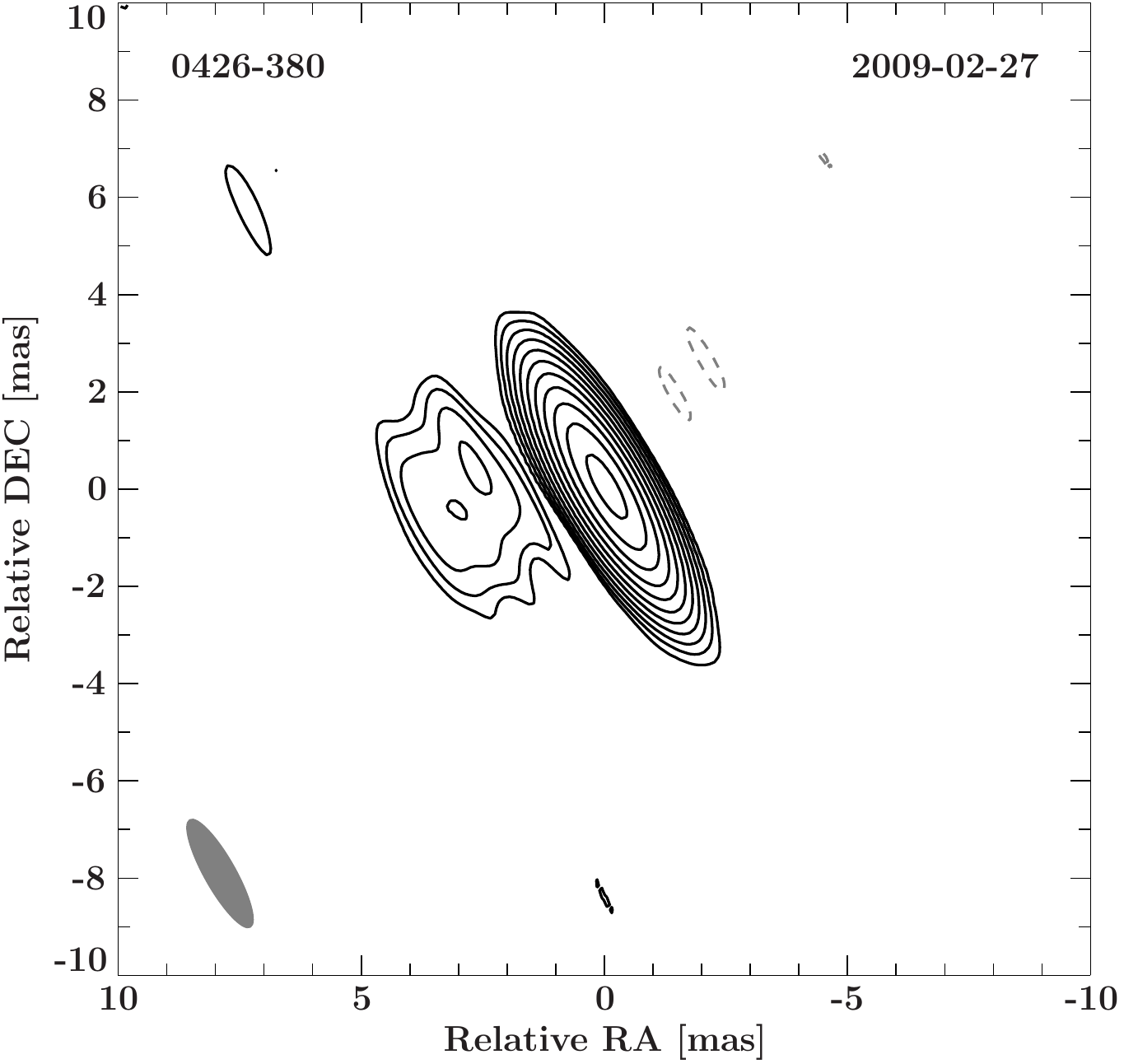}
\includegraphics[width=0.32\textwidth]{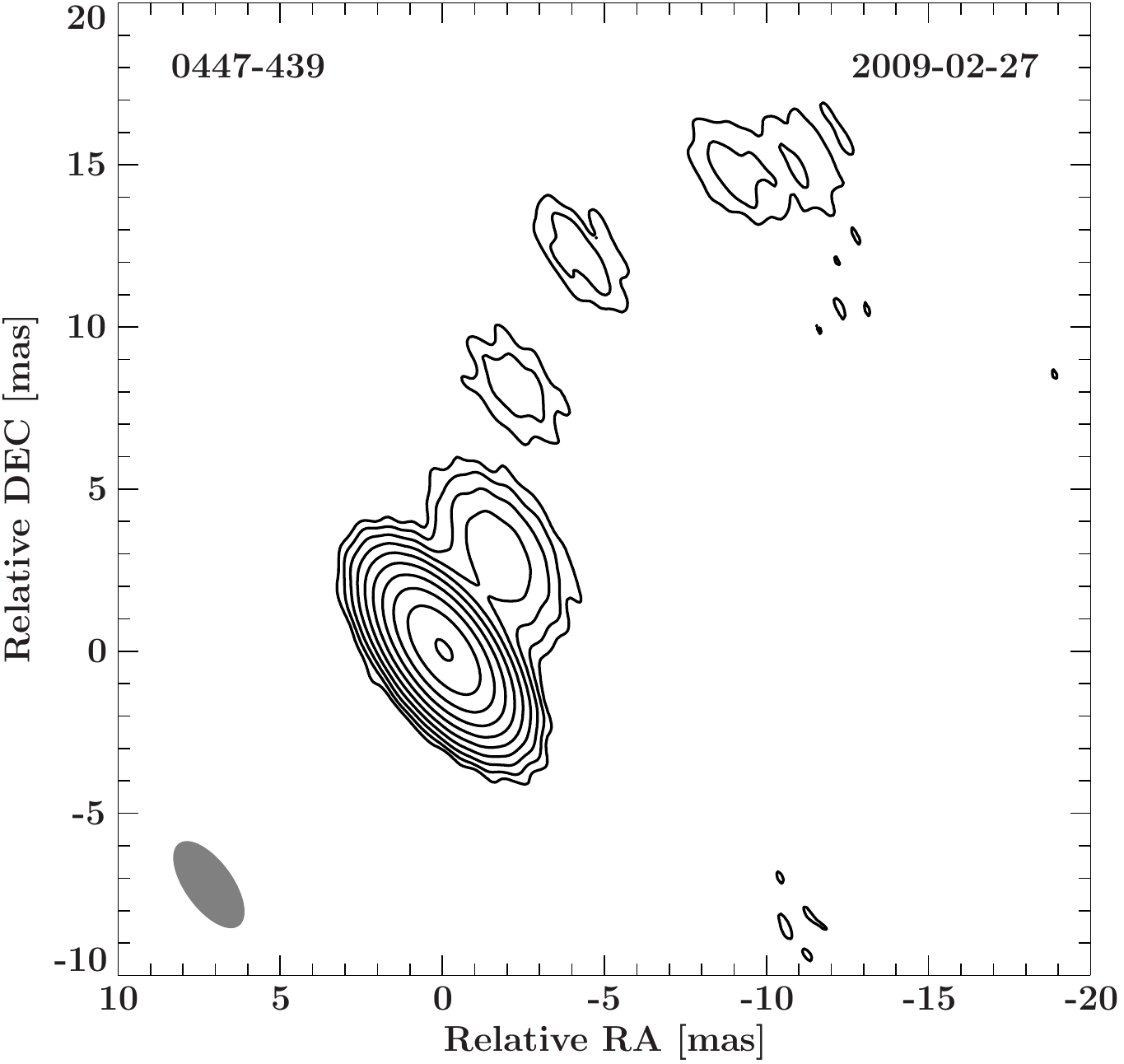}
\caption{First epoch 8.4\,GHz \texttt{clean} images of additional TANAMI sources. The
black contours indicate the flux density level (dashed gray contours
are negative), scaled logarithmically
and separated by a factor of 2, with the lowest level set to the
3$\sigma$ noise level (for more details see
Table~\ref{table:images}). 
The size of the synthesized
beam for each observation is shown as a gray ellipse in the lower left
corner of each image. The IAU B1950 name and the date of the observing epoch is given in the
upper left and right corner, respectively. }
\label{fig:newimages1}
\end{figure*}

\begin{figure*}
\centering
\includegraphics[width=0.32\textwidth]{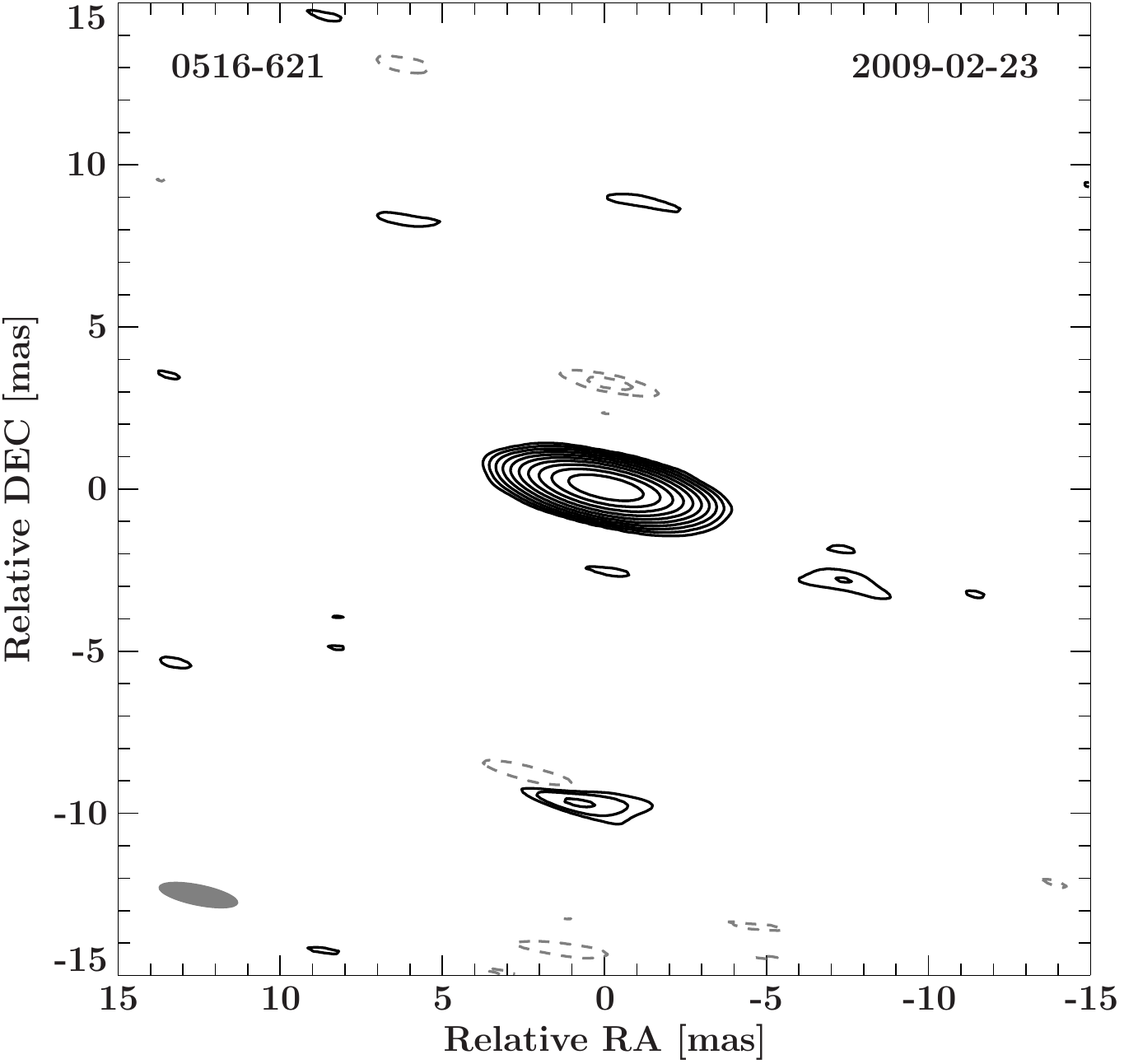}
\includegraphics[width=0.32\textwidth]{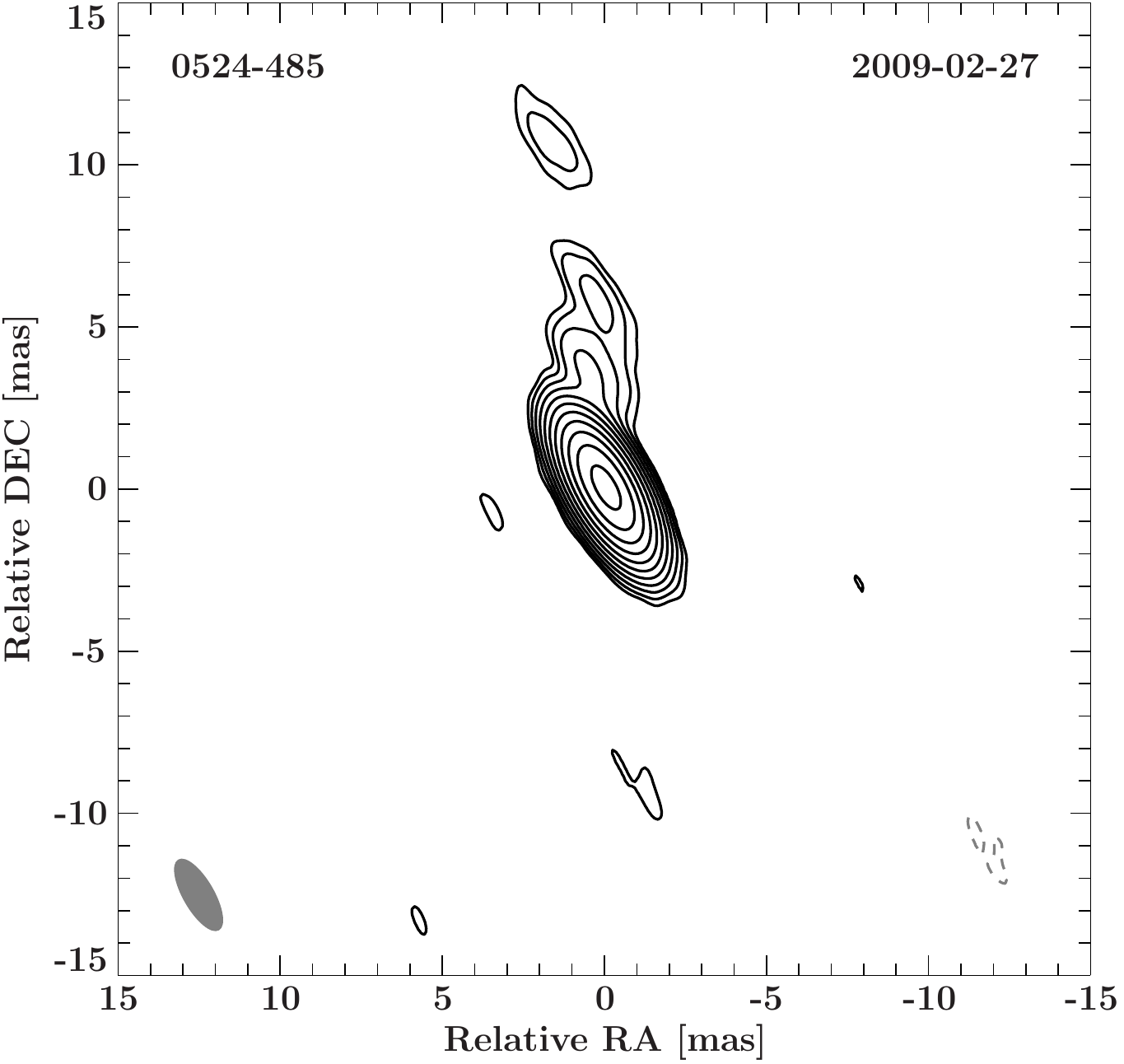}
\includegraphics[width=0.32\textwidth]{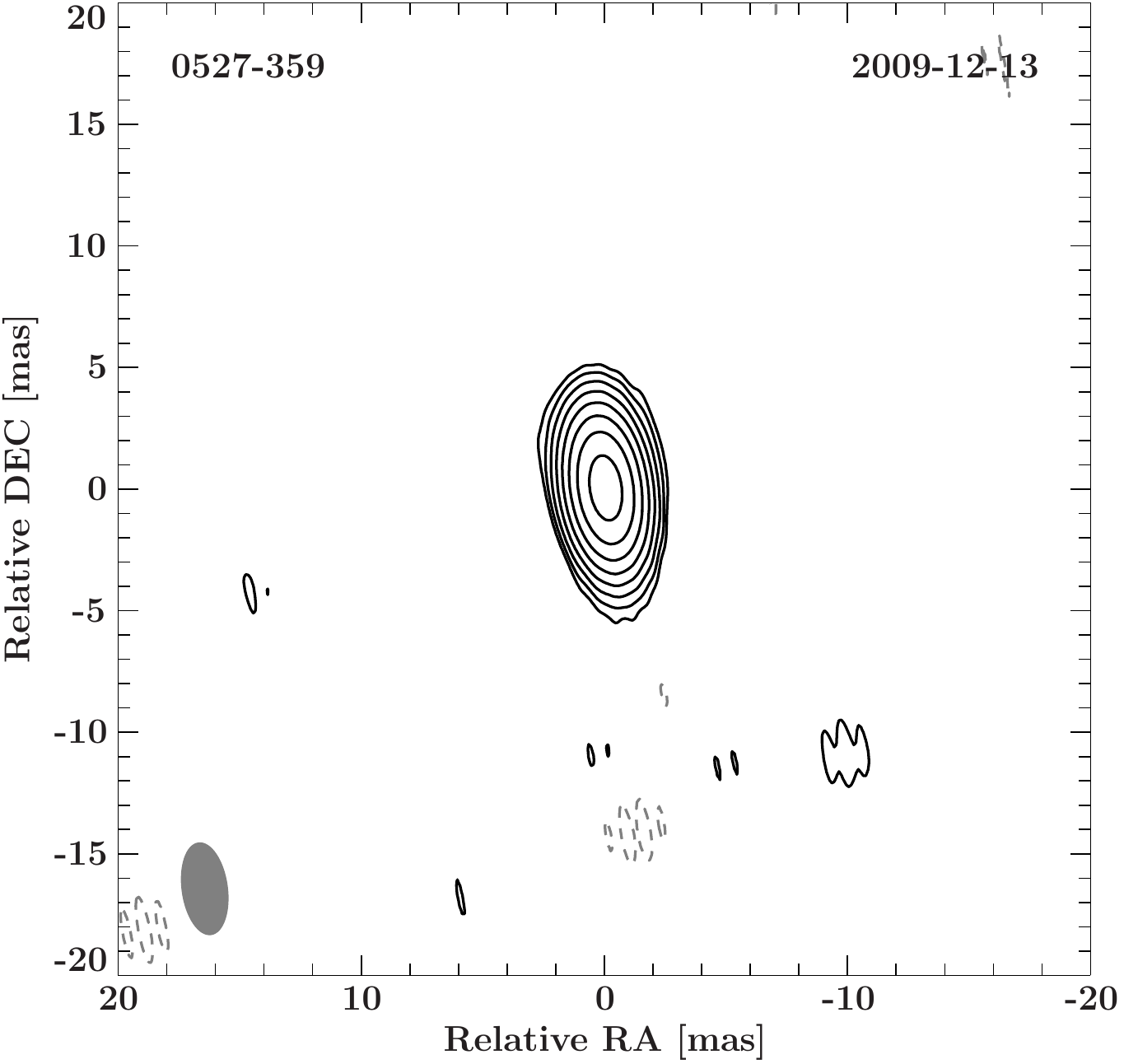}

\includegraphics[width=0.32\textwidth]{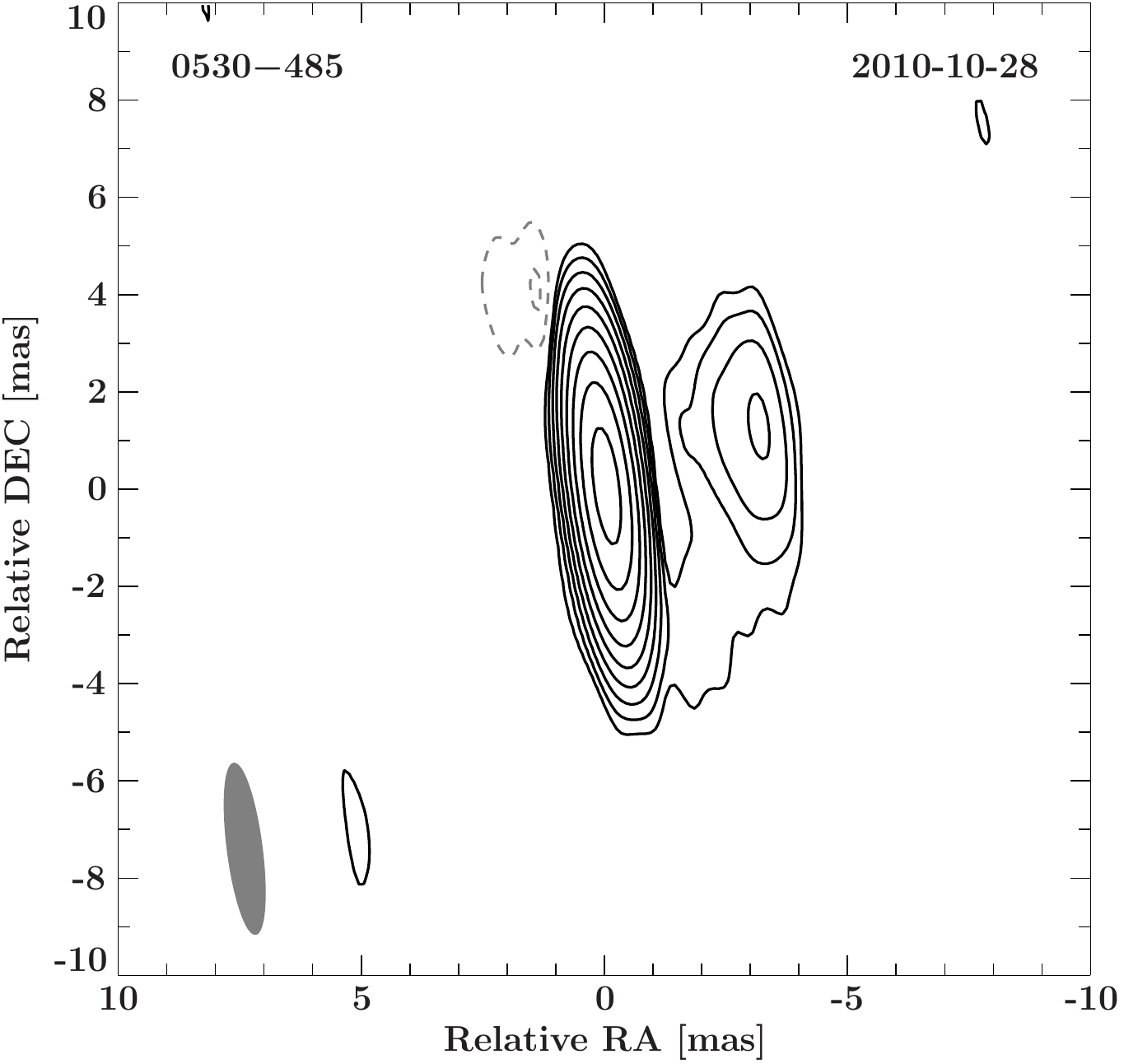}
\includegraphics[width=0.32\textwidth]{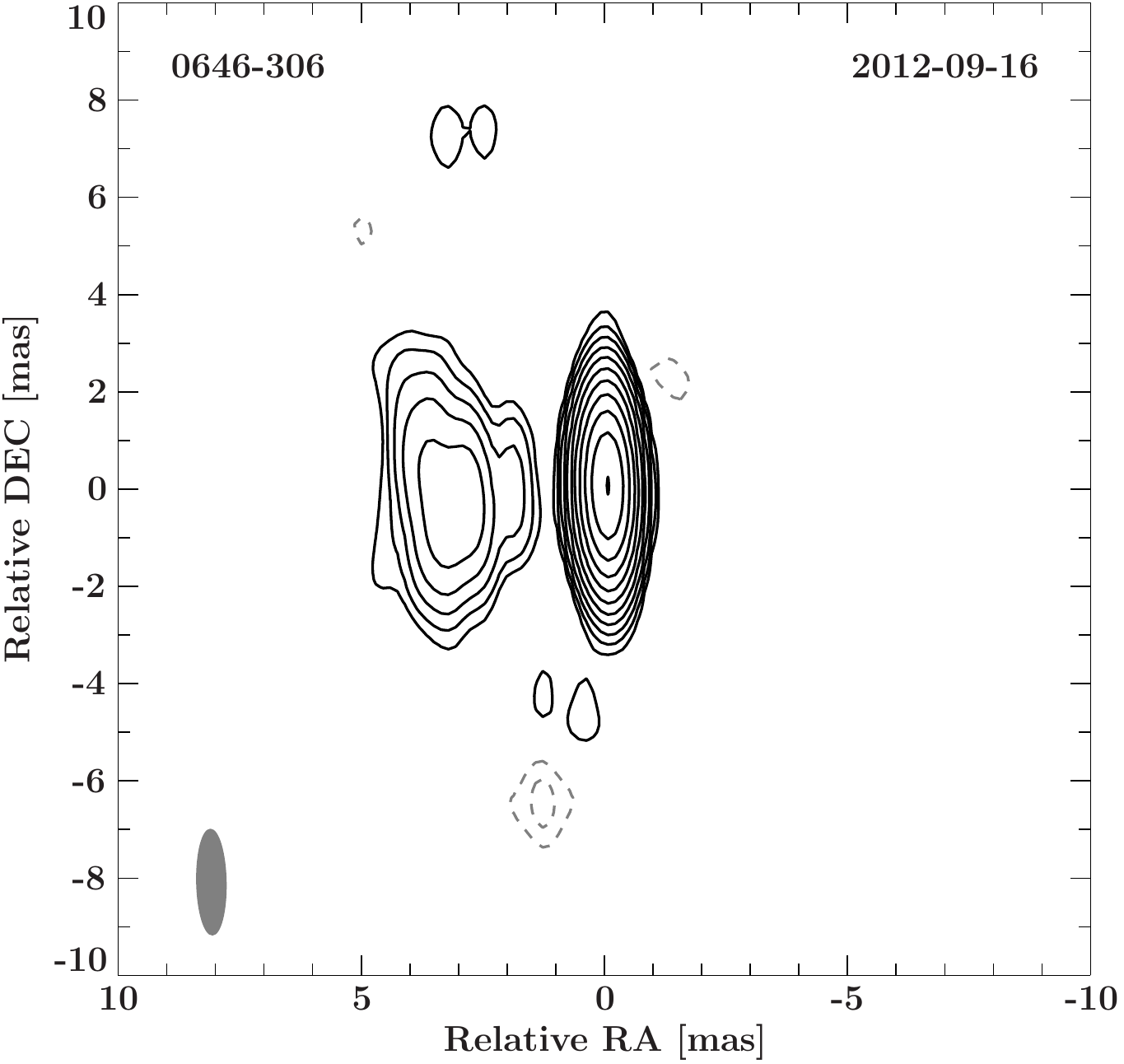}
\includegraphics[width=0.32\textwidth]{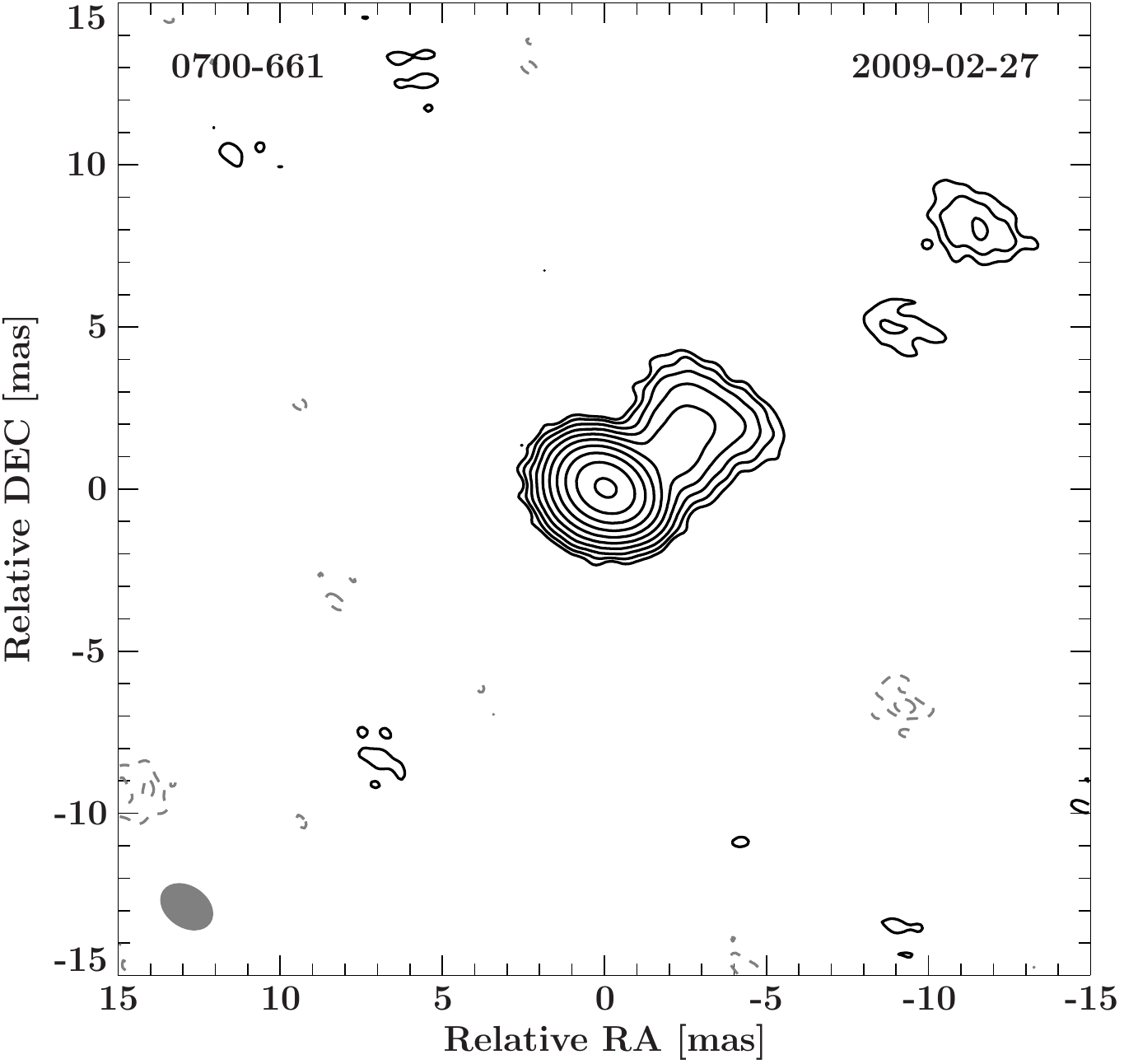}

\includegraphics[width=0.32\textwidth]{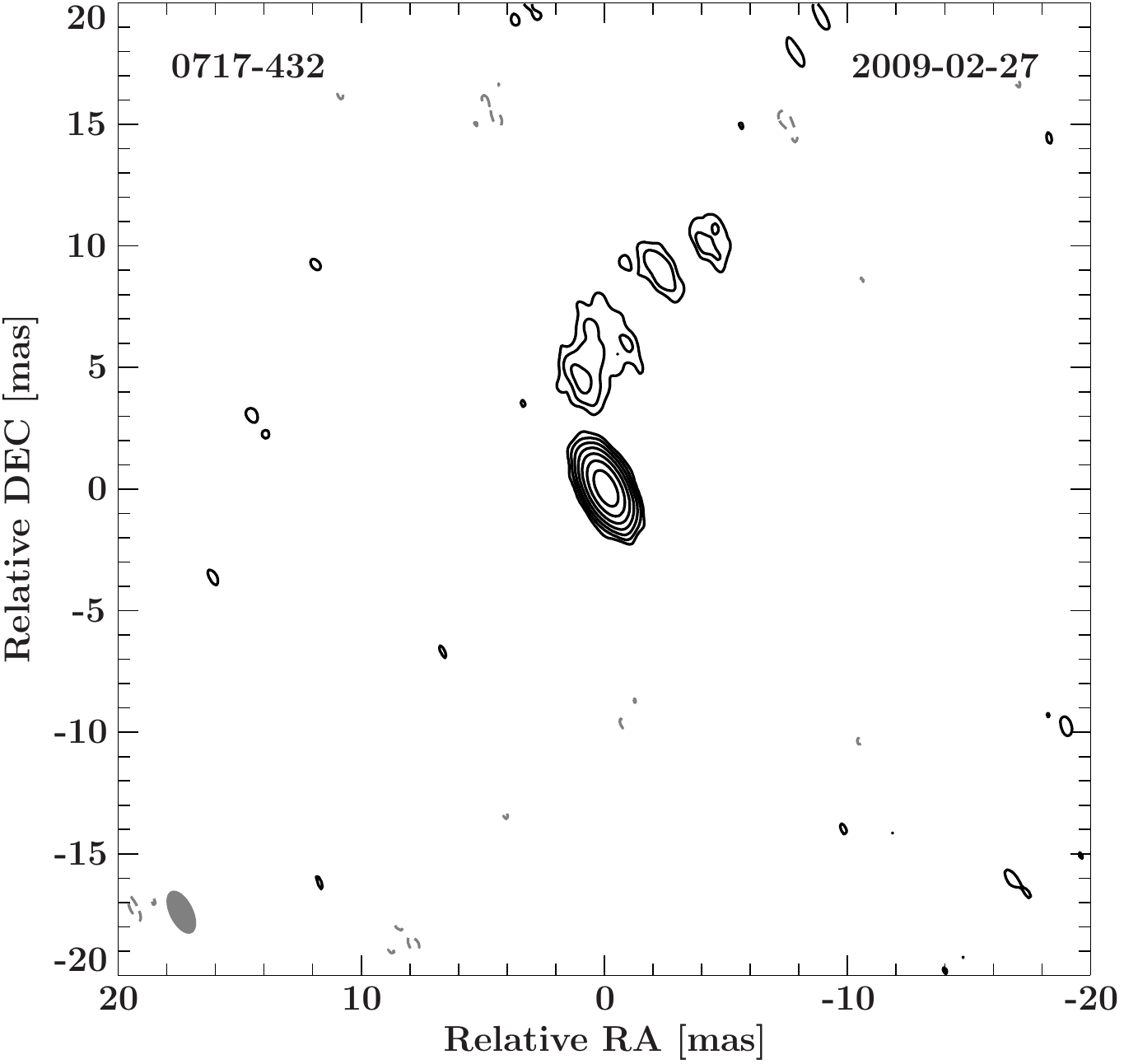}
\includegraphics[width=0.32\textwidth]{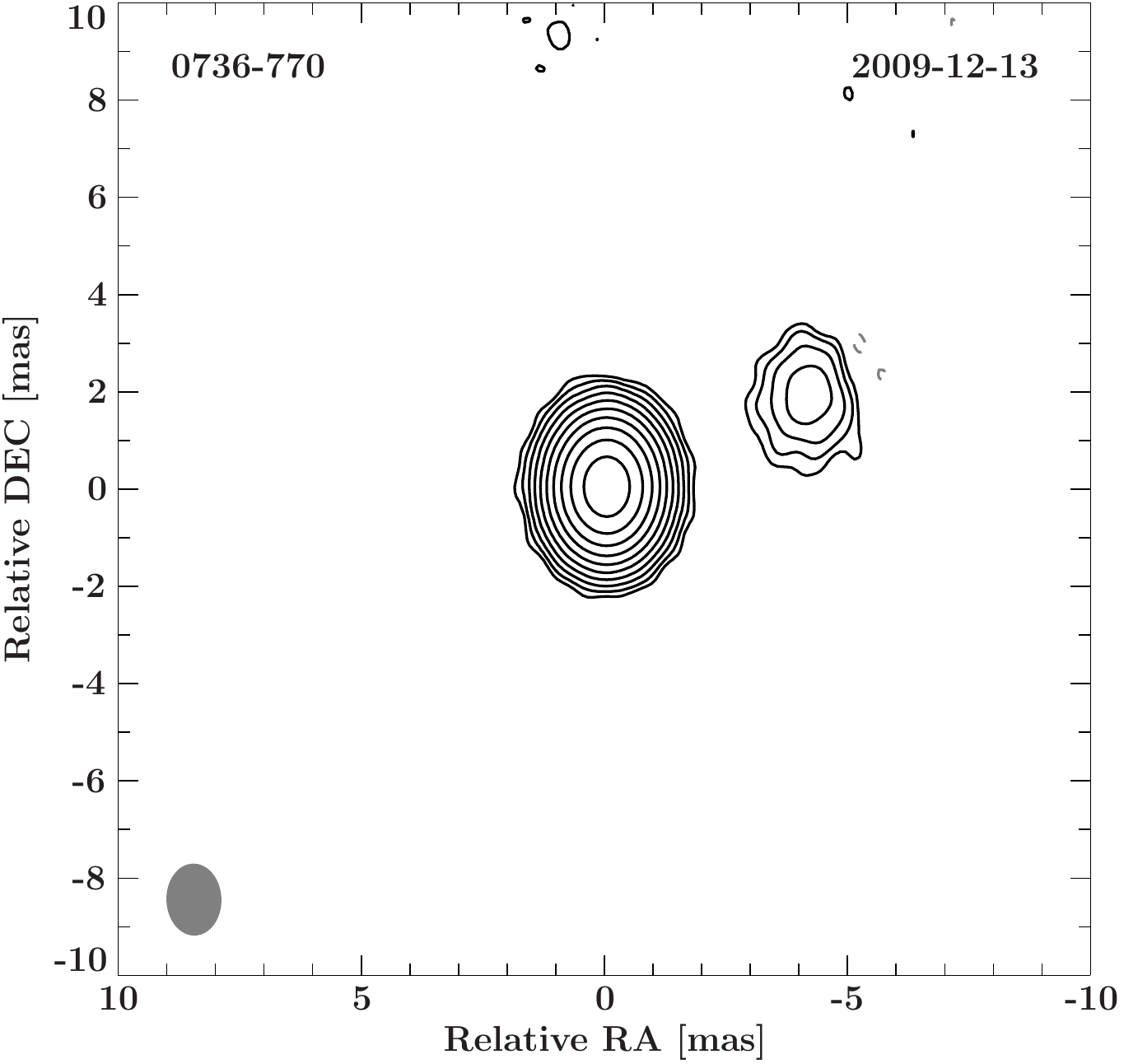}
\includegraphics[width=0.32\textwidth]{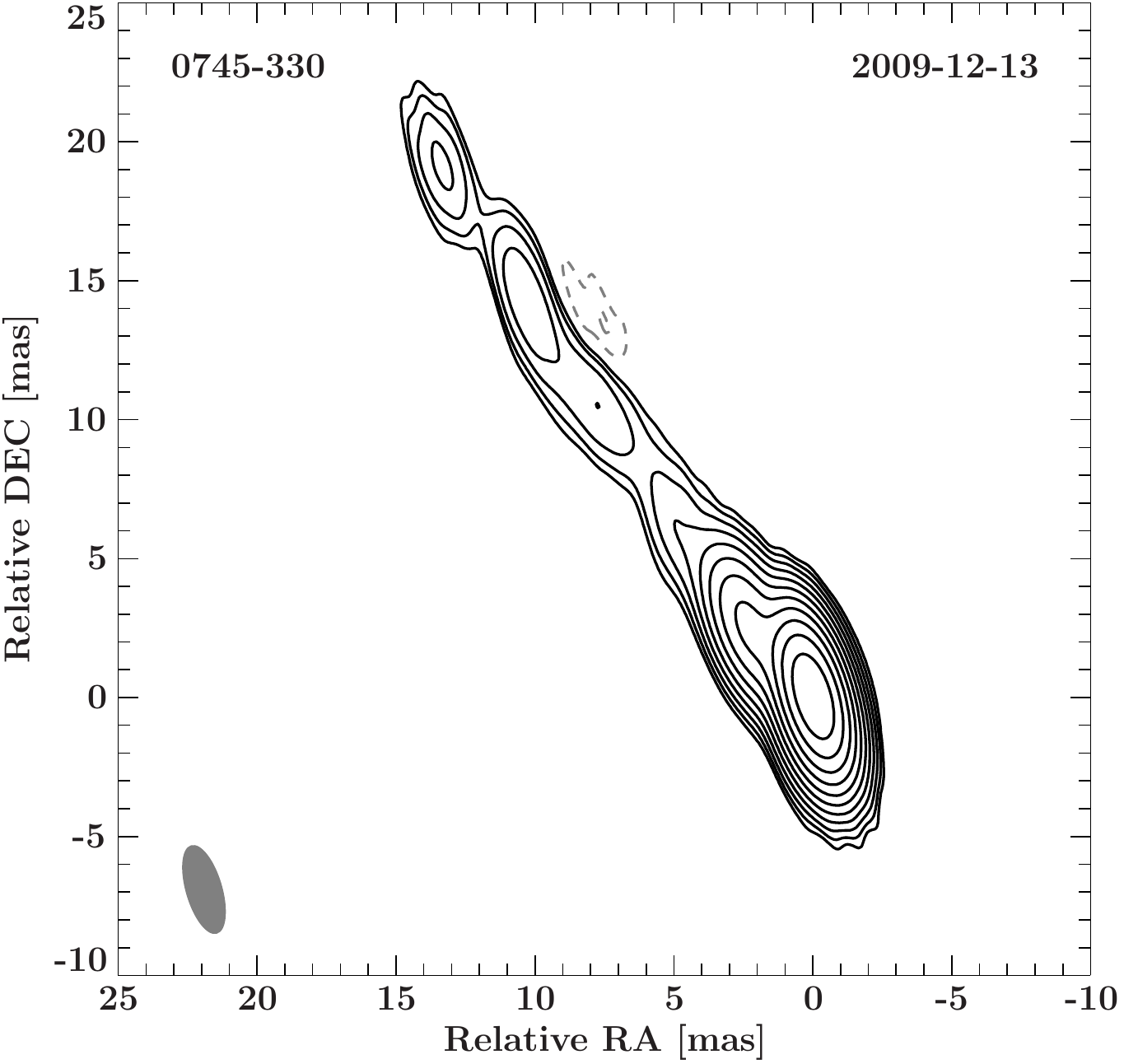}

\includegraphics[width=0.32\textwidth]{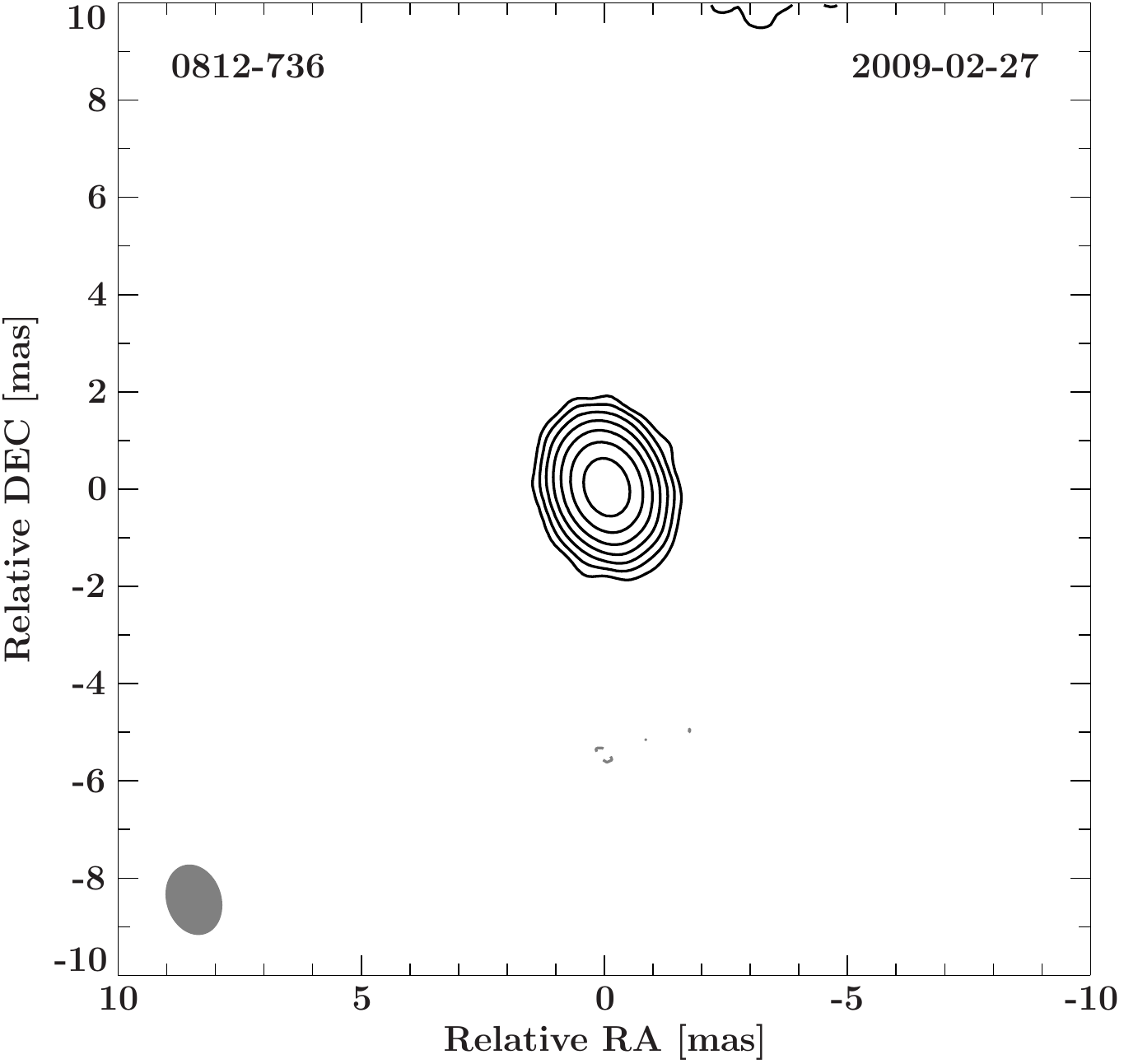}
\includegraphics[width=0.32\textwidth]{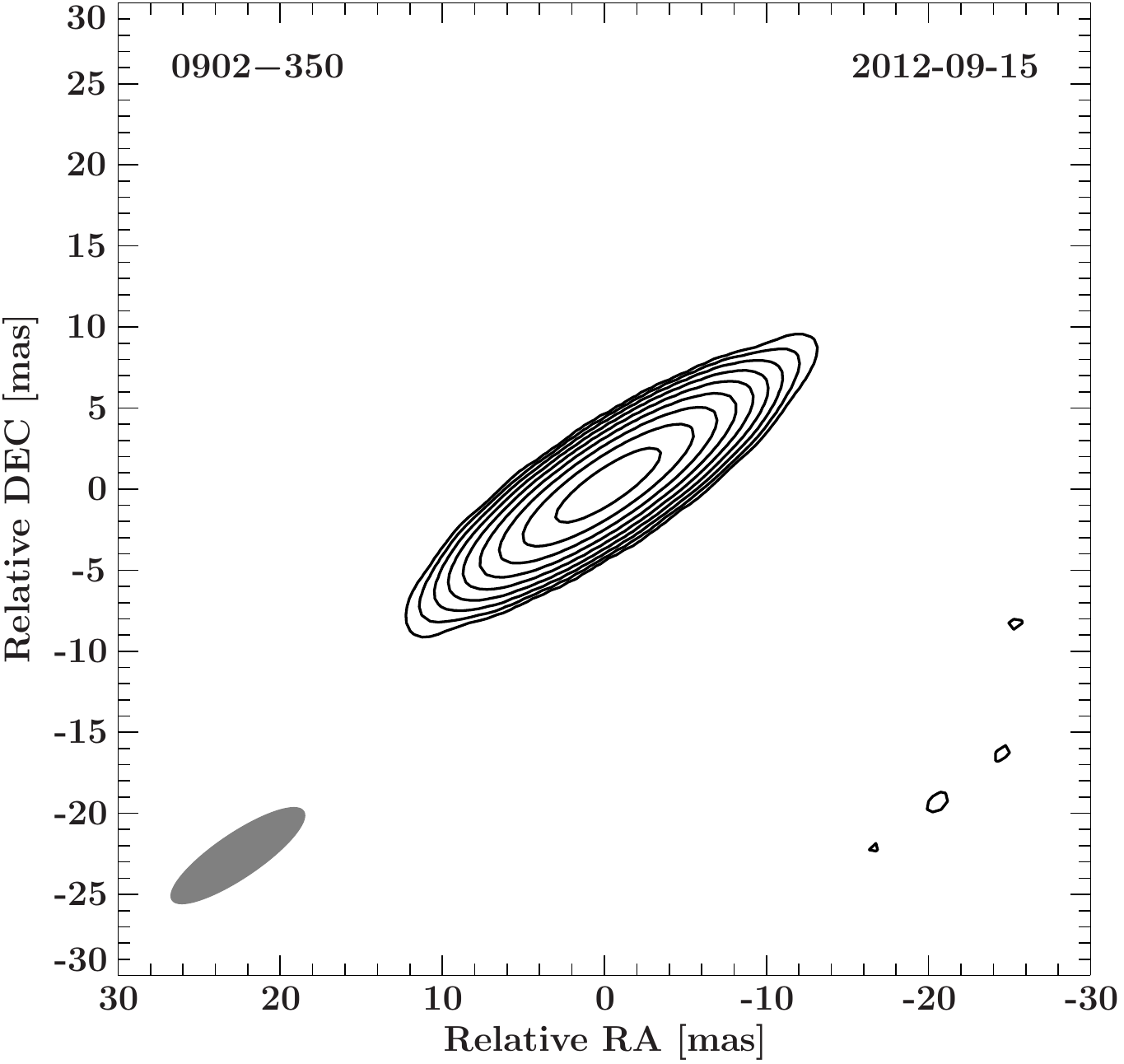}
\includegraphics[width=0.32\textwidth]{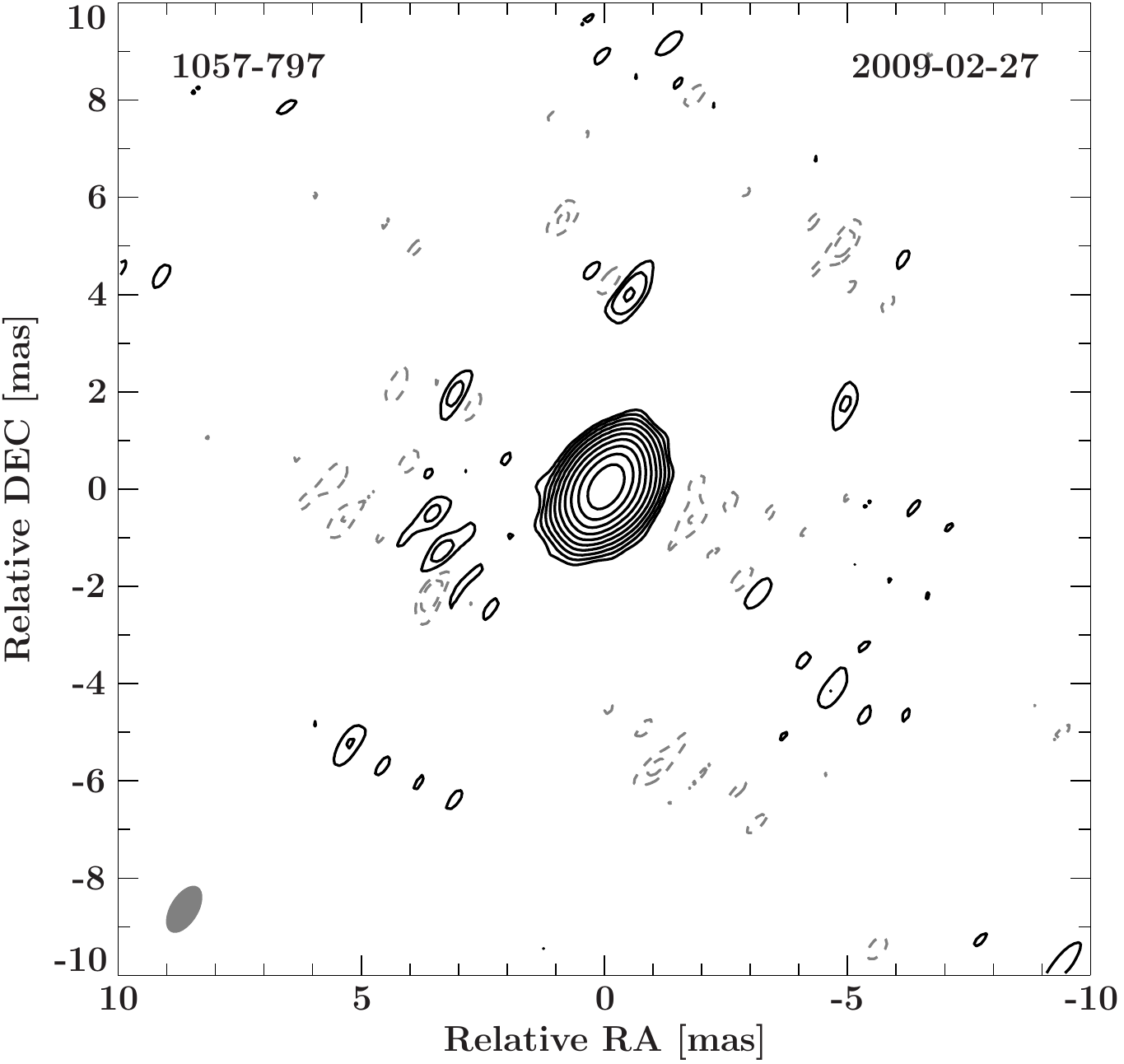}
\caption{First epoch 8.4\,GHz \texttt{clean} images of additional TANAMI
  sources --  Fig.~\ref{fig:newimages1} continued. Source parameters
  are provided in Table~\ref{table:images}.}
\label{fig:newimages2}
\end{figure*}

\begin{figure*}
\centering
\includegraphics[width=0.32\textwidth]{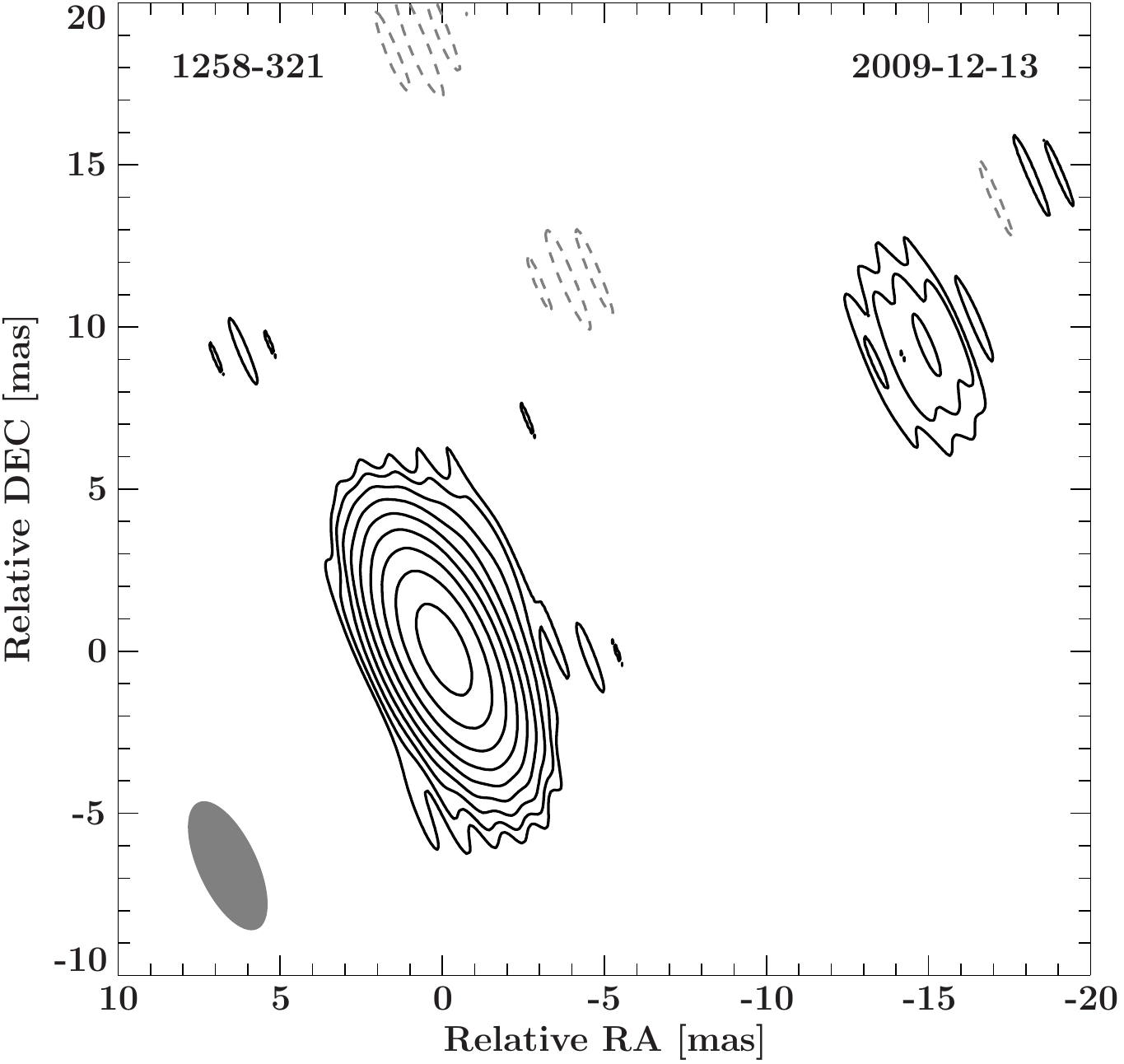}
\includegraphics[width=0.32\textwidth]{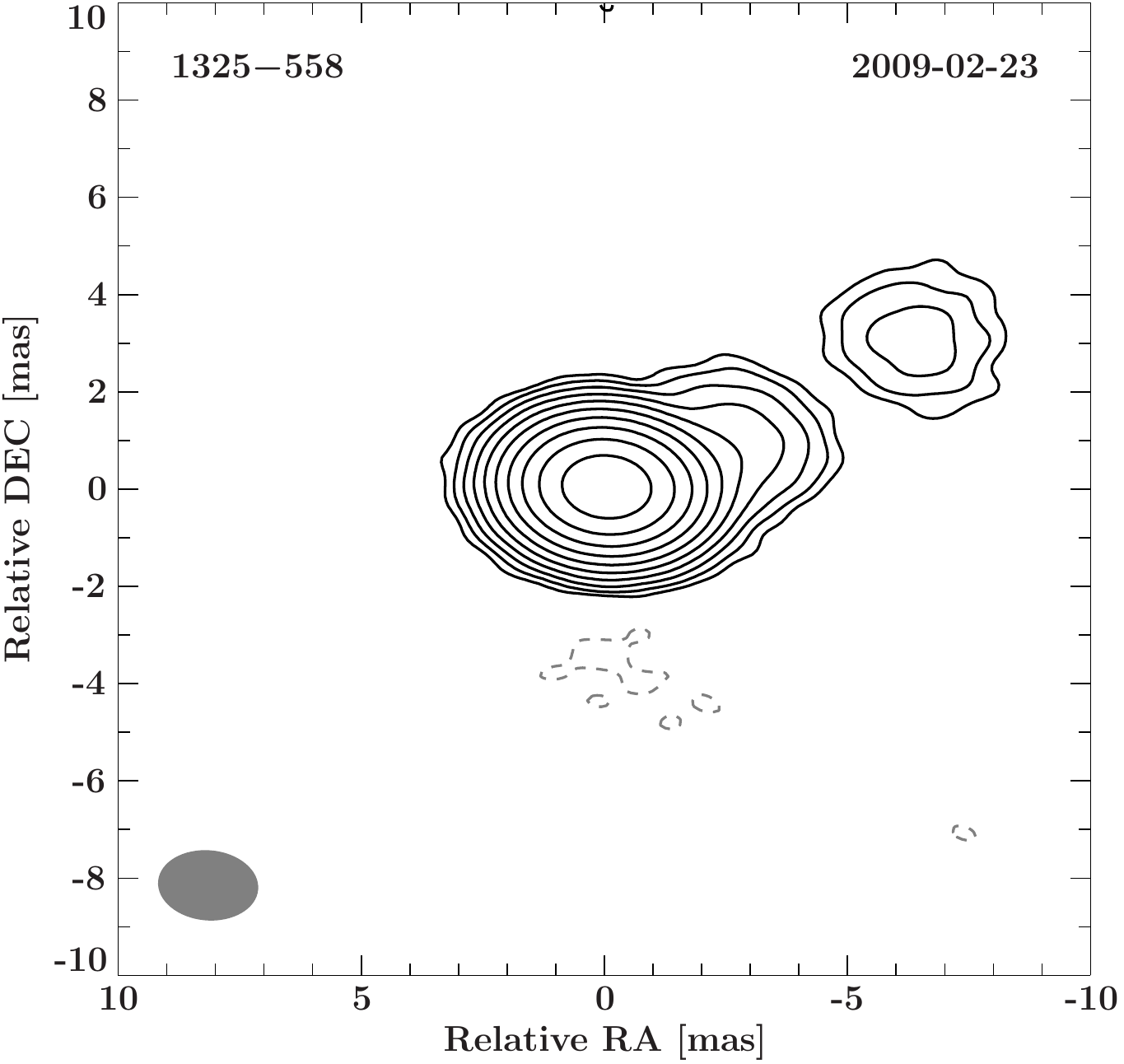}
\includegraphics[width=0.32\textwidth]{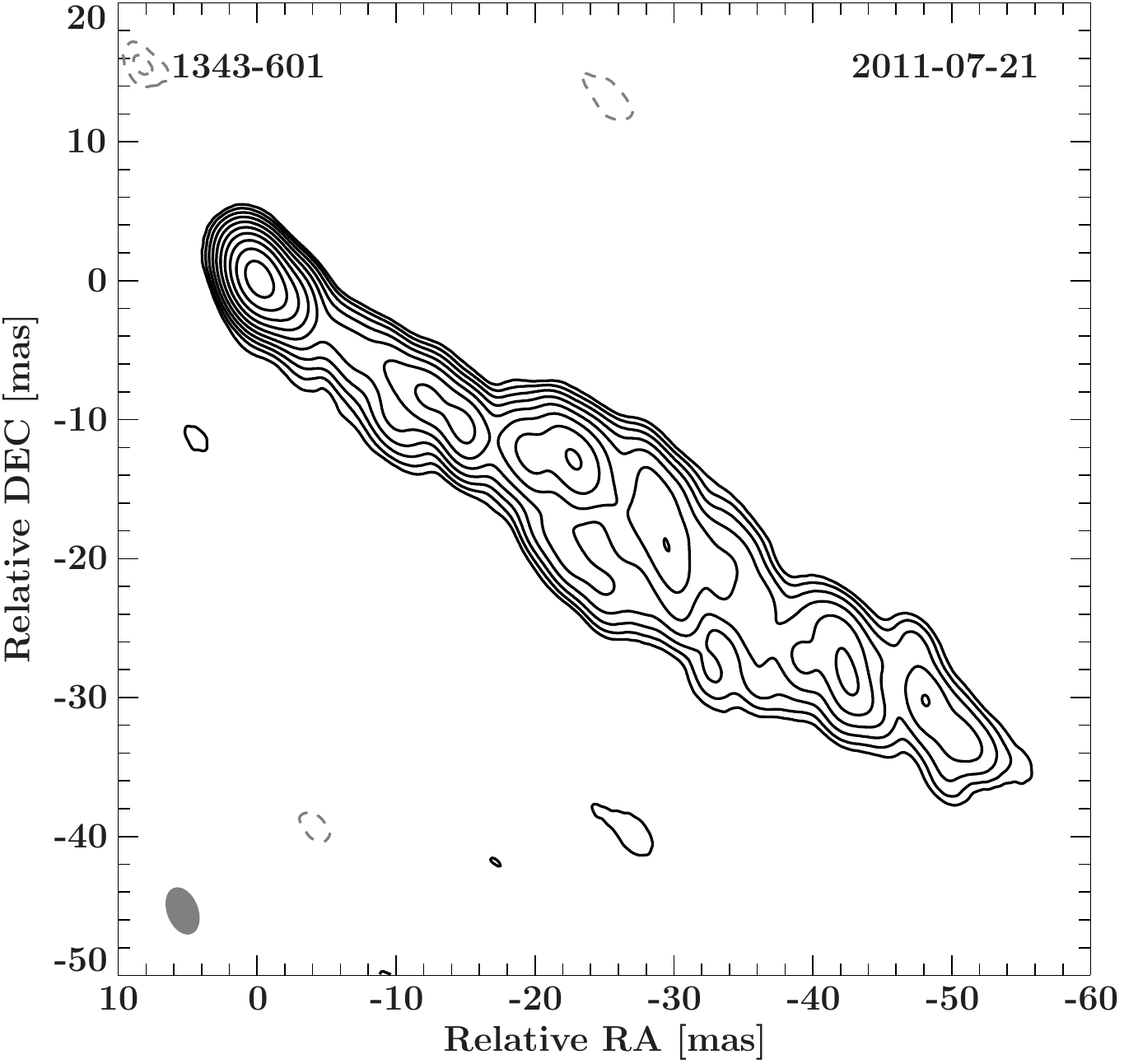}

\includegraphics[width=0.32\textwidth]{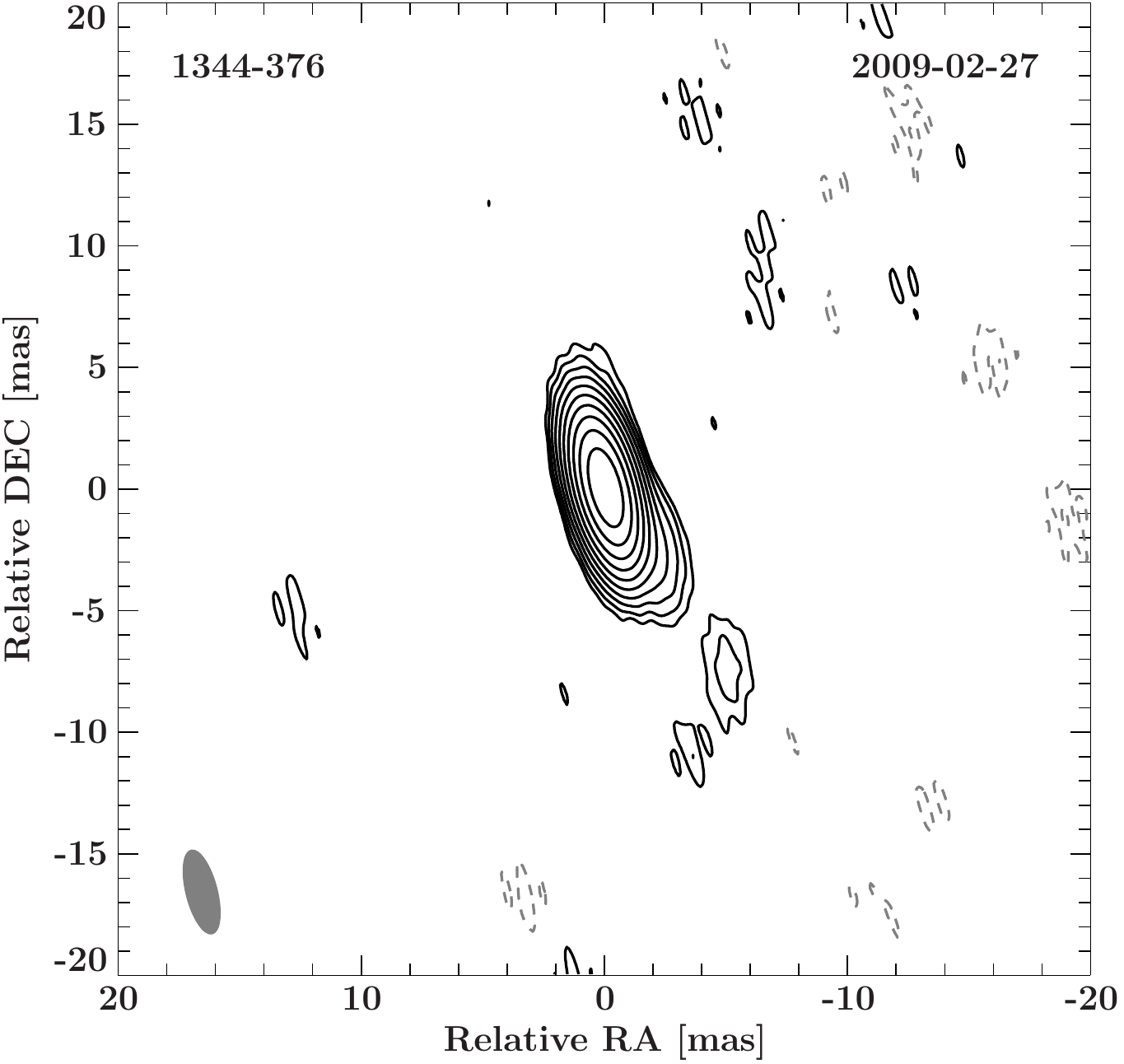}
\includegraphics[width=0.32\textwidth]{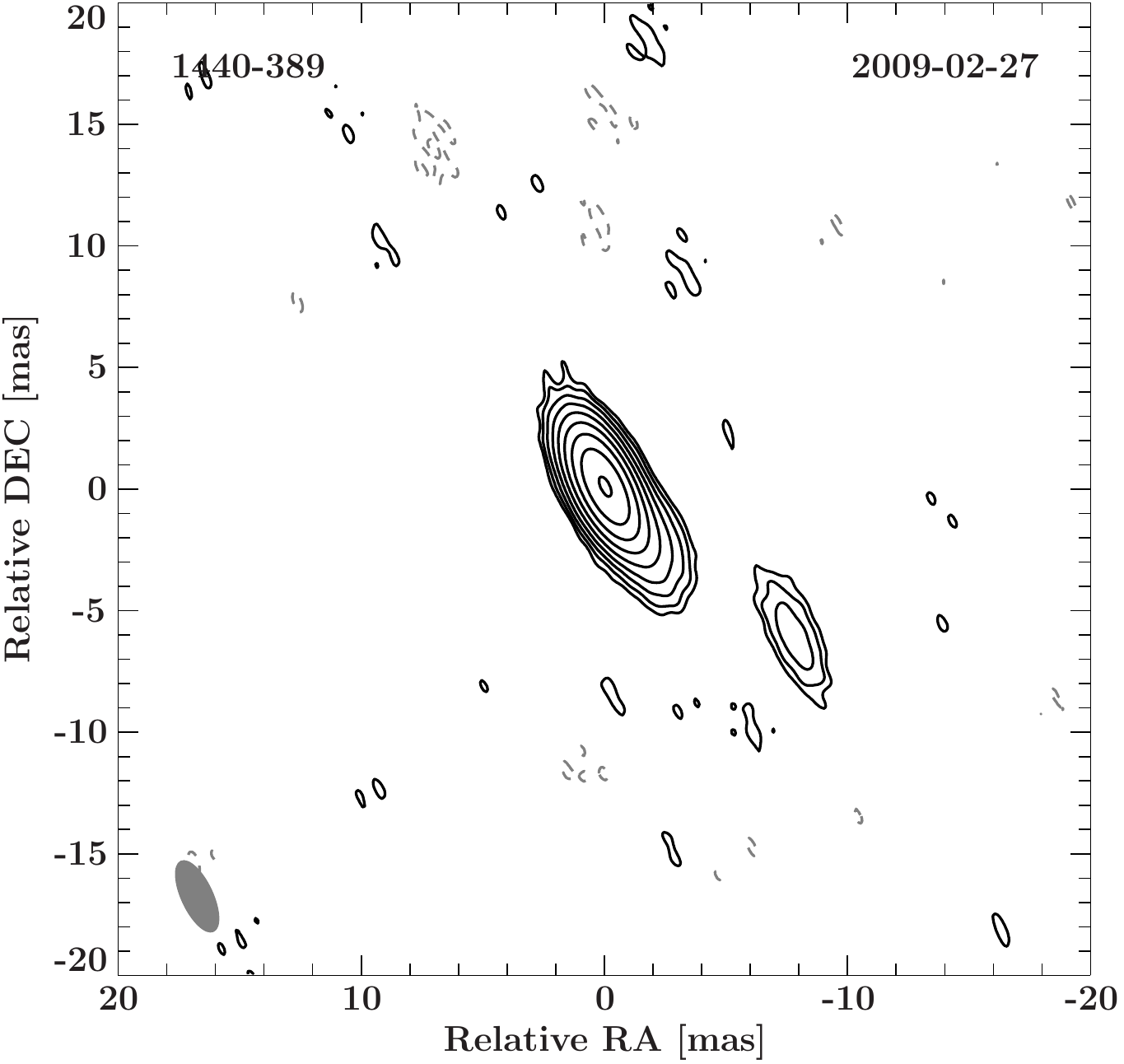}
\includegraphics[width=0.32\textwidth]{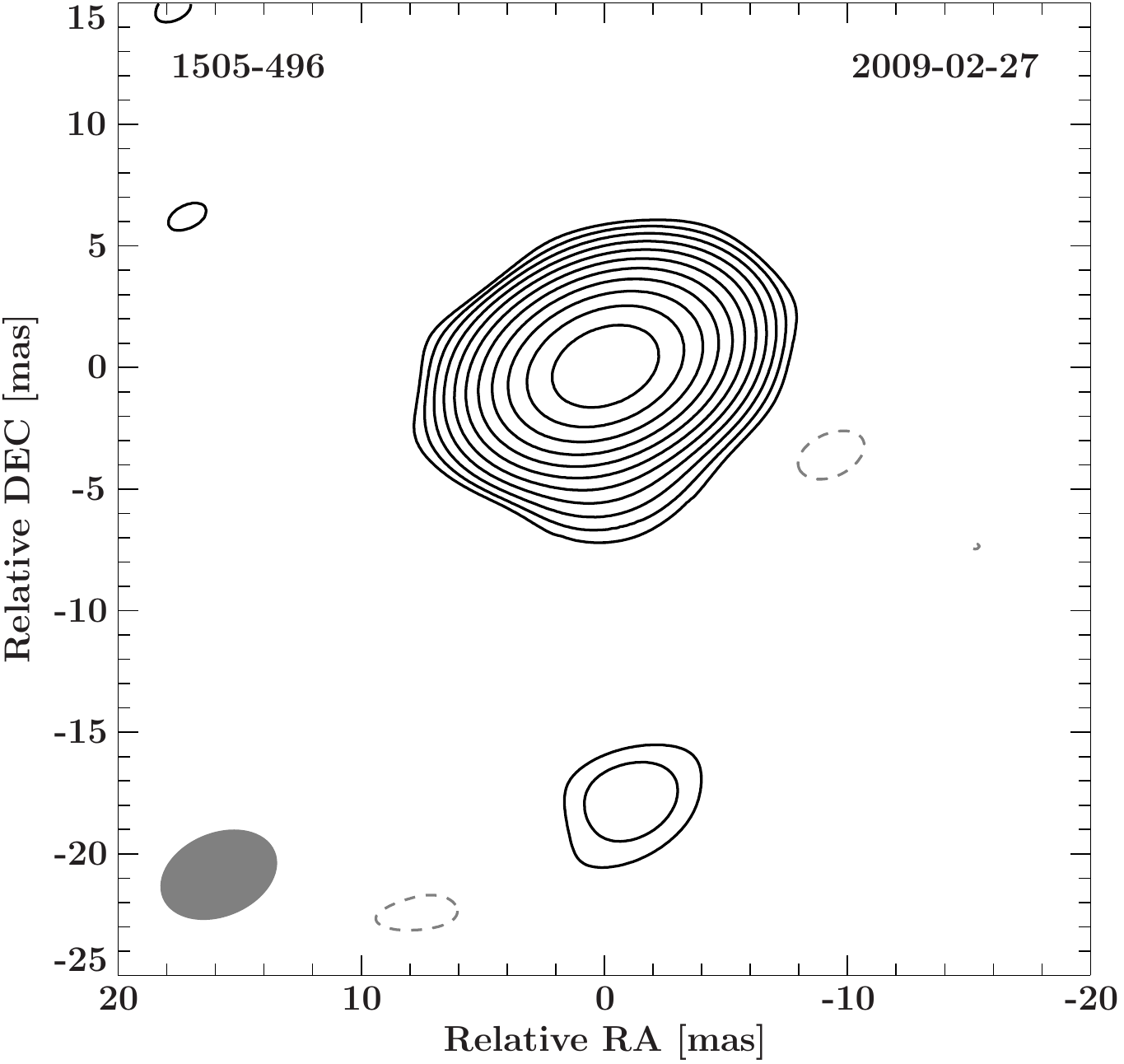}

\includegraphics[width=0.32\textwidth]{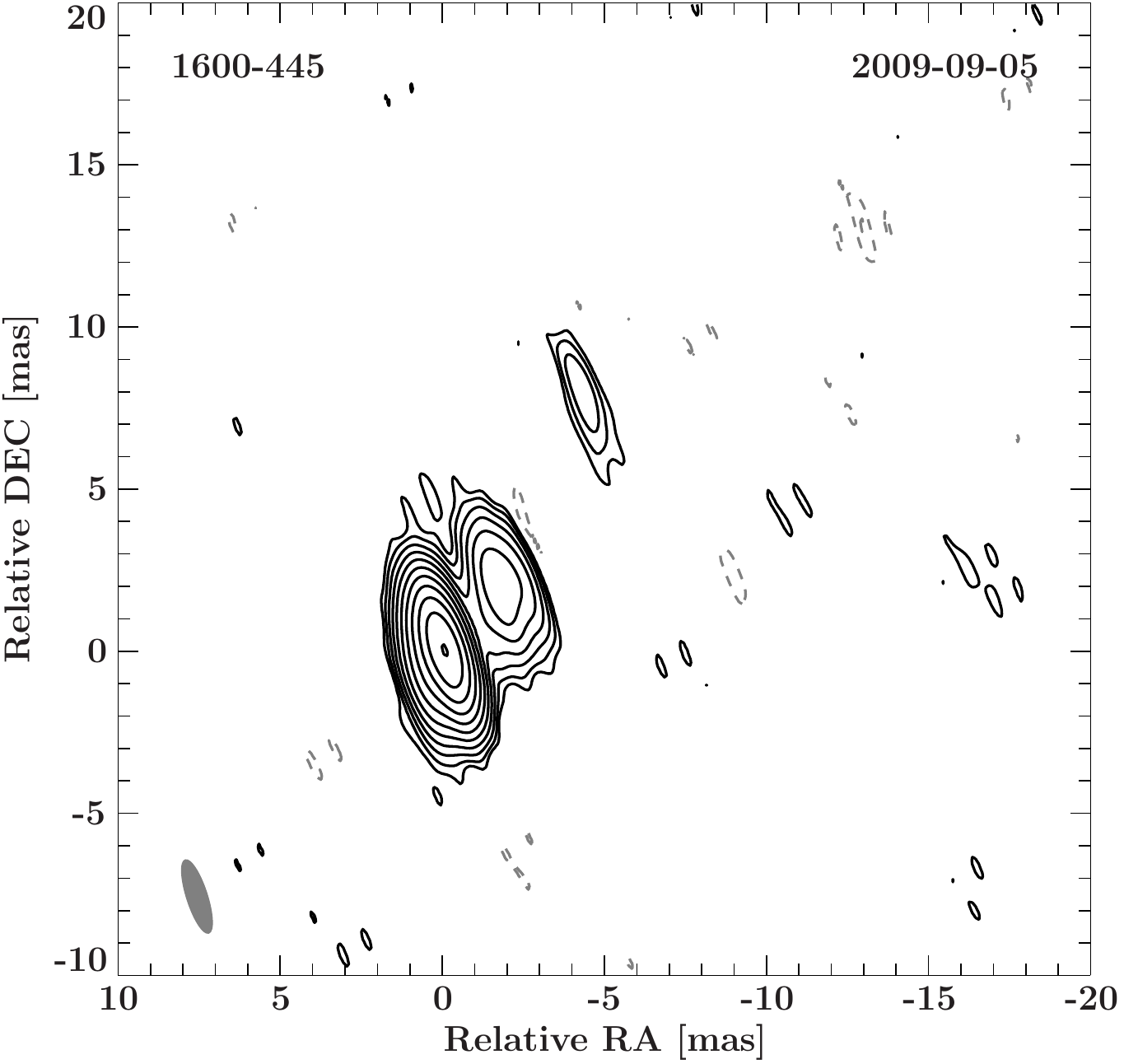}
\includegraphics[width=0.32\textwidth]{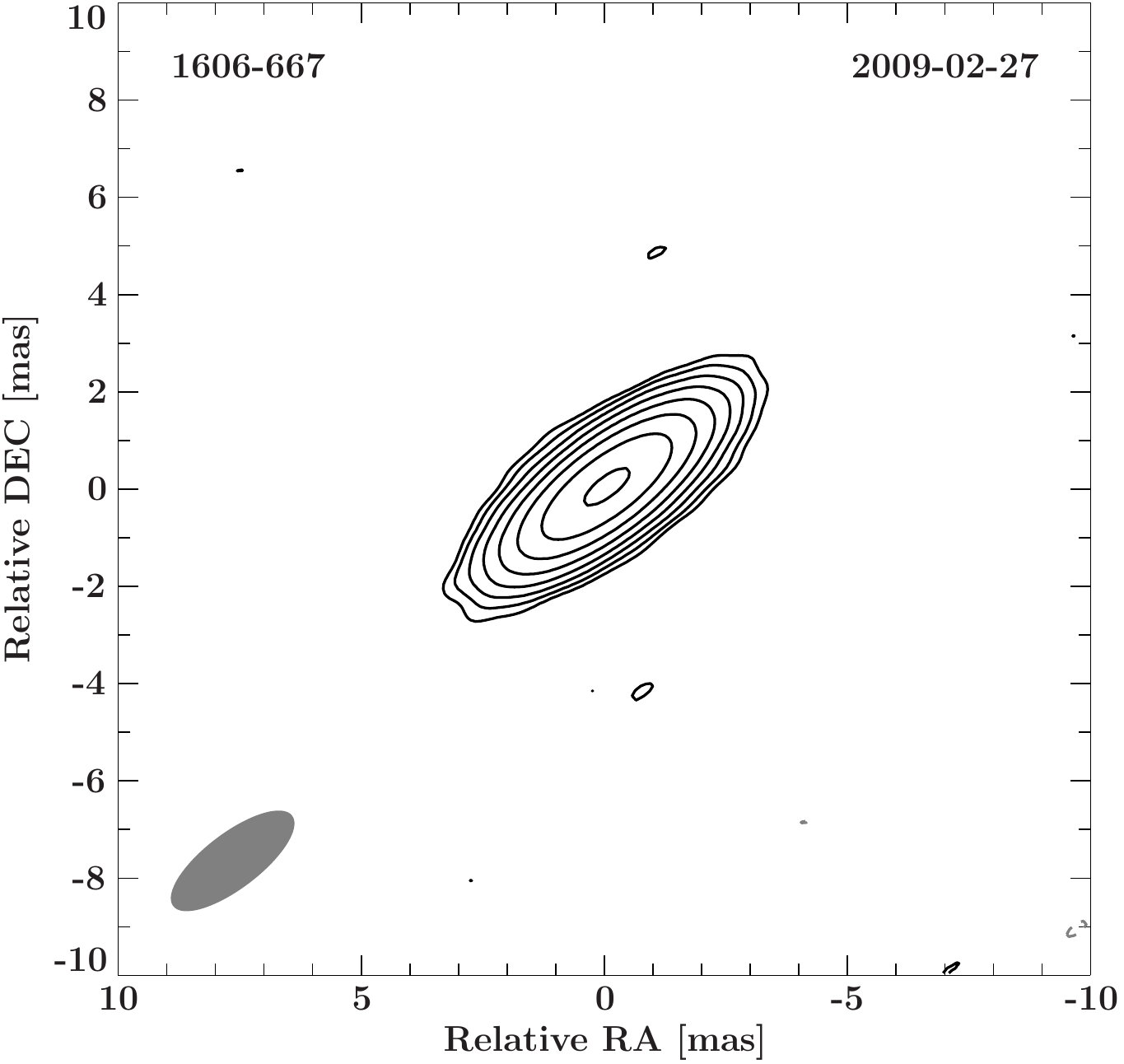}
\includegraphics[width=0.32\textwidth]{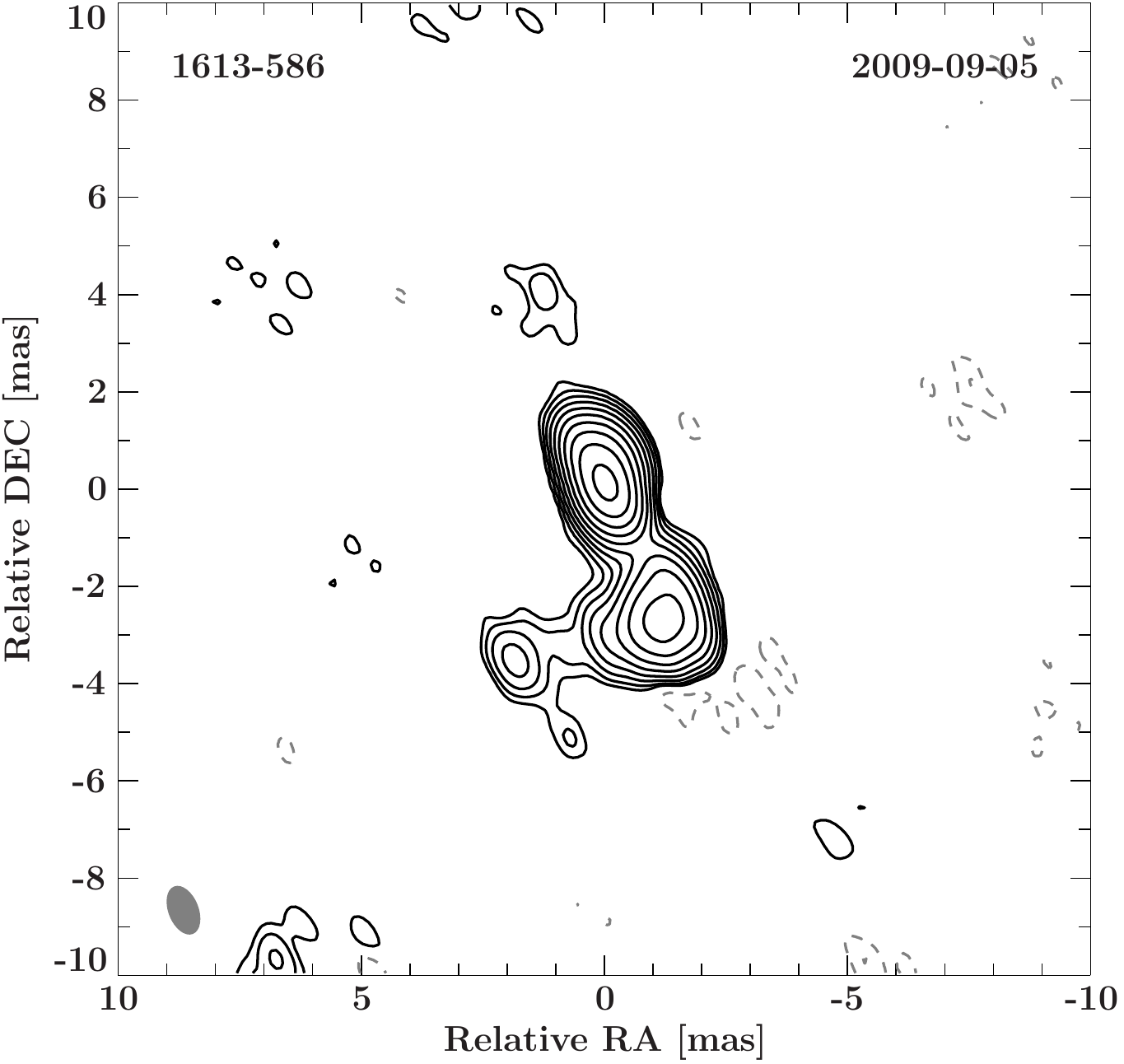}

\includegraphics[width=0.32\textwidth]{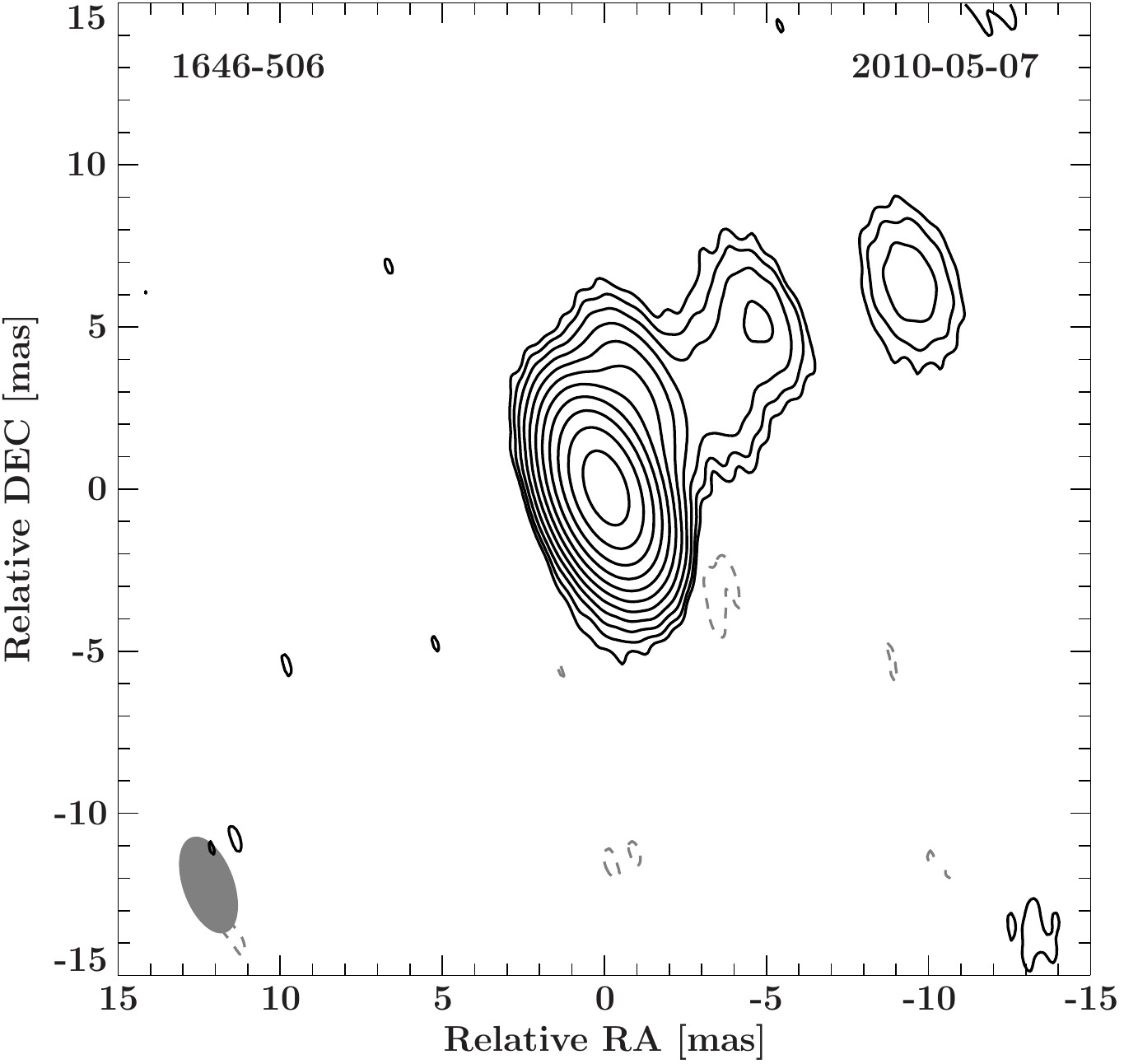}
\includegraphics[width=0.32\textwidth]{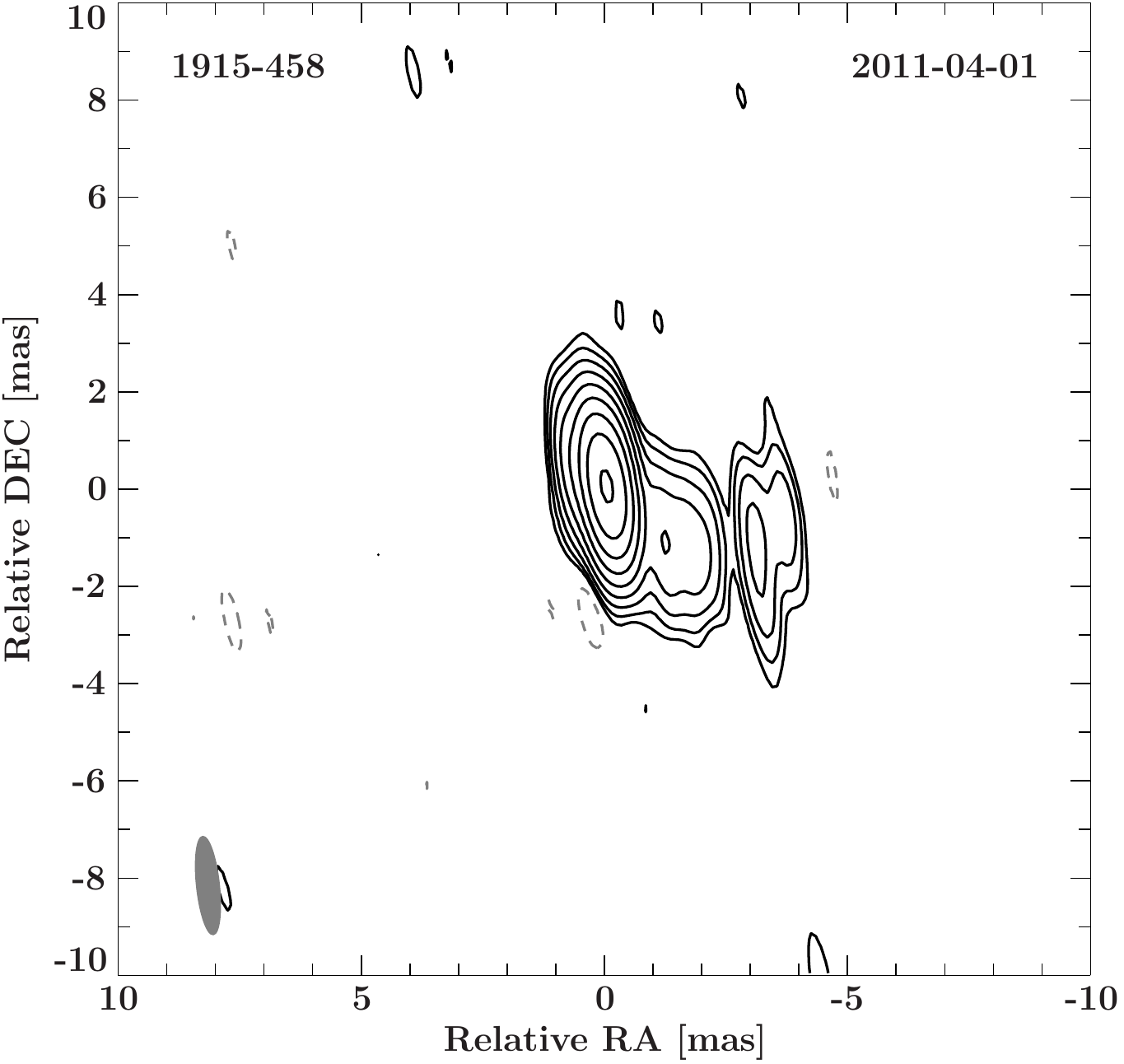}
\includegraphics[width=0.32\textwidth]{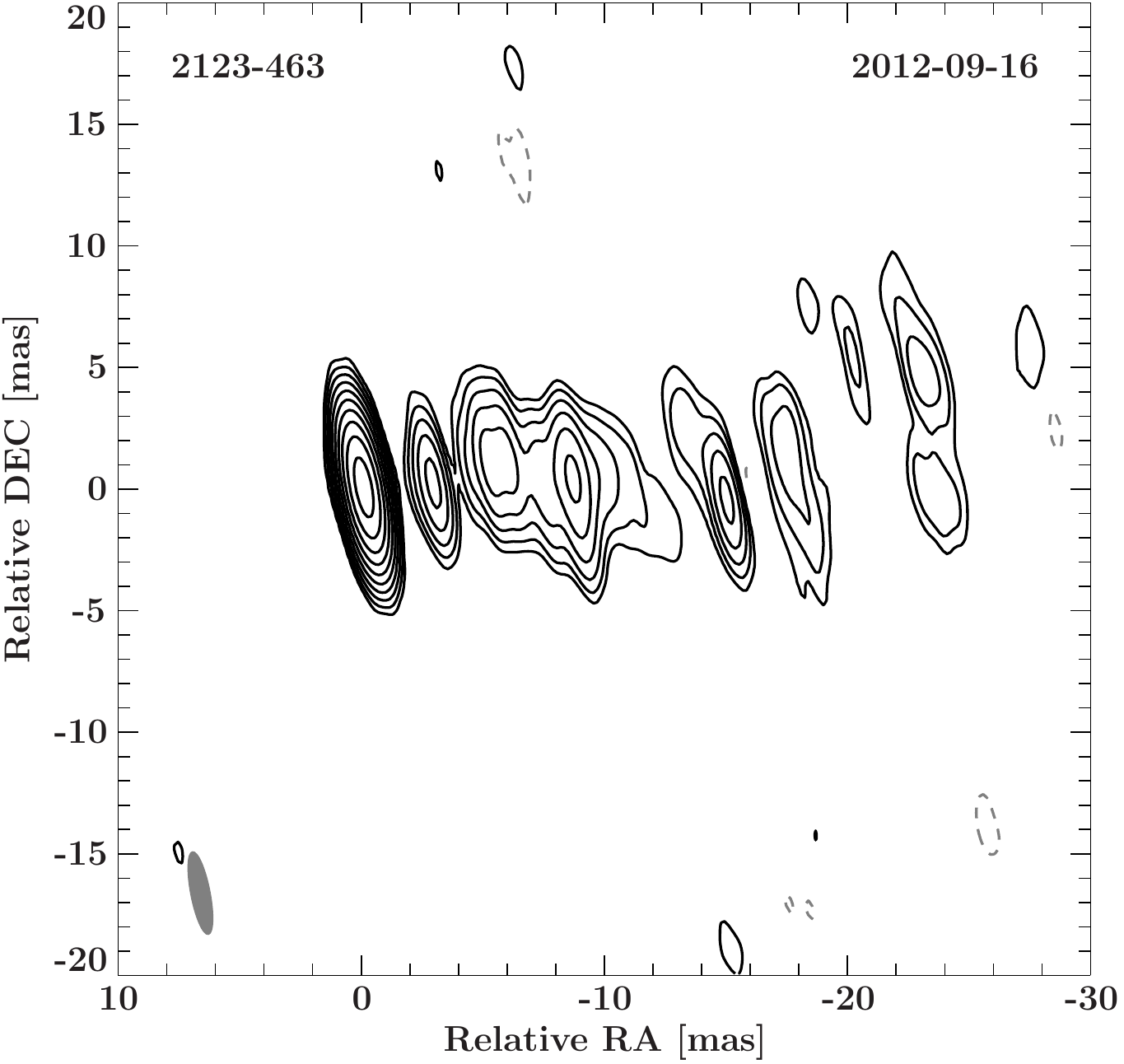}
\caption{First epoch 8.4\,GHz \texttt{clean} images of additional TANAMI
  sources --  Fig.~\ref{fig:newimages1} continued. Source parameters
  are provided in Table~\ref{table:images}.}
\label{fig:newimages3}

\end{figure*}

\begin{figure*}
\centering
\includegraphics[width=0.32\textwidth]{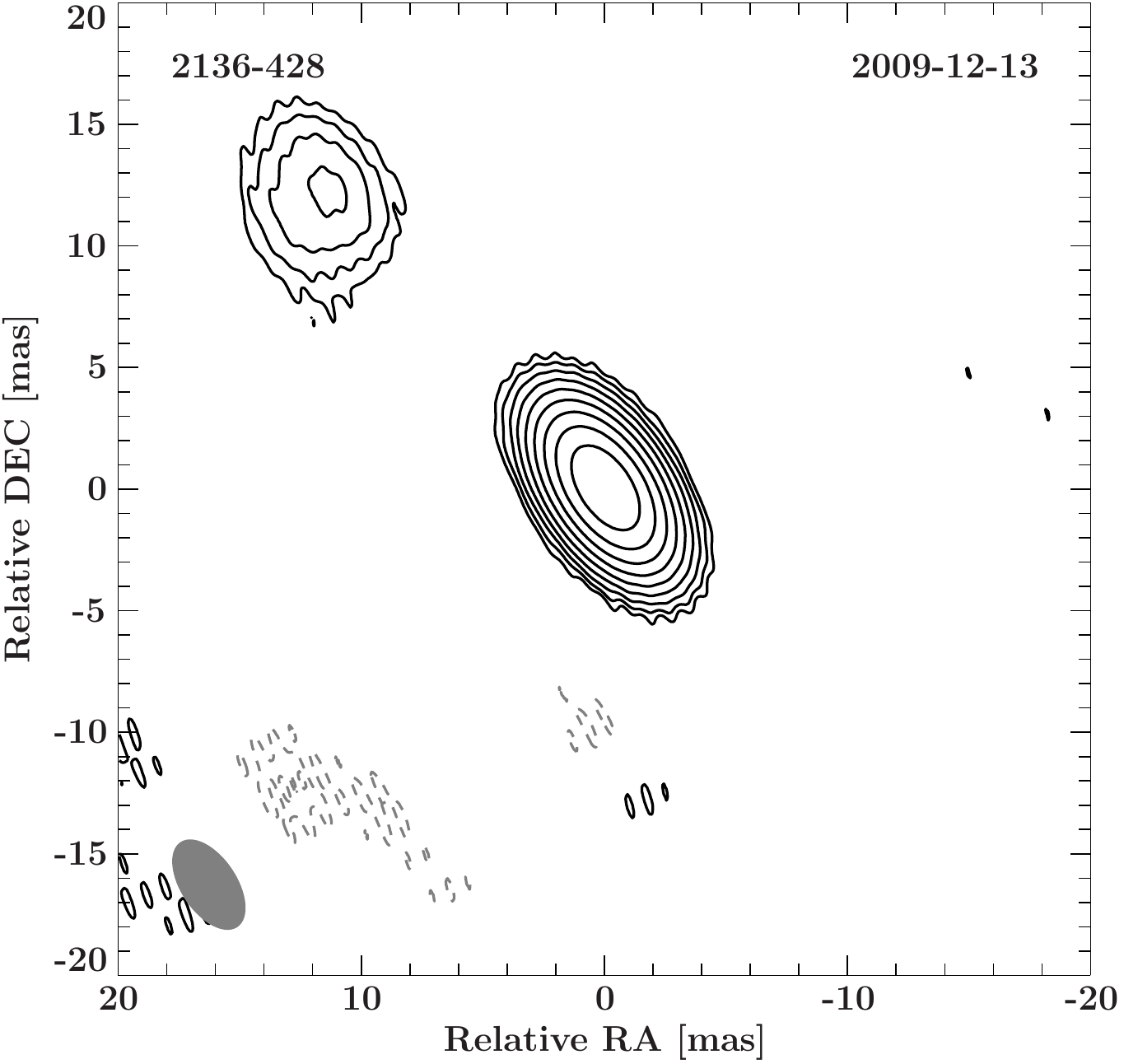}
\includegraphics[width=0.32\textwidth]{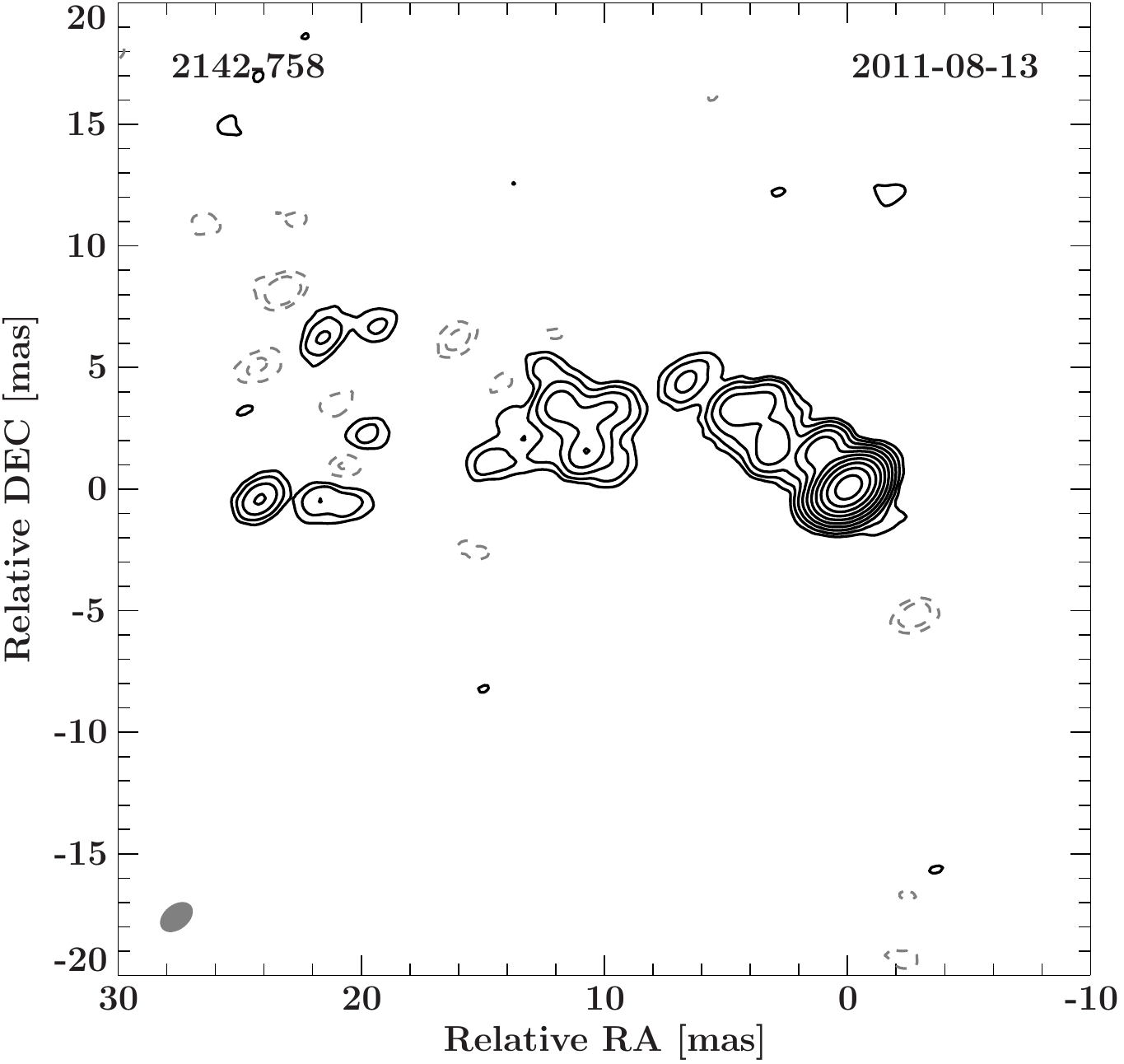}
\includegraphics[width=0.32\textwidth]{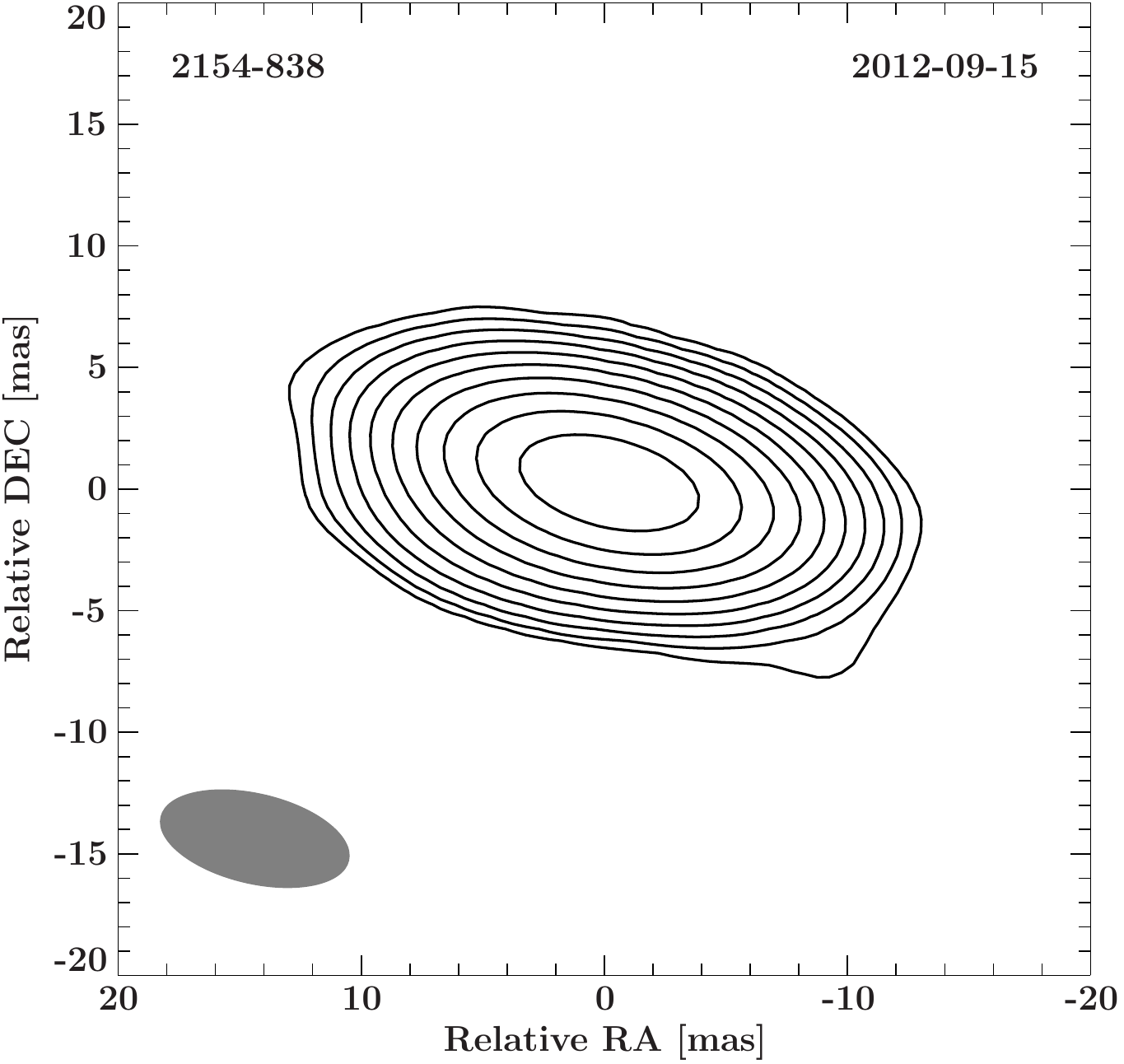}
\caption{First epoch 8.4\,GHz \texttt{clean} images of additional TANAMI
  sources --
  Fig.~\ref{fig:newimages1} continued. Source parameters
  are provided in Table~\ref{table:images}.}
\label{fig:newimages4}
\end{figure*}

\begin{figure}
\centering
\includegraphics[width=0.4\textwidth, angle=270]{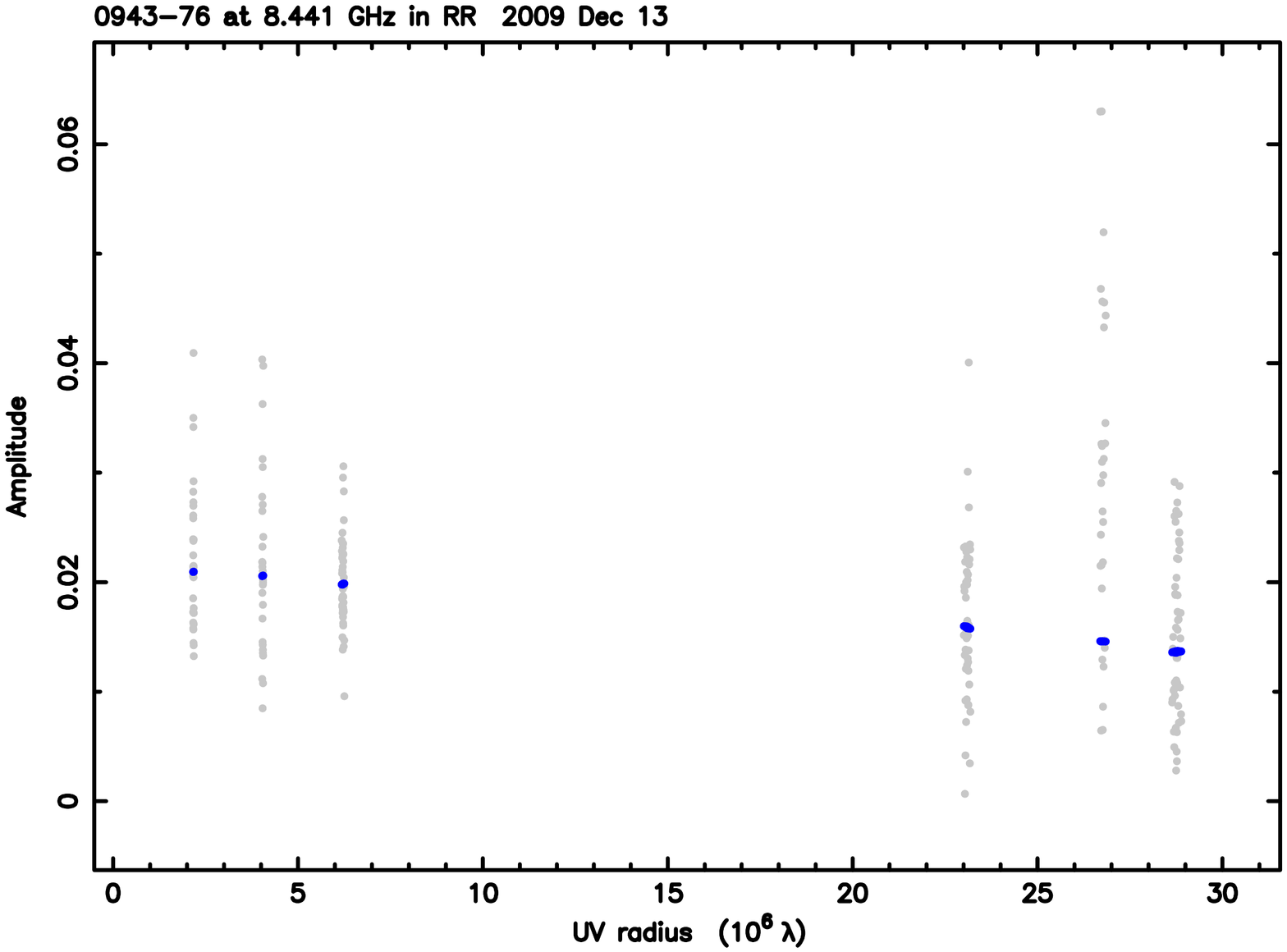}
\includegraphics[width=0.4\textwidth, angle=270]{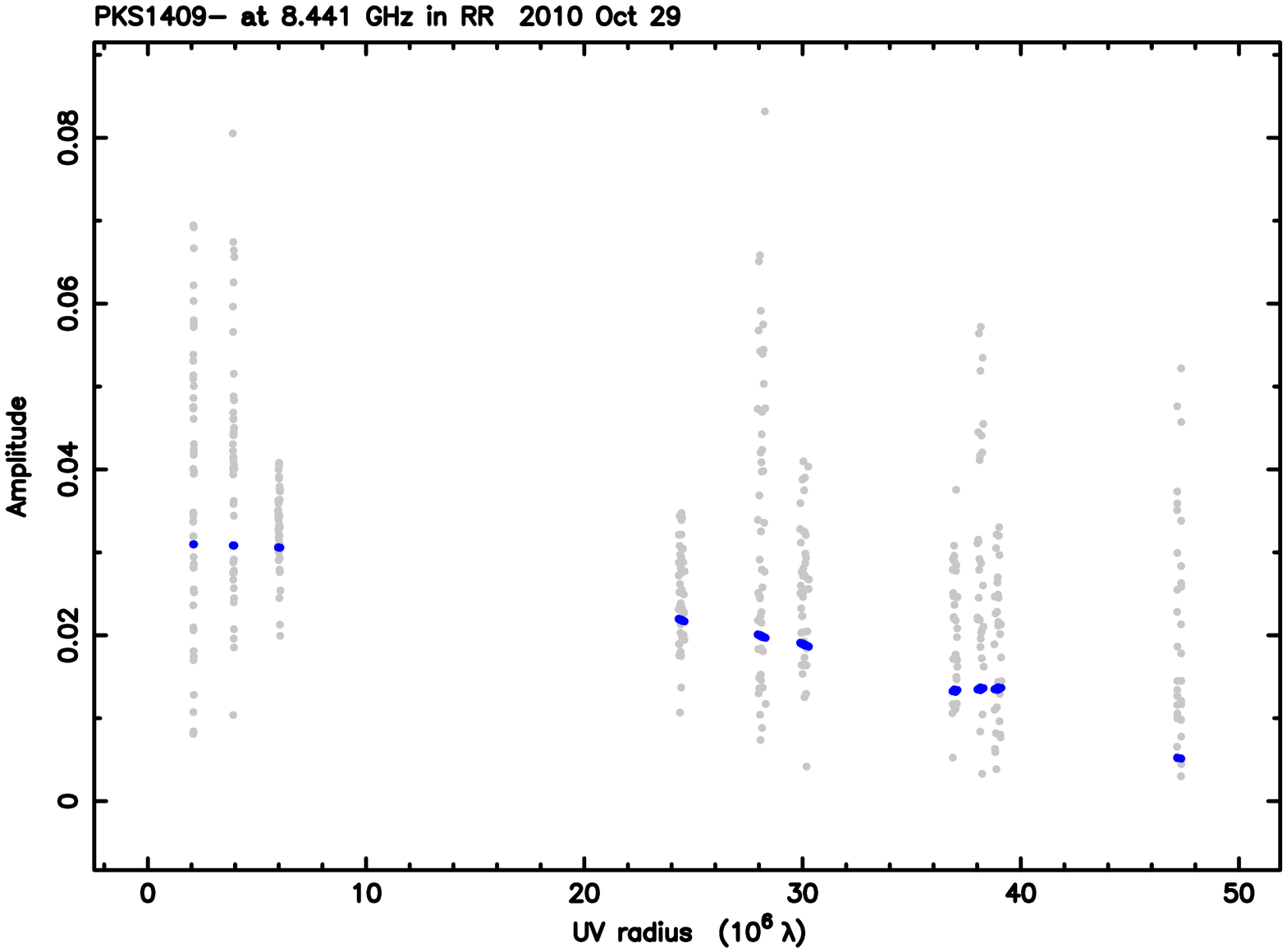}
\caption{8.4\,GHz visibility amplitude versus $(u,v)$-distance
  plot, showing the measured visibilities (gray) in the single-scan
  observations of 0943$-$761 (top) in 2009.95 and of 1409$-$651
  (Circinus galaxy; bottom) in 2010.82. The data are averaged over
  32\,s. A simple two-dimensional Gaussian model (blue) is shown overlaid on
  the data.}
\label{fig:radplot}
\end{figure}

Figures~\ref{fig:newimages1} to~\ref{fig:newimages4} present the
TANAMI first-epoch 8.4\,GHz images of the 39 sources that  were added
after the start of \lat $\gamma$ ray observations in 2008 August,
extending the sources discussed in Paper~I.  Image properties are listed
in Table~\ref{table:images}.  The first-epoch images of PKS\,1101$-$536,
PMN\,J1603$-$4904, Swift\,J1656.3$-$3302, and PKS\,2004$-$447 have
been already published in dedicated papers (see
Sect.~\ref{sec:individual}).  Furthermore, we provide information on
the VLBI detections of PKS\,0943$-$76 and PKS\,1409$-$651 (Circinus Galaxy).
Both of these sources have only been observed with a single short scan. Hence, we show a
visibility amplitude versus $(u,v)$-distance plot of the data in Fig.~\ref{fig:radplot}.

Table~\ref{table:all}  (column 6) includes the sources whose VLBI core flux densities and brightness
temperatures were studied along with their $\gamma$-ray properties in
\citet{Boeck2016}. To complement that work, we include here the images
from which those VLBI properties were extracted (and discuss the
properties in Sect.~\ref{sec:Boeck}). We further note that 
images for other epoch observations of PKS\,0235$-$618,
PKS\,0302$-$623, and PKS\,0308$-$611 have already been published in \citet{Krauss2014a}.

The total flux densities are derived directly from the models of the brightness distributions,
  which are determined in the hybrid-imaging process (see Paper~I).
  The core flux density is obtained by fitting an
elliptical Gaussian model to the core (defined as the brightest, most
compact feature) following the approach described in Paper~I.

The brightness temperature for the radio core at 8.4\,GHz of each
source was determined using
%%%%
\begin{equation}\label{eq:tb}
T_\mathrm{B} = \frac{2\ln 2}{\pi
k_\mathrm{B}}\frac{S_\mathrm{core} \lambda^2
(1+z)}{\theta_\mathrm{maj}\theta_\mathrm{min}}\quad,
\end{equation}
%%%%%
where $S_\mathrm{core}$, $\theta_\mathrm{maj}$, and
$\theta_\mathrm{min}$ are respectively the flux density, the major axis,  and minor axis of a two-dimensional Gaussian model component for the core in the radio image (see Table~\ref{table:images}); $k_\mathrm{B}$ is the Boltzmann
constant; $z$ the redshift of the source; and $\lambda$ the observing
wavelength. 
If the size of the fitted model component for the core
emission falls below the resolution limit, very high brightness
temperatures are indicated, which cannot be constrained with
ground-based VLBI arrays \citep[see][for recent space-based VLBI
measurements of compact AGN cores with the RadioAstron
antenna]{Kovalev2016}. In these cases,
we calculated lower limits
for $T_\mathrm{B}$ following \citet{Kovalev2005}.

\begin{figure*}
\centering
\includegraphics[width=0.32\textwidth]{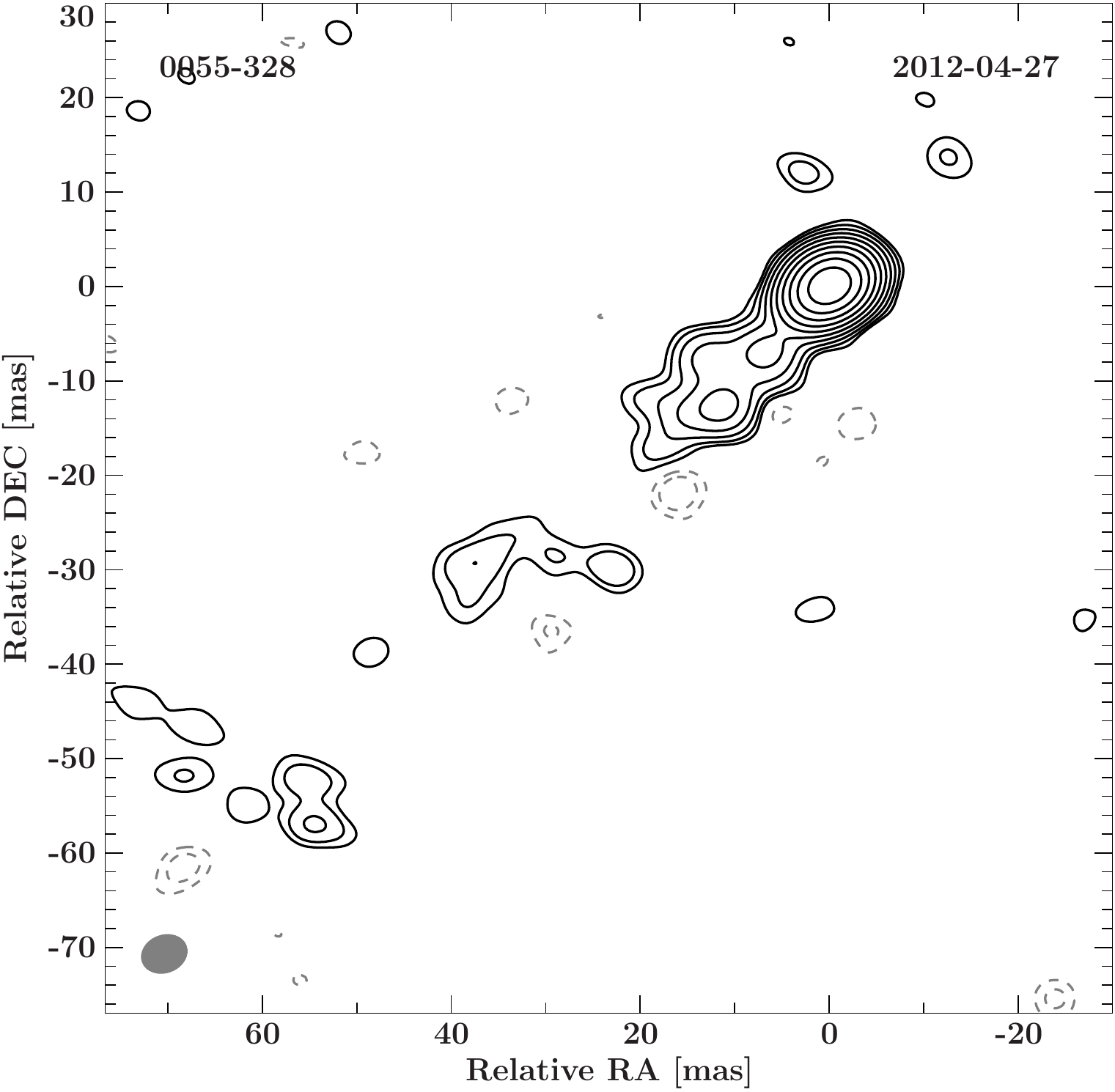}
\includegraphics[width=0.32\textwidth]{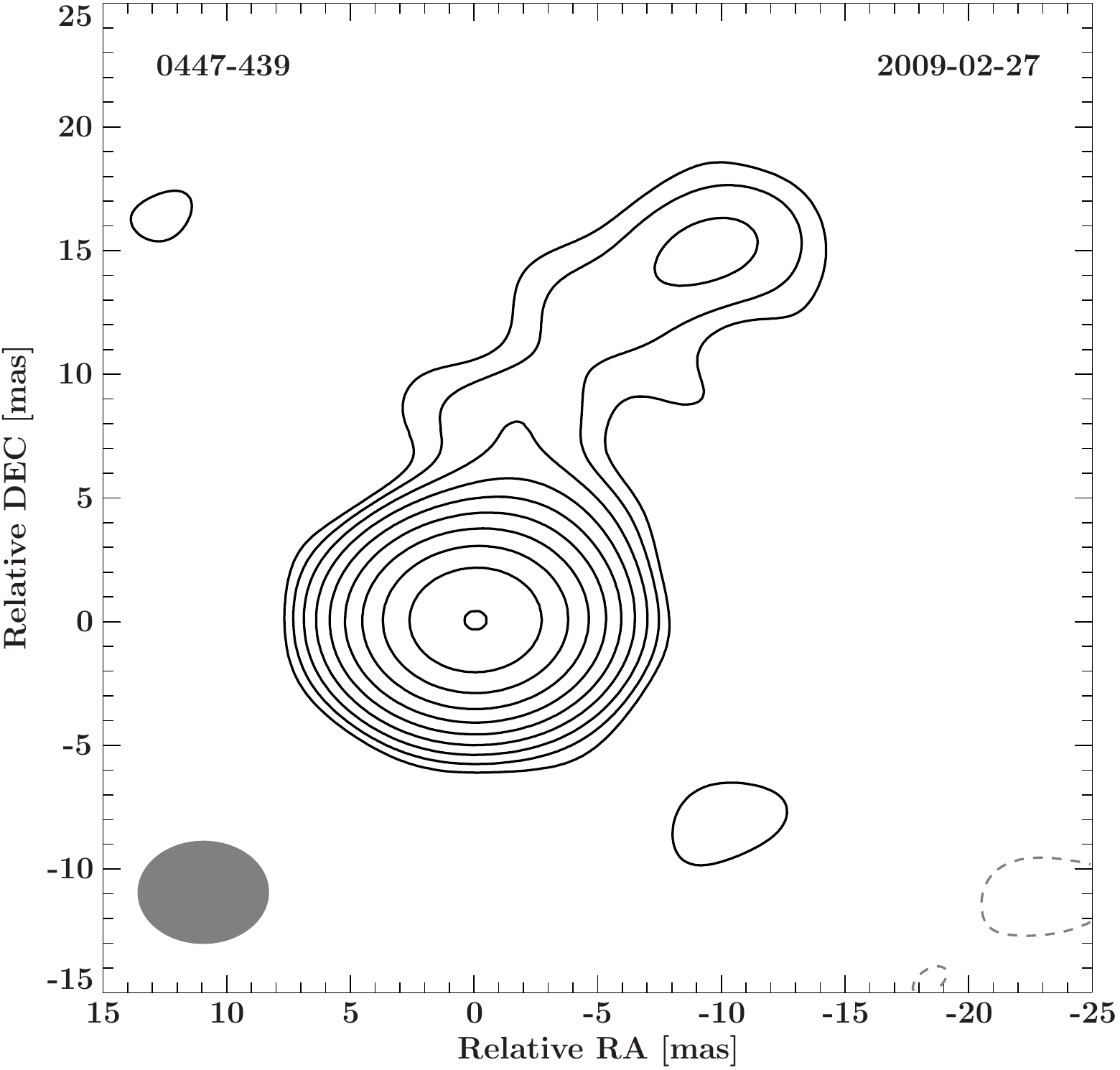}
\includegraphics[width=0.32\textwidth]{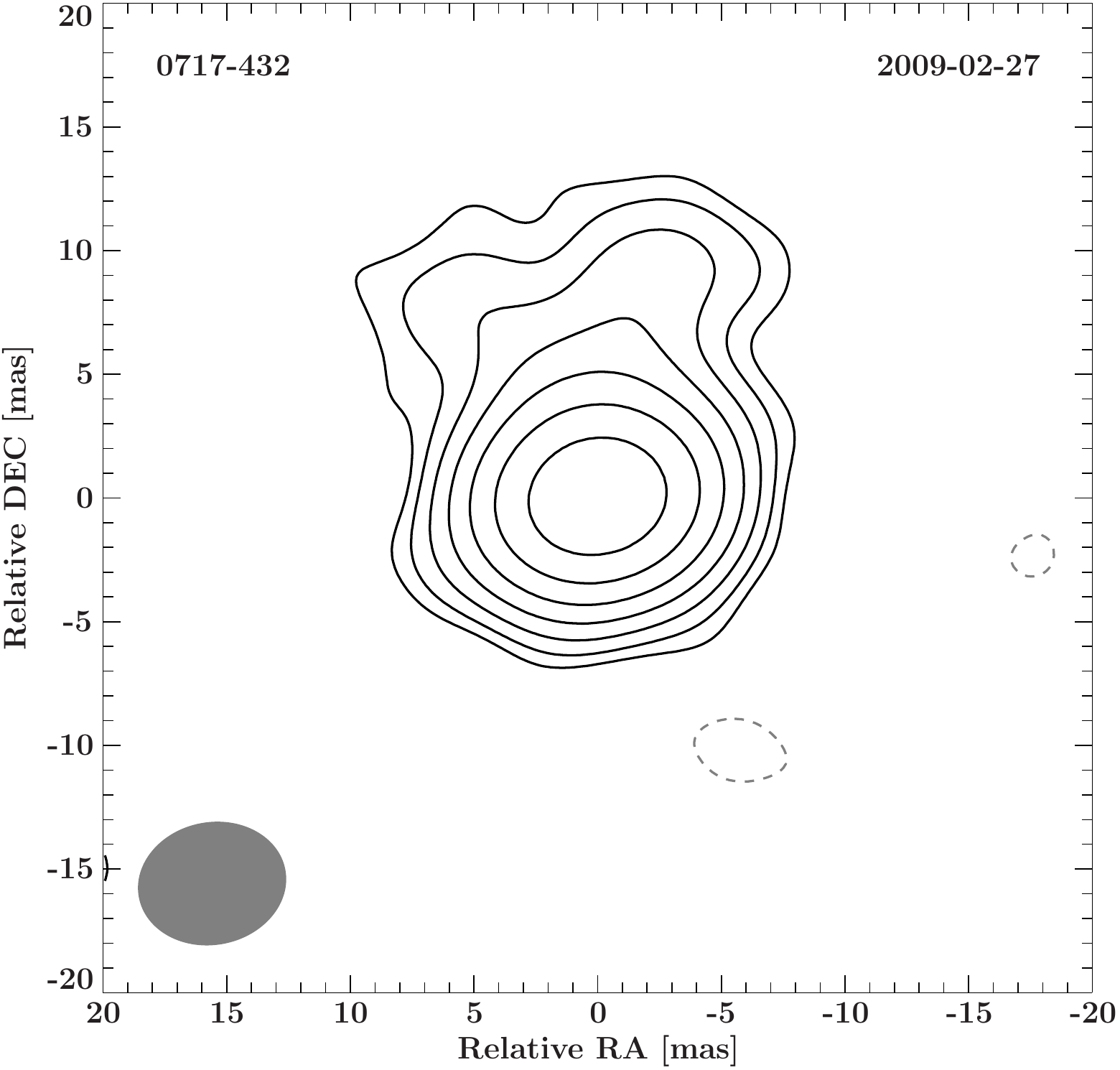}

\caption{Tapered 8.4\,GHz \texttt{clean} images of selected TANAMI
  sources revealing more extended emission. See Table~\ref{table:tapered} for image parameters.}
\label{fig:taperedimages}
\end{figure*}

Most of the sources whose images are shown in this work, exhibit 
clearly resolved mas-scale structures. Only
for seven sources is the size of the core component not
constrained. For most sources, we  typically find single-sided, blazar-like morphologies.
Figure~\ref{fig:taperedimages} shows tapered images of three sources
to better display their extended emission. The corresponding image parameters can be
found in Table~\ref{table:tapered}.

%-------------------------------------------------------------------

%% %%%-----------------------------------------------------------------------%%%

%%%%%%%%%%                                                                                                                       

%__________________________________________________________________

\section{Notes on individual sources}\label{sec:individual}
Here we comment on the radio properties of individual sources,
complementing the discussion in \citet{Boeck2016} and Paper~I.

\paragraph{\object{PKS\,0055$-$328}:} 

In epoch 2009 Feb\ 27, this high-redshift BL\,Lac object appears very compact. 
In Fig.~\ref{fig:taperedimages} we show the tapered image from a later
TANAMI VLBI epoch (2012 Apr~27), which was more sensitive to diffuse extended 
emission, revealing a larger scale jet extending over more than 60\,mas.

\paragraph{\object{PKS\,0235$-$618}:} This quasar is one of three
TANAMI blazars that are positionally consistent with the IceCube PeV neutrino
event HESE-20 \citep[see][where the
inner structure of the second 8.4\,GHz TANAMI VLBI observation -- quasi-simultaneous
to the IceCube event -- is presented]{Krauss2014a}. Here, we show the full source
extension at mas-scales from the first TANAMI VLBI observation.

\paragraph{\object{PKS\,0302$-$623}:} This source is positionally consistent with the PeV 
neutrino event HESE-20 \citep{Krauss2014a}. The mas-scale morphology is classified as
irregular due to the halo-like emission around the core.

\paragraph{\object{PKS\,0402$-$362}:}
This object is a $\gamma$-ray loud quasar \citep{Nesci2011ATel0402} with slightly extended morphology on
mas-scales.

\paragraph{\object{PKS\,0426$-$380}:} This source is one of the brightest $\gamma$-ray
loud blazars in the southern sky and has not been imaged before at comparable
angular resolution and image fidelity. It shows a compact jet structure with a faint
extension to the east.

\paragraph{\object{PKS\,0447$-$439}:} This $\gamma$-ray blazar 
\citep[see also][]{Nesci2011ATel0447} is
detected up to TeV energies \citep{HESS2013PKS0447}. The redshift
measurements are contradictory as discussed in \citet{Pita2014}. The TANAMI
mas-scale image shows a faint jet to the northwest. In
Fig.~\ref{fig:taperedimages} we show the tapered image to better
display the faint extended emission.

\paragraph{\object{PMN\,0529$-$355} (PKS\,0527$-$359):} This source
was included in the initial TANAMI sample as a
candidate EGRET $\gamma$-ray blazar, but was not observed with VLBI due to
scheduling problems (Paper~I). Here, we present the first-epoch image. It is
classified as a non-BL\,Lac object by \citet{Landoni2013} and it is
not detected by \lat.

\paragraph{\object{PKS\,0717$-$432}: } This is a faint source with a total flux of $S_\mathrm{8\,GHz}=0.03$\,Jy. We
show a tapered image in
Fig.~\ref{fig:taperedimages} displaying the faint extended emission.

\paragraph{\object{PKS\,1101$-$536}:} This is one of  four
$\gamma$-ray blazars listed in the \fermi 2LAC catalog \citep{2lac}
that is positionally consistent with the 165\,TeV neutrino event
HESE-4 \citep{Ice14}. The first-epoch TANAMI VLBI image is
from \citet{Krauss2015}.

\paragraph{\object{PKS\,1258$-$321} (ESO\,443-G024):} The ATCA
8.4\,GHz image of this radio galaxy
shown by \citet{Marshall2005} shows a very symmetric, double-sided FR~I
morphology. The TANAMI VLBI image reveals a one-sided jet to
  the northwest with a position angle consistent with the kpc-scale jet
  structure. No \lat detection has been reported so far \citep{3fgl}.

\paragraph{\object{PKS\,1343$-$601} (Centaurus~B):} Our VLBI image
shows an extended jet of $\sim 50$\,mas extension to the southwest with
the same position angle as the large-scale jet-structure at 8\,GHz
shown by \citet{Marshall2005}. A counterjet is not significantly
detected at VLBI scales in contrast to the large-scale structure at
843\,MHz \citep{Jones2001}. \citet{Katsuta2013} report on the \lat
detection of Cen\,B discussing both the innermost jet region and the extended
radio lobes as possible $\gamma$-ray emission zones. In the former 
scenario, broadband SED modeling of the unresolved core data results in a
required inclination angle of the jet of $\sim$$20^\circ$--$25^\circ$. From our VLBI
image, we can use the ratio of the surface brightness of the jet and
image noise ($R\sim61$) to constrain the inclination angle to $\lesssim
20^\circ$.

\paragraph{\object{PKS\,1409$-$651} (Circinus Galaxy):}
The source was added to the TANAMI sample due to its
$\gamma$ ray detection by \lat\ \citep{Hayashida2013}. We performed a
snapshot observation in 2010, detecting compact
emission of $\sim 30$\,mJy (see Fig.~\ref{fig:radplot}).

\paragraph{PKS\,1600$-$489 (\object{PMN\,J1603-4904}):} 
This object was classified as a BL\,Lac object by
\citet{Shaw2013a}. Extensive TANAMI related multiwavelength
observations revealed broadband properties which are difficult to
reconcile with the blazar classification, but favor a $\gamma$-ray
loud young radio galaxy
\citep{Mueller2014a,Mueller2015a,Mueller2016}. This conclusion is also supported
by the first epoch image published by \citet{Mueller2014a}. This image
shows a
double-sided morphology.  The redshift of $z=0.23$ was determined by
the detection of optical emission lines by \citet{Goldoni2016}.  Here,
we treat this object as a galaxy (``G'').

\paragraph{\object{Swift\,J1656.3$-$3302} (1653$-$329):} 
This high-redshift ($z=2.4$) blazar has the highest predicted PeV-neutrino flux inside
the median-positional-uncertainty field of the IceCube PeV neutrino
event HESE-14 \citep[see][where the first-epoch TANAMI mas-scale image
is shown]{Krauss2014a}.  This source also has a neutrino-signal flux
fitted by the ANTARES likelihood analysis and based on six years of
data corresponding to one event, although an atmospheric origin cannot
be excluded \citep{AntaresTanami2015}.

\paragraph{\object{PKS\,2004$-$447}:} 
This source is the only known $\gamma$-ray detected radio-loud narrow-line Seyfert~1
($\gamma$RL-NLS1) galaxy in the southern hemisphere \citep{3fgl}. A
comprehensive TANAMI multiwavelength study has been presented by
\citet{Kreikenbohm2016} and by \citet{Schulz2016}, who published  the
first epoch image. It shows
the inner 10\,mas of a collimated jet ending in a region of enhanced
flux density at about 45\,mas from the core. The source shows a persistent
steep radio spectrum and a low limit on its large scale size, which are both consistent
with a compact steep spectrum object. This is unusual for $\gamma$-RL-NLS1
and $\gamma$-ray detected AGN in general.

\paragraph{\object{PKS\,2123$-$463}:}
Based on multiwavelength variability \citet{DAmmando2012} firmly
identify this radio source with the \textsl{Fermi}/LAT detection. They
provide a new photometric redshift of $z=1.46$.

\paragraph{\object{PKS\,2142$-$758}:} In \citet{Dutka2013}, the
multiwavelength properties of this FSRQ during
flaring and quiescent $\gamma$-ray states are extensively discussed. The
first-epoch VLBI image shows a core-dominated structure with a faint
jet extension to the east.

%__________________________________________________________________
\section{Discussion}\label{sec:discussion}

\subsection{VLBI source properties}\label{sec:Boeck}
In \citet{Boeck2016}, we discussed the 0.1-100\,GeV $\gamma$ ray
properties of 75 AGN jets over a one-year period of \lat integration
\citep[2008 August through 2009 September, corresponding to the 1FGL
period;][]{1fgl}, including a \lat upper limit analysis for all
non-detected sources.  To investigate the radio-$\gamma$ connection,
we used the VLBI core-flux densities and core brightness temperatures
(extracted from Gaussian model components)
from the first VLBI observation during that one-year period.  Here, we
discuss in detail the corresponding mas-scale radio properties for the
same set of observations.  We consider only the 67 TANAMI sources for
which a VLBI observation during the considered one-year period is
available to ensure \mbox{(quasi-)}simultaneity and thus reduce possible 
variability effects\footnote{The check source
  PKS\,1934$-$638 is not included in this VLBI analysis.}.  

The ratios of the $\gamma$-ray flux to radio core flux density at \X
($r = S_\mathrm{0.1-100\,GeV}/S_\mathrm{\X,core}$) are calculated to
define four subsamples: $r<10^{-8}\,\mathrm{ph/cm^{2}/s/Jy}$
corresponds to all non-detected sources (i.e., $\gamma$-ray faint),
$10^{-8}\,\mathrm{ph/cm^{2}/s/Jy}<r\leq0.6\times10^{-7}\,\mathrm{ph/cm^{2}/s/Jy}$
to low,
$0.6\times10^{-7}\,\mathrm{ph/cm^{2}/s/Jy}<r\leq2\times10^{-7}\,\mathrm{ph/cm^{2}/s/Jy}$
to medium, and $r>2\times10^{-7}\,\mathrm{ph/cm^{2}/s/Jy}$ to
high $\gamma$-ray brightness. The division was chosen such that each
bin includes approximately the same number of sources.  In the
following discussion we investigate the source properties of these
four distinct subsamples.

We note that the $\gamma$-ray bright sample is strongly biased towards
low radio fluxes, as newly detected $\gamma$ ray sources  without
extensive previous radio coverage were added to the TANAMI program
with high priority in the years after the launch of the \textsl{Fermi}
telescope.
For this reason it is not surprising that a Kolmogorov-Smirnov (KS)
test 
finds rather large differences between the $\gamma$-ray faintest and
brightest sources (p-value of 0.87 and 0.10 for the core
and total flux distributions, respectively).
Therefore, for further discussions, we concentrate on the distributions
of the ratio of core to total radio flux density 
and the brightness temperatures.

\begin{figure}
\includegraphics[width=\columnwidth]{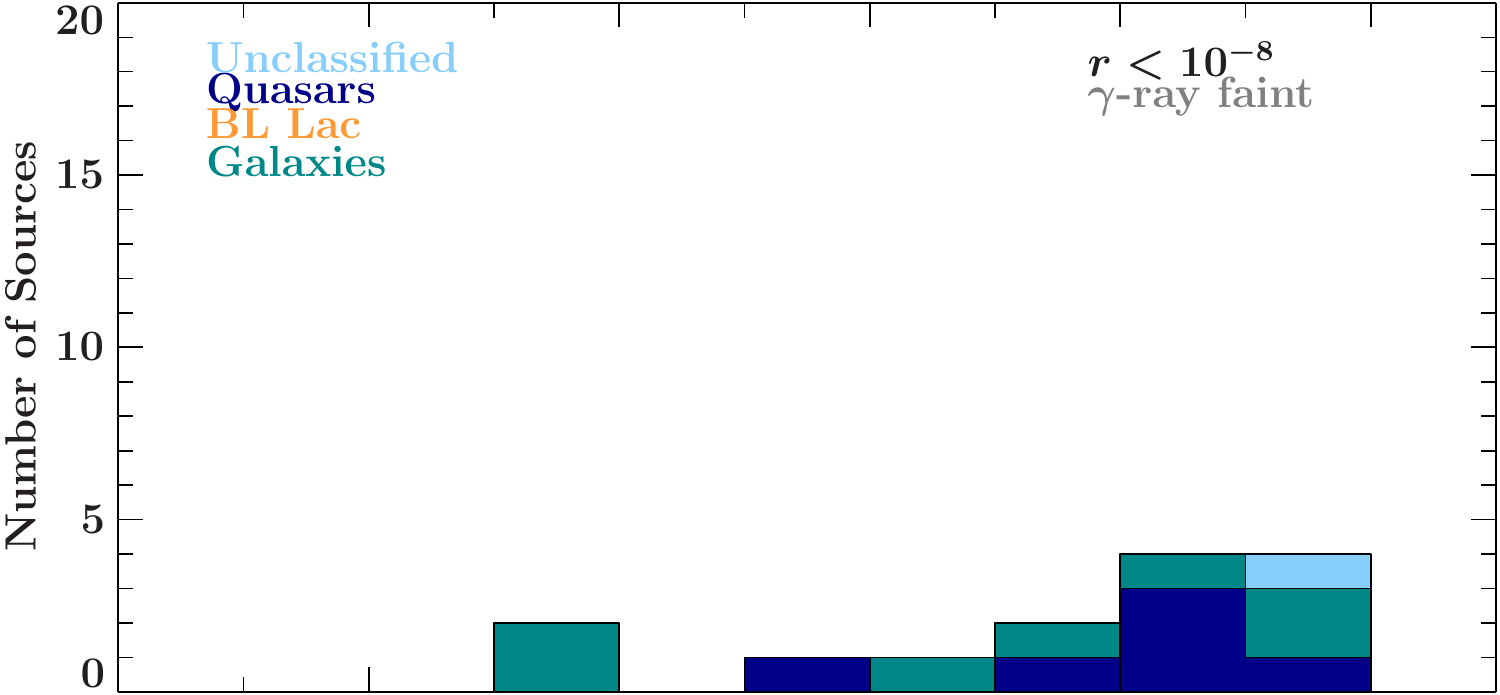}
\includegraphics[width=\columnwidth]{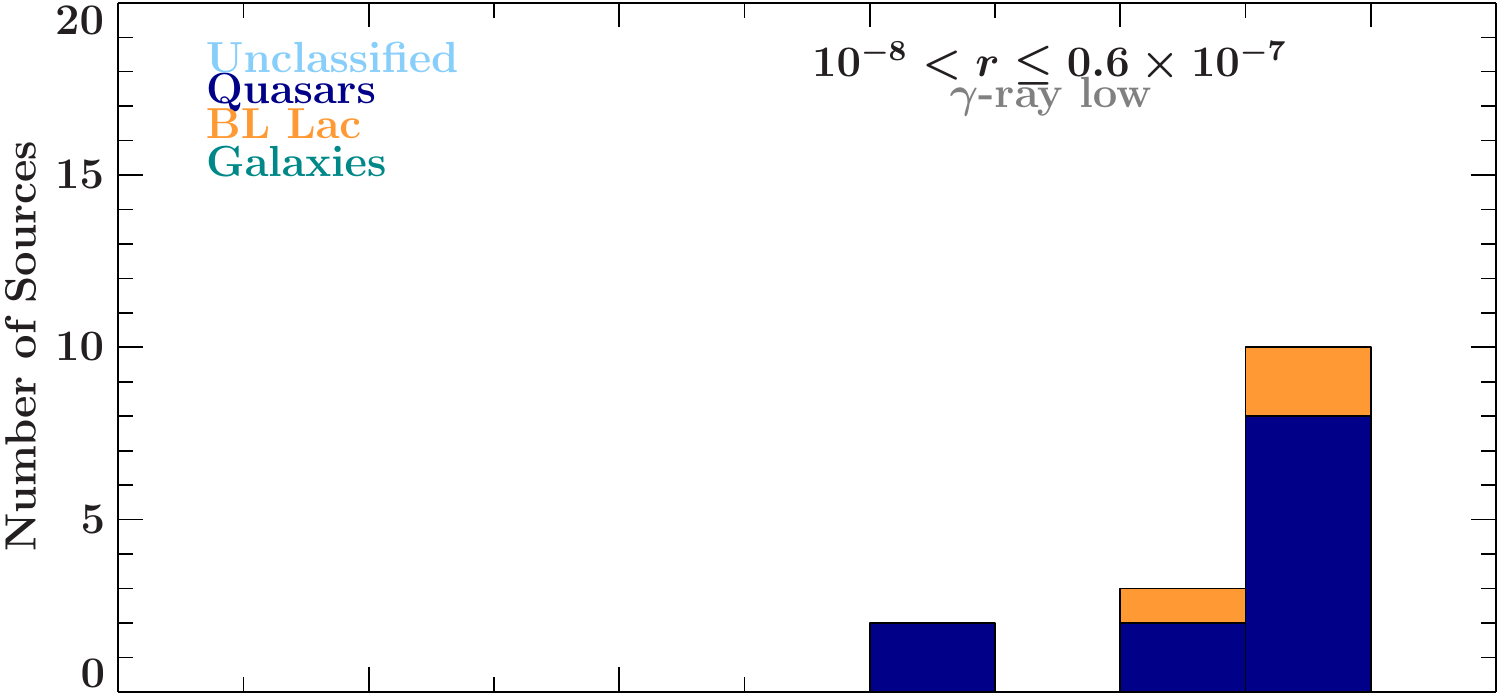}
\includegraphics[width=\columnwidth]{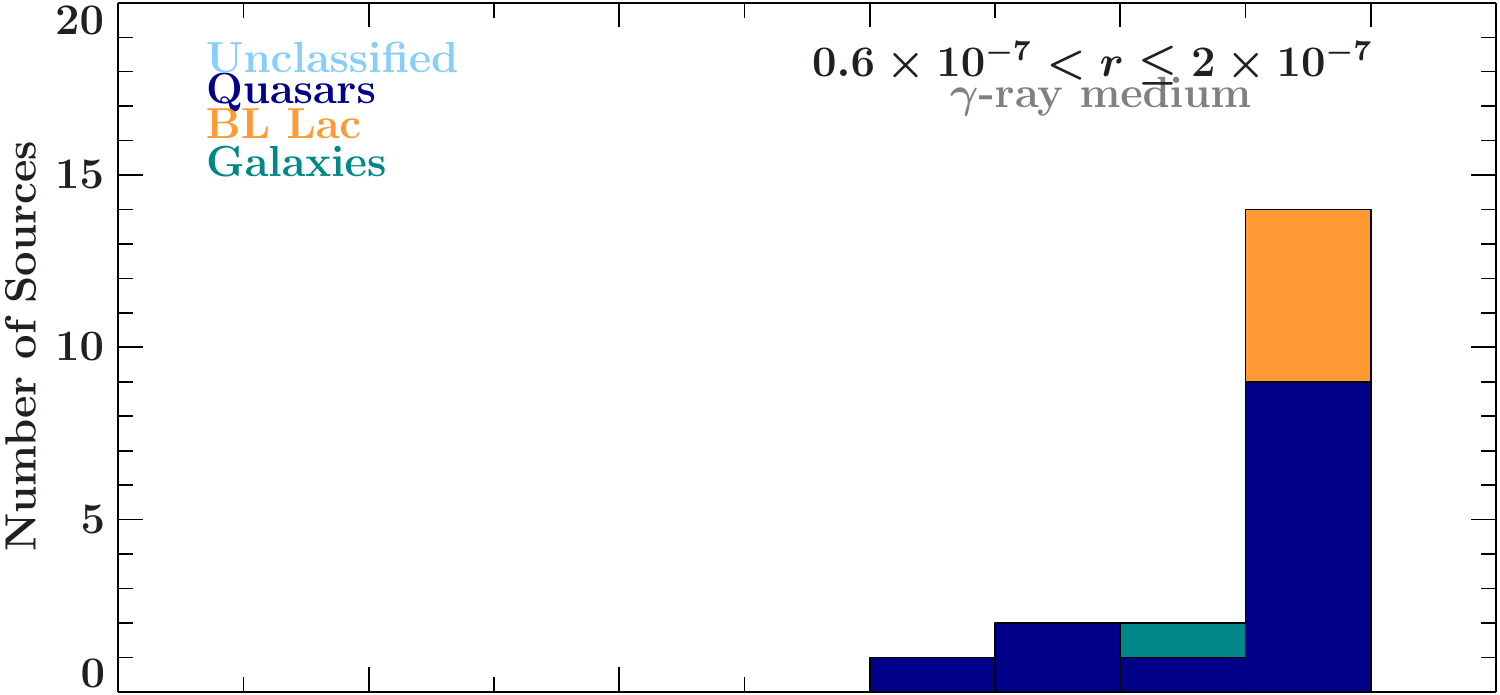}
\includegraphics[width=\columnwidth]{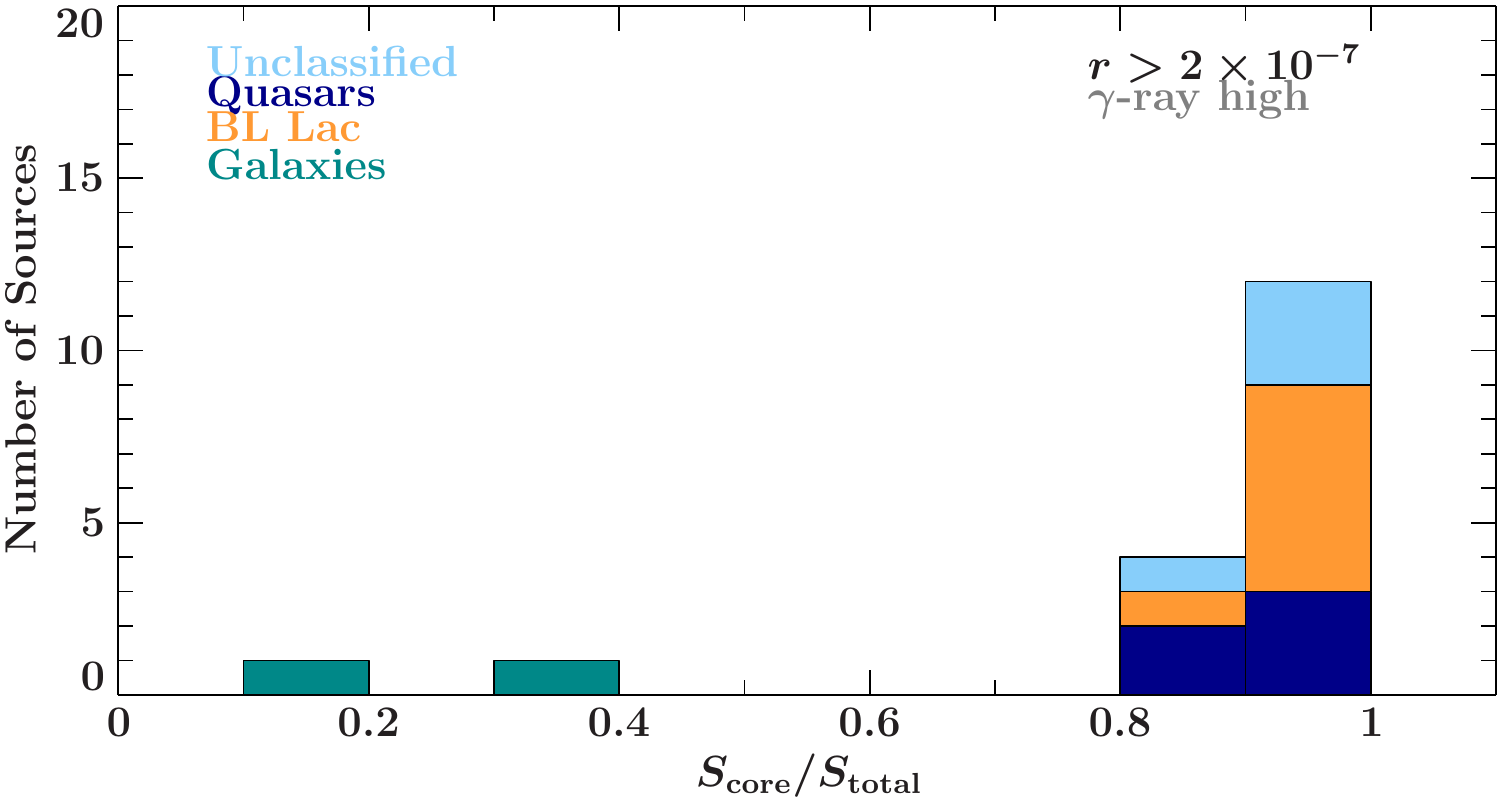}
 \caption{Distribution of the core dominance
   $\kappa = S_\mathrm{core}/S_\mathrm{total}$ for the four subclasses of 
$\gamma$-ray faint, low, medium, and high sources. Flux density ratio $r$ is in 
units of $\mathrm{ph/cm^{2}/s/Jy}$.}
\label{fig:histo_ratio}
\end{figure}

The $\kappa =
S_\mathrm{core}/S_\mathrm{total}$ ratio (see
Fig.~\ref{fig:histo_ratio}), shows that in most sources the bulk radio
emission ($\geq 80$\%) originates from the unresolved sub-mas scales,
although most of the sources clearly have a resolved (not compact)
mas-scale structure.  The $\gamma$-ray faint class 
shows several sources with a rather low $\kappa$ ratio, which implies
that the $\gamma$-ray detected sources tend to be more core dominated.
This is in agreement with the $T_\mathrm{B}$-distribution \citep[see
Fig.~10 of][]{Boeck2016}, implying that the $\gamma$-ray detected
sources tend to have higher core brightness temperatures
\citep[compare also to, e.g.,][]{Kovalev2009}.

\begin{figure}
\includegraphics[width=\columnwidth]{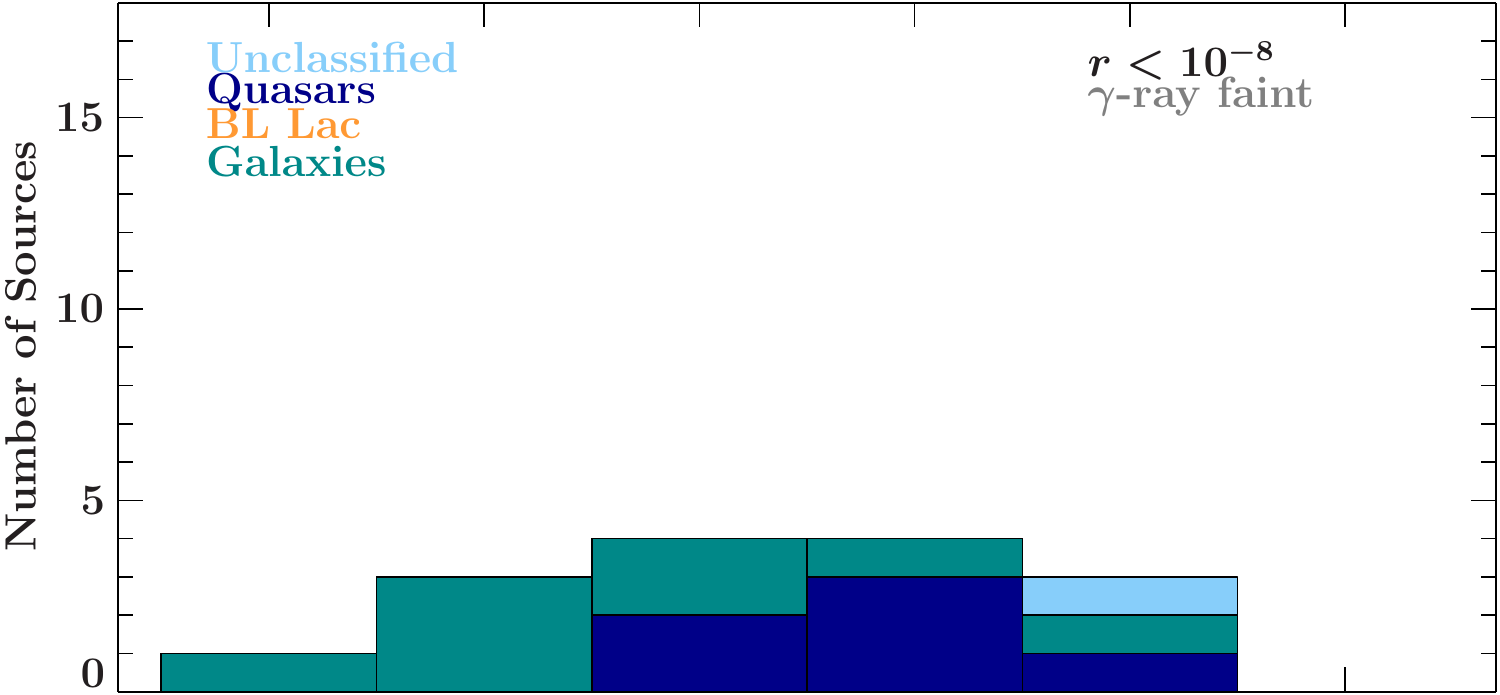}
\includegraphics[width=\columnwidth]{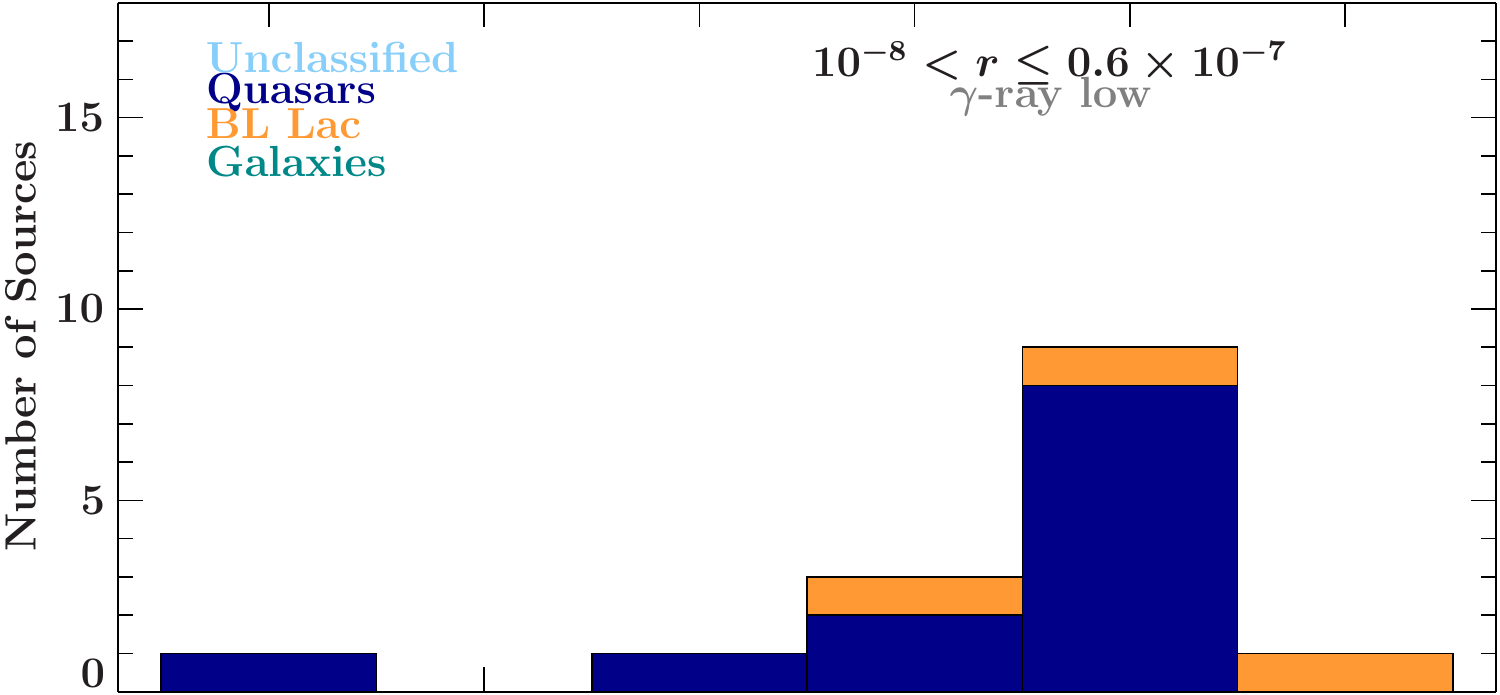}
\includegraphics[width=\columnwidth]{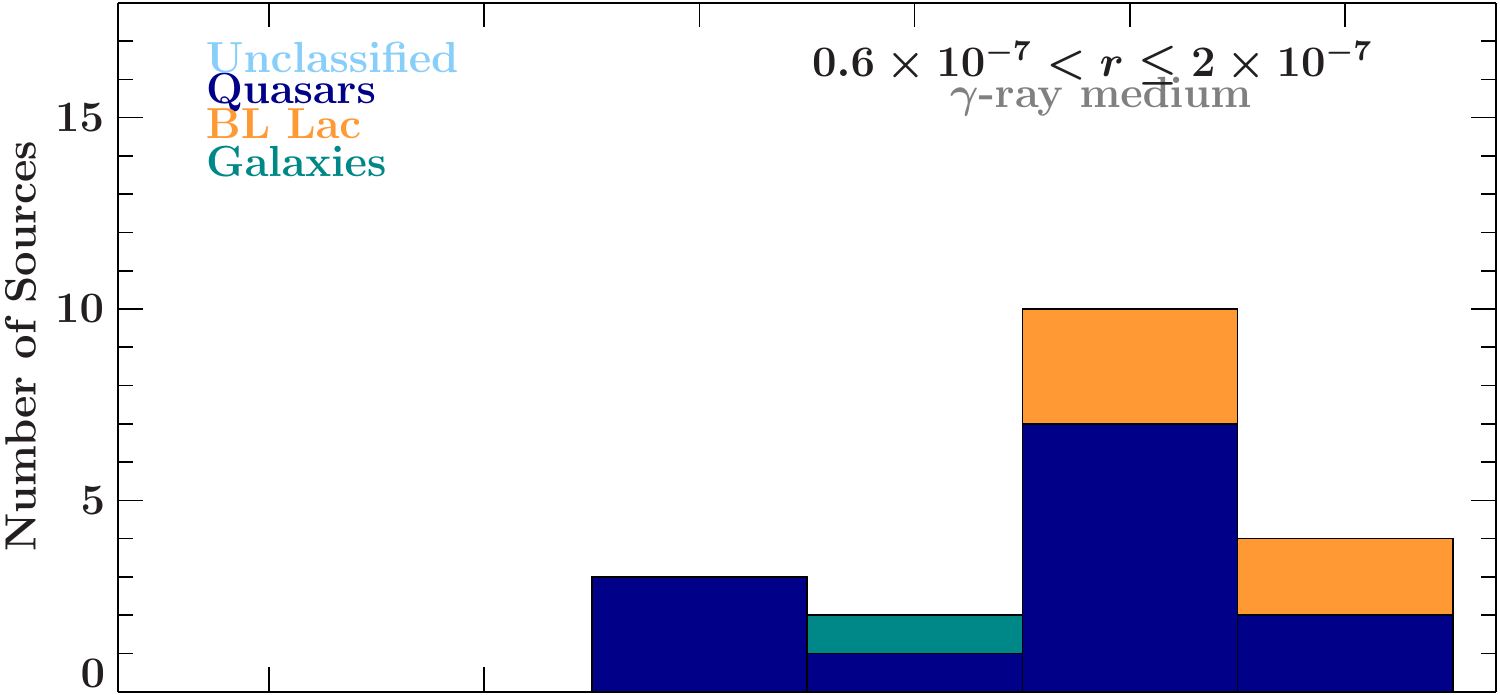}
\includegraphics[width=\columnwidth]{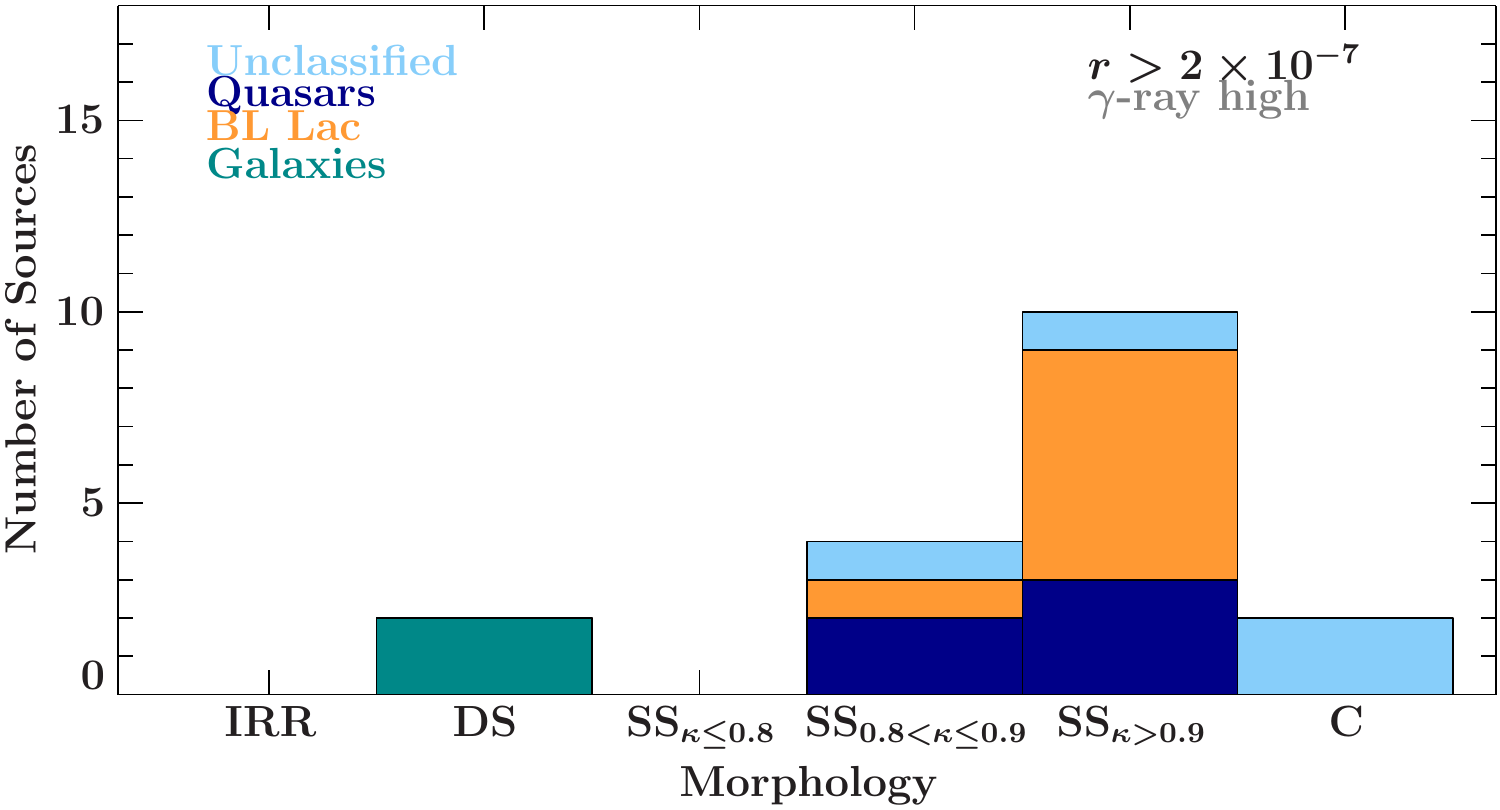}
  \caption{Distribution of the \X morphologies (C: compact, SS:
    single-sided, DS: double-sided, IRR: irregular), subdividing the
    single-sided sources depending on the core dominance
    $\kappa=S_\mathrm{core}/S_\mathrm{total}$. Flux density ratio $r$ is
    in units of $\mathrm{ph/cm^{2}/s/Jy}$. 
}
\label{fig:histo_morph}
\end{figure} 

Our morphology classification scheme is adopted from
\citet{Kellermann1998} as used in Paper~I: barely resolved
sources are classified as compact (C); those with a core-jet like
structure, i.e., with the most compact feature at either end of the
source structure as single-sided (SS); those with the most compact
component in the middle as double-sided (DS); and those with an irregular
two-dimensional structure  as irregular  (IRR). 
This morphology
scheme takes no spectral information into account.

Of the 67~objects considered in this analysis, 60 are found to be resolved, while 7 are
classified as compact. Only two of the $\gamma$-ray bright sources are
classified as double-sided.
The comparison between total flux densities of jet sources measured by
single-dish observations with VLBI measurements shows that their
compactness on arcsecond scales is typically $\geq 90$\%; i.e., most
of the flux density is contained within \mbox{(sub-)}mas scales
\citep{Kovalev2005}. This compactness-measure can be taken as an
indicator for Doppler boosting and hence as a rough estimator of the
jet-inclination angle.  \citet{Kovalev2005} further showed that the
VLBI core dominance (i.e., the ratio of the core to the total flux
density on VLBI scales, here $\kappa$) of galaxies is often small, while the quasars
and BL\,Lacs are far more compact on \mbox{(sub-)mas} scales,
consistent with the AGN unification model.  Using the $\kappa$-value
of every source, we can expand the morphology scheme for the
single-sided sources as follows.  We define three subcategories:
$\kappa \leq 0.8$ for the single-sided sources with the most
significant jet contribution, $0.8\leq \kappa <0.9$ for the
intermediate, and $\kappa > 0.9$ for the most compact sources.

Figure~\ref{fig:histo_morph} shows that the $\gamma$-ray loud objects
are generally more core dominated than the faint objects, while the
distributions for low, medium, and high $\gamma$-ray brightness show
no significant differences. This result
is consistent with the picture that $\gamma$-ray loud blazars are
pointing closer to the line of sight and are more strongly Doppler boosted
\citep[see Paper~I and, e.g.,][]{Lister2015}. The unclassified objects
in our sample all show a high core dominance typical for blazars.

The $\kappa$-distribution of quasars is quite broad, ranging down to
$\kappa$-values typical for radio galaxies.  There is tentative
evidence for a dichotomy between $\gamma$-ray faint and loud
quasars, in the sense that 
the VLBI morphologies of $\gamma$-bright quasars show a strong
peak at the most compact bins, while the (one-sided) jets of $\gamma$-faint 
quasars show values of $\kappa$ typically in the range $0.8$ to $0.9$, 
similar to the one-sided jets of radio galaxies.  
In this context, it is interesting to note that the $\kappa$-values of all BL\,Lac objects are high.  This supports recent findings
by \citet{Lister2015}, which suggest that the $\gamma$-ray
non-detection of some of the radio-brightest blazars is in part due to
their lower peaked spectral energy distributions and in part due to
their lower Doppler factors. The morphological manifestation of the
latter effect, tentatively seen in our data, should be studied further
using an even larger jet sample.

\subsection{Properties of sources in positional agreement with
  $>$100\,TeV neutrino events}
\label{sec:neutrinos}

\begin{figure*}
\includegraphics[width=\textwidth]{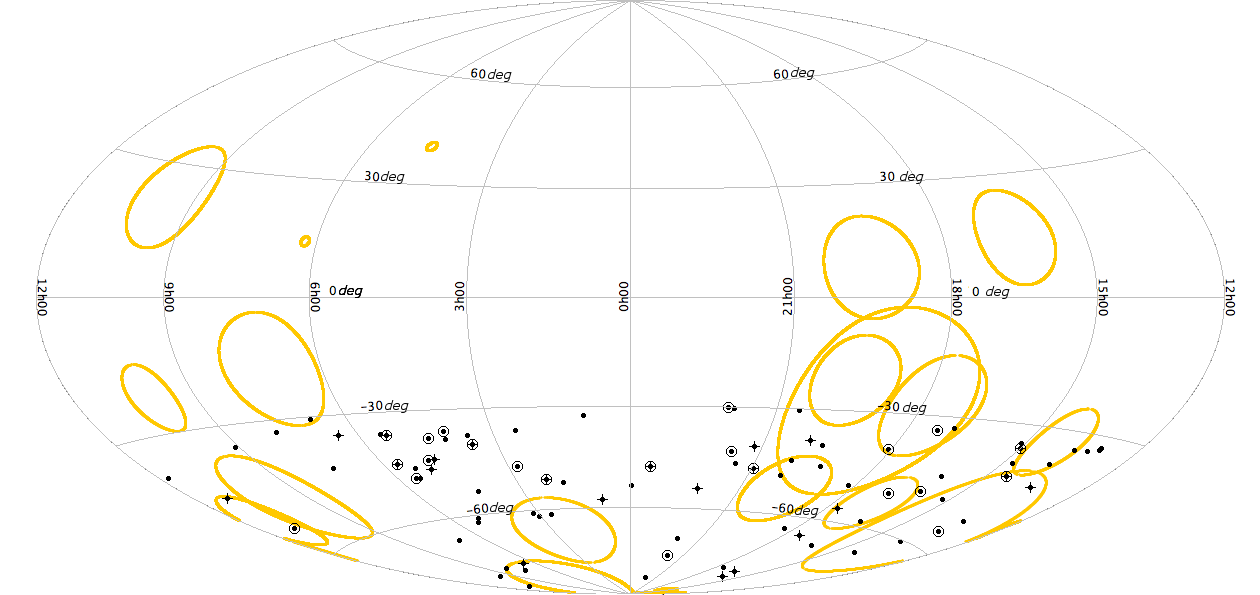}
  \caption{Sky distribution of IceCube HESE neutrino events with
    energies in excess of 100\,TeV (yellow solid closed lines) in
    Aitoff projection using equatorial coordinates. The position of
    TANAMI sources is marked with black dots, and sources from the radio sample
    and the $\gamma$ ray 
    sample are  marked with
    plus signs and circles, respectively.}
\label{fig:hese_equat_ait}
\end{figure*}

We have also attempted to compare the VLBI properties of AGN jets in
positional coincidence with neutrinos from the IceCube HESE
analysis\footnote{We use the merged HESE data sets by \citet{Ice14}
  and \citet{Ice15a}.}  at energies above 100\,TeV.  The HESE regions
are large ($\sim$200 to $\sim$2000 square degrees) due to the poor
angular resolution of cascade events in the IceCube detector. Most
sources in these fields are therefore expected to be serendipitous
coincidences. 
However, any putative rare source properties could 
 stand out clearly against the sample properties of the non-coincident sources.
The 13 known events at energies above 100\,TeV\footnote{At these high energies, fewer than 0.1 neutrino events are expected 
from a non-astrophysical origin \citep{Ice15a}. }, which  at least
partially cover the TANAMI part of the sky south of
$-30^\circ$~declination, are in agreement with the quasar-photopion
scenario \citep{Kadler2016,Mannheim1995,Mannheim1992a}, which predicts a peaked
neutrino spectrum at energies between $\sim100$\,TeV and
10\,PeV. 
These events cover
about 50\,\% of the TANAMI sky region
(thus, about 1/8 of the full sky).  However, the TANAMI sources are
not uniformly distributed over the sky because only a few sources are
located near the Galactic plane and because the overall number of
sources is not large enough to smooth out apparent inhomogeneities. 
As
a result, only 31 out of 88 TANAMI sources are located inside the
median-positional-uncertainty regions $R_{50}$ of
neutrino events of $>100$\,TeV (see Fig.~\ref{fig:hese_equat_ait} and
column 9 of
Table~\ref{table:all}).

We divide our sample into sources located within the median
positional-uncertainty radius $R_{50}$ of the IceCube HESE neutrino
events with energies in excess of 100\,TeV and outside of $1.5\,R_{50}$
to compare the properties of sources whose positions in the sky are
consistent with  possible neutrino associations and sources which are
clearly inconsistent with these high-energy neutrinos because of their
large positional offsets.

\begin{figure}
\includegraphics[width=0.5\textwidth]{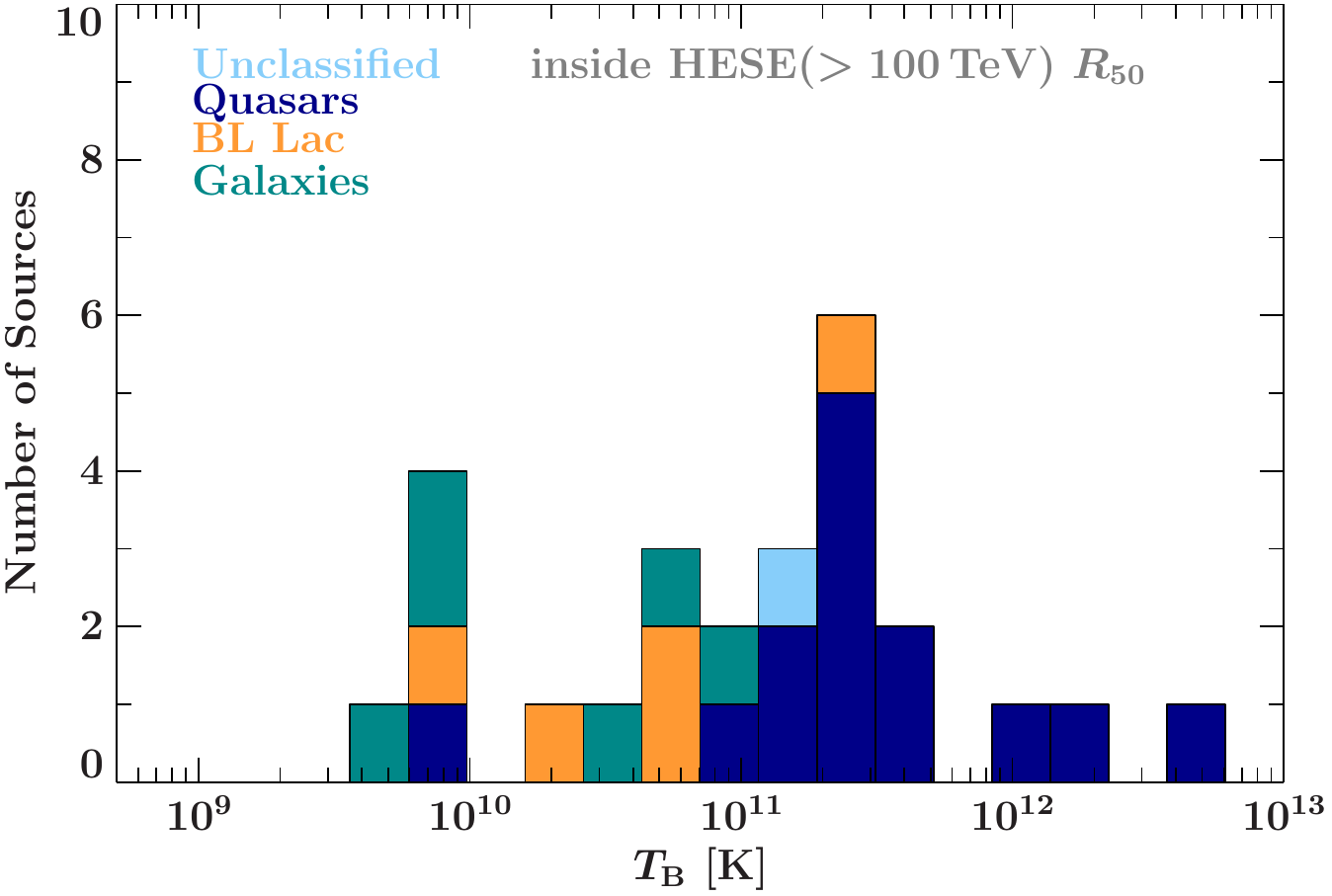}
\includegraphics[width=0.5\textwidth]{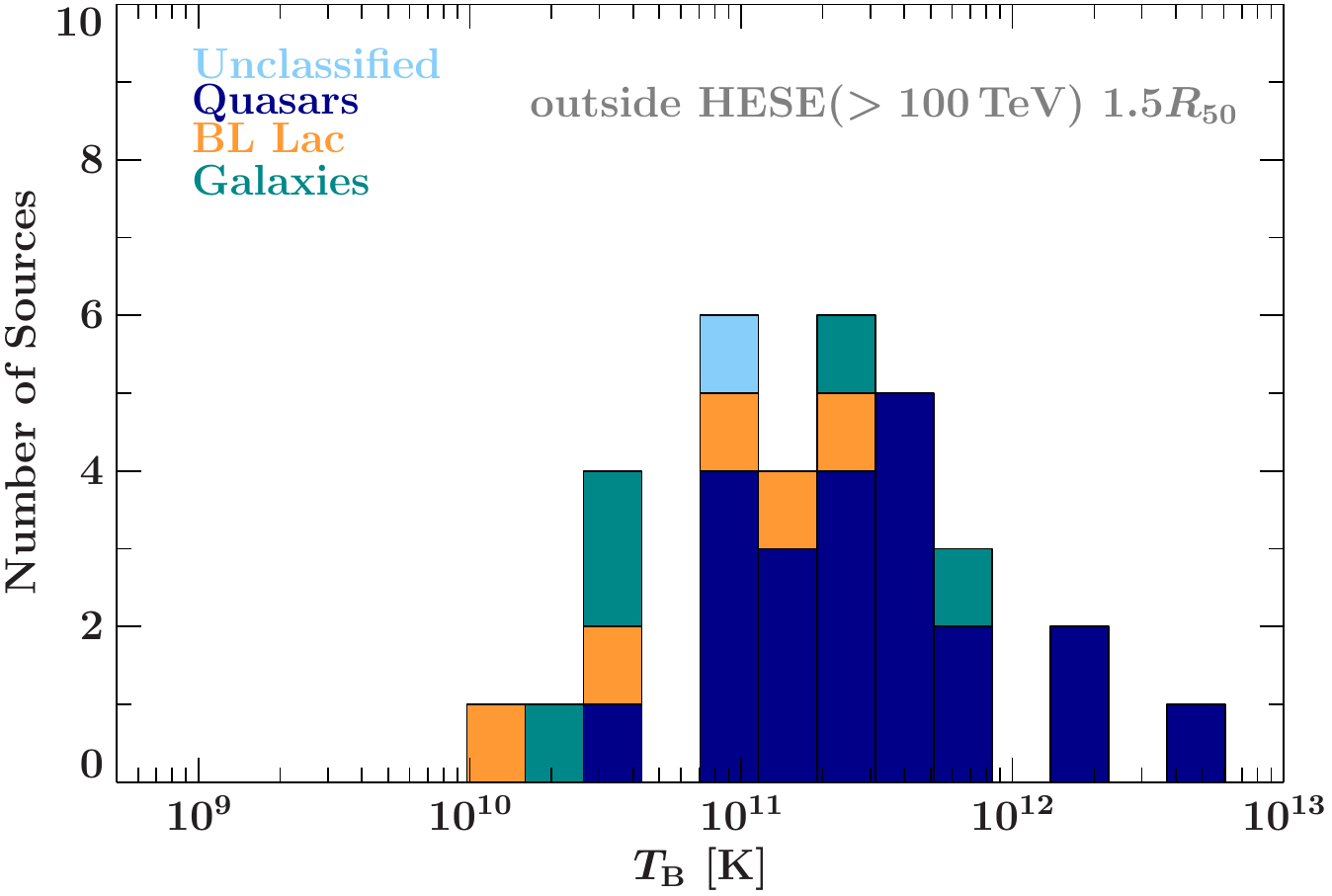}
  \caption{Brightness temperature distribution (redshift corrected)  for sources inside (\textsl{top}) and
    outside (\textsl{bottom}) of the error radius $R_{50}$ of the IceCube HESE neutrino events with
    energies in excess of 100\,TeV. 
}
\label{fig:neutrino_tb}
\end{figure}

\begin{figure}
\includegraphics[width=0.5\textwidth]{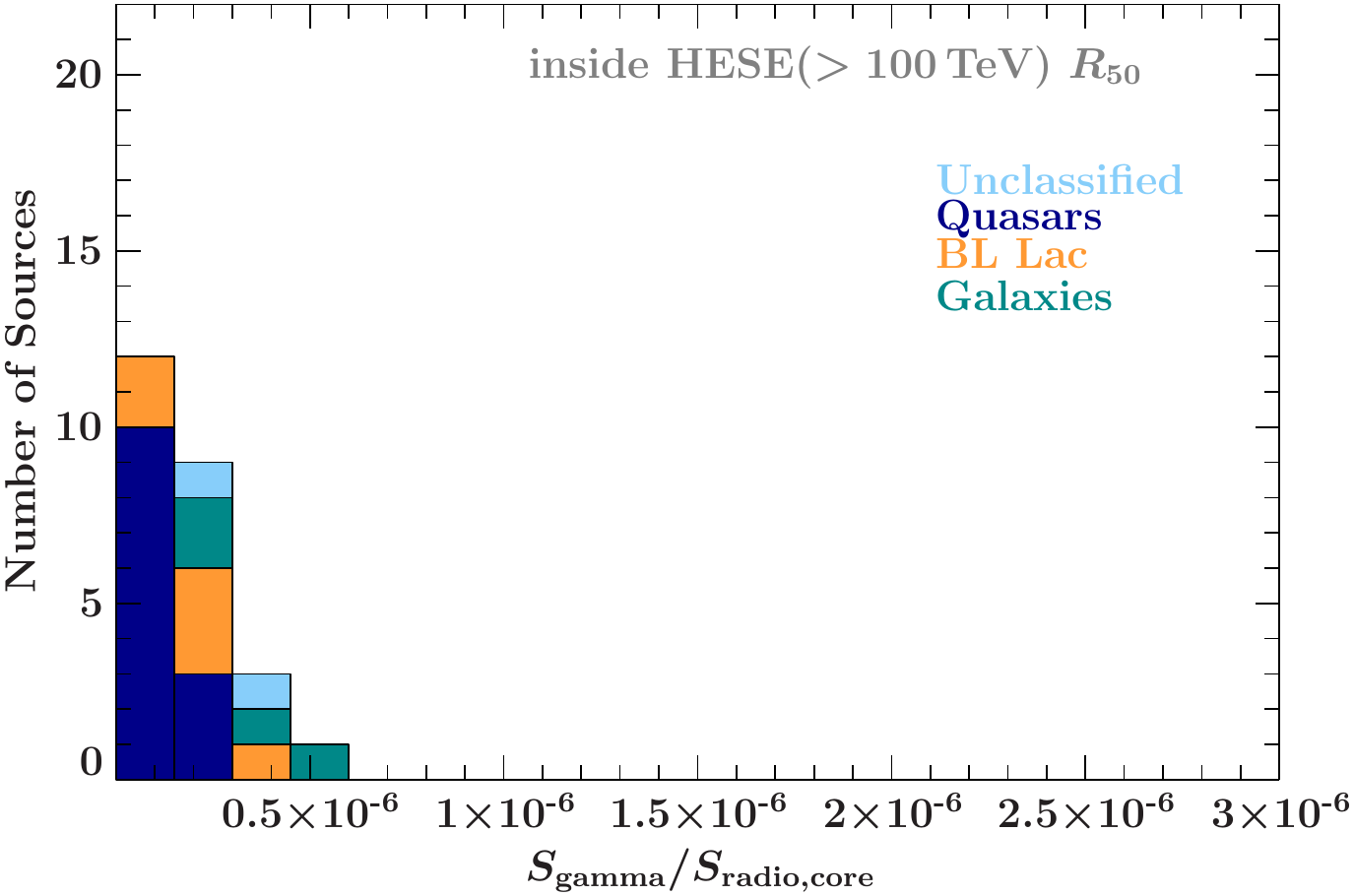}
\includegraphics[width=0.5\textwidth]{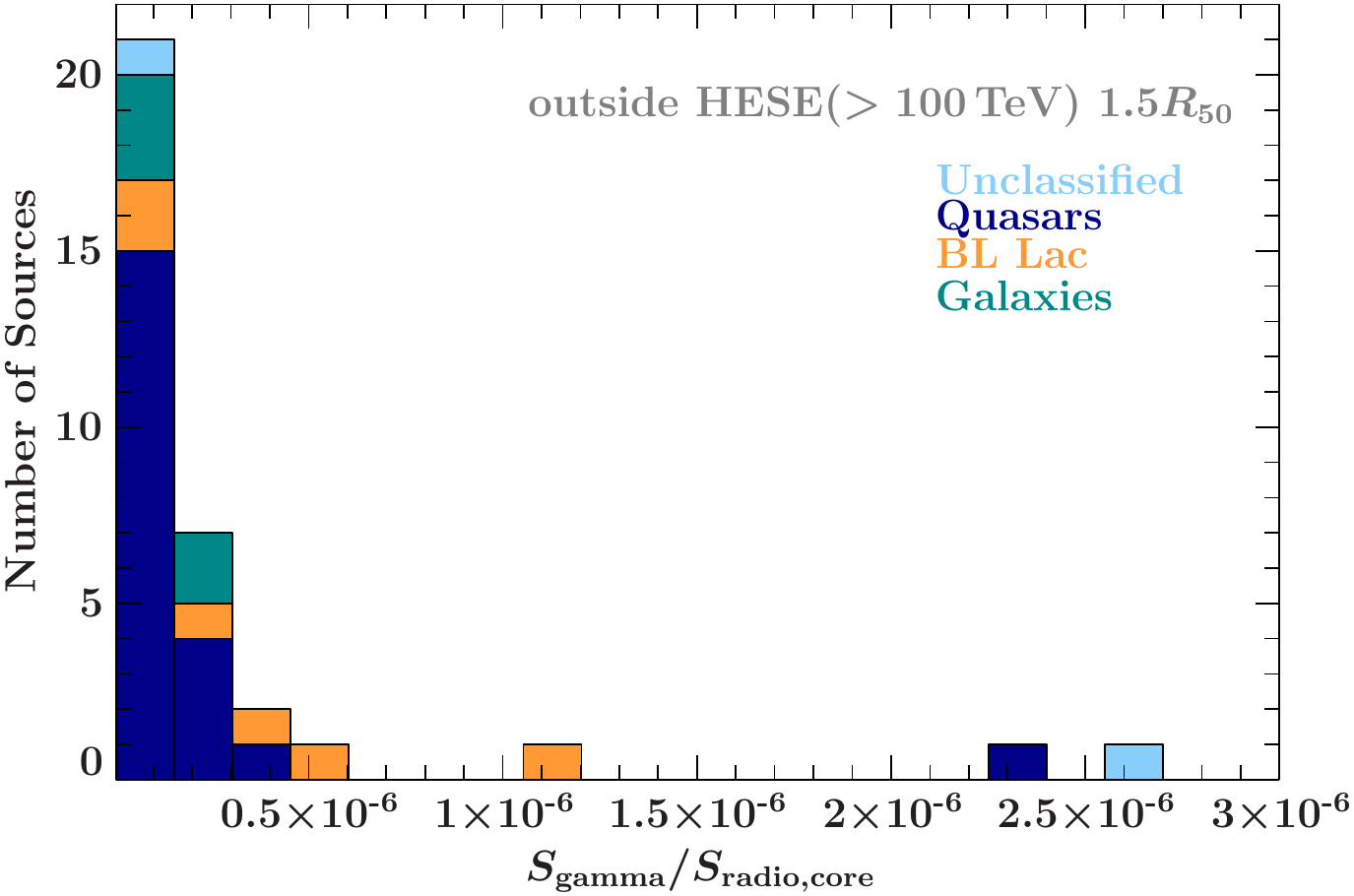}
  \caption{Distribution of $\gamma$-ray to 8.4\,GHz radio core flux for sources inside (\textsl{top}) and
    outside (\textsl{bottom}) the median positional uncertainty radius $R_{50}$ of the IceCube HESE neutrino events with
    energies in excess of 100\,TeV. }
\label{fig:neutrino_radiogammafluxratio}
\end{figure}

\begin{figure}
\includegraphics[width=\columnwidth]{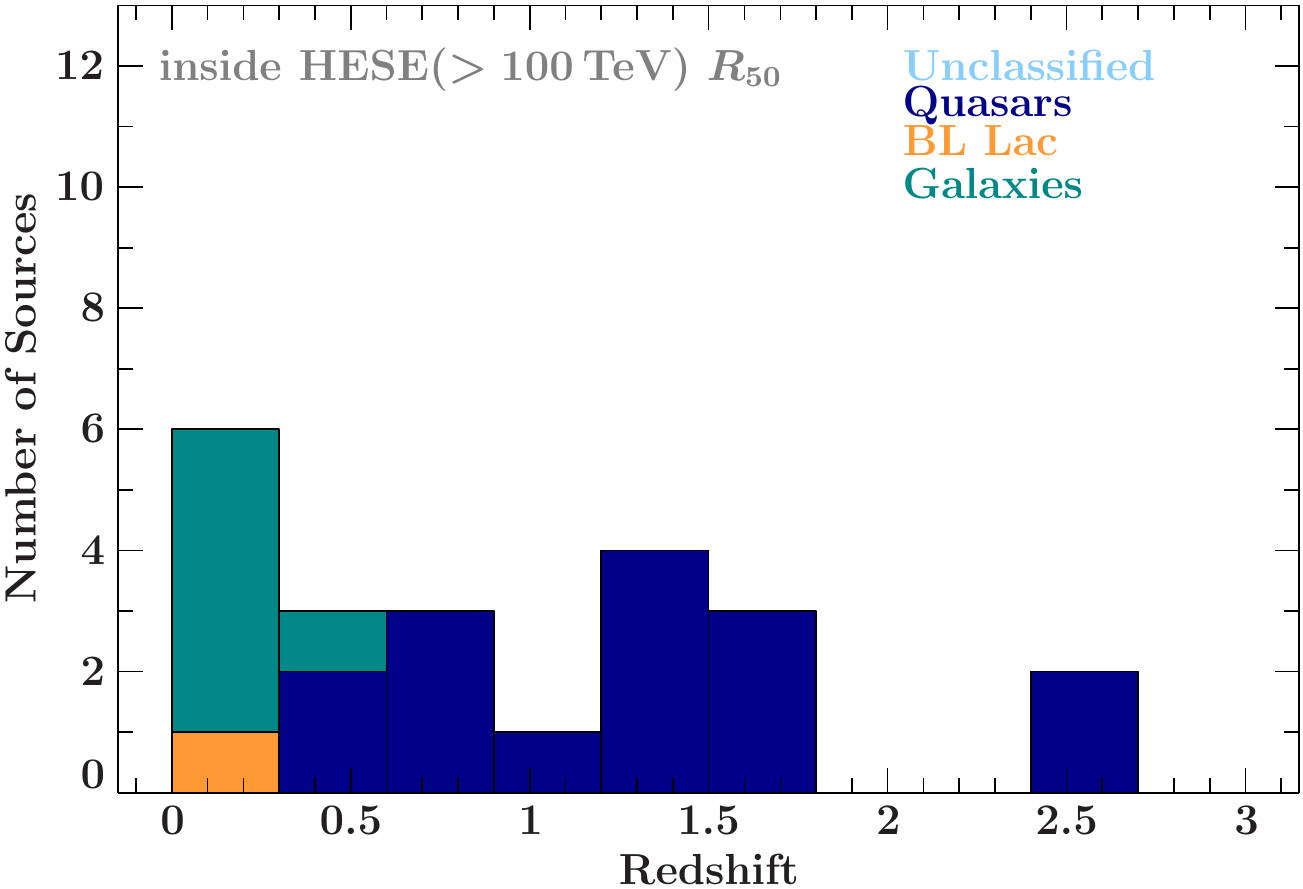}
\includegraphics[width=\columnwidth]{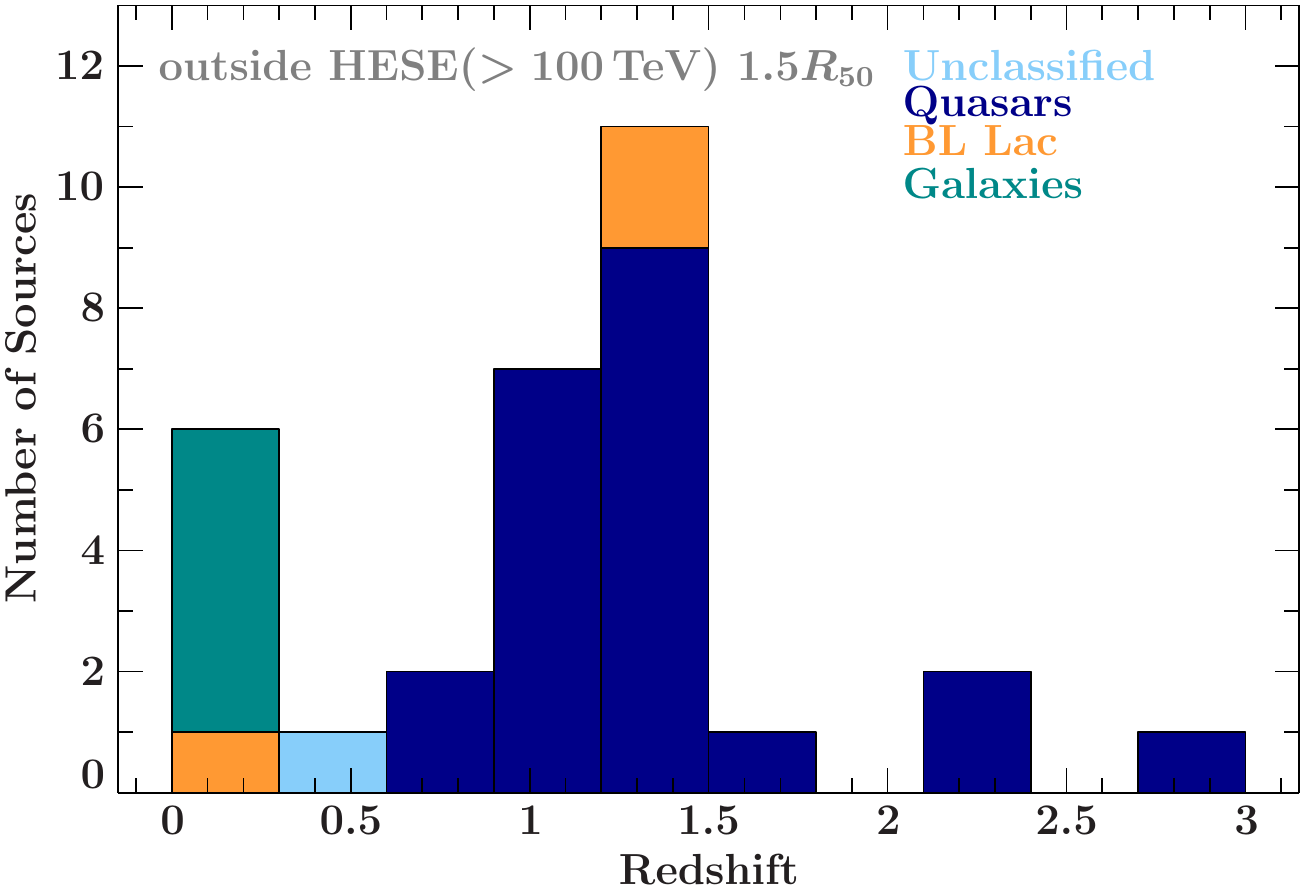}
  \caption{Redshift distribution for sources inside (\textsl{top}) and
    outside (\textsl{bottom}) of the error radius $R_{50}$ of the IceCube HESE neutrino events with
    energies in excess of 100\,TeV.
There is no apparent peak in the distribution for the subsample in
coincidence with the $R_{50}$ radii of $>100$\,TeV IceCube HESE
neutrino events, but a KS test yields a p-value of $0.11$ that both samples were drawn from the same
underlying distribution.
}
\label{fig:neutrino_redshift}
\end{figure}

The VLBI images of the AGN in both subsamples do not show any obvious
morphological differences.  We find no clear difference in
compactness, 8.4\,GHz core and total flux density, and brightness
temperature (Fig.~\ref{fig:neutrino_tb}). 
Sources within $R_{50}$ tend
to show lower brightness temperatures, although this trend is not
statistically significant (KS test: 46\%).  A comparison of the
$\gamma$ ray dominance (Fig.~\ref{fig:neutrino_radiogammafluxratio})
shows no major difference, although the few strongly $\gamma$-dominated
sources are from the subsample outside $1.5\,R_{50}$.  The redshift
distribution (Fig.~\ref{fig:neutrino_redshift}) of these two
subsamples shows a peak between $z \gtrsim 1$ and $z\lesssim 1.5$ for
the subsample outside $1.5\,R_{50}$, which is well known for the full
TANAMI sample.  This peak is not apparent for the subsample in
coincidence with the $R_{50}$ radii of $>100$\,TeV IceCube HESE
neutrino events but a two-sample KS test yields a p-value of $0.11$
 that both samples were drawn from the same
underlying distribution. In conclusion, the TANAMI VLBI data show no significant
  difference in parsec-scale properties for the sources inside and
  outside the HESE $>100$\,TeV $R_{50}$ radii.

Several authors \citep[e.g.,][]{BeckerTjus2014,Krauss2014a,Hooper2016,Kadler2016,Padovani2016} have proposed specific subclasses of AGN as the
dominant sources of the IceCube neutrino signal. It is thus
interesting to look at the spatial distribution of these AGN subclasses in
the TANAMI sample. Table~\ref{tab:count_statistics} shows the fraction
of quasars, BL\,Lac objects, radio galaxies, and unclassified AGN of the full
TANAMI sample,
which are positionally consistent with $> 100$\,TeV IceCube neutrinos
and the comparison sky regions clearly inconsistent with these events.
We also include the statistically complete subsamples of the radio-
and $\gamma$-ray loudest TANAMI AGN in this consideration.  Consistent
with the sky region covered by the median-positional-uncertainty
regions of $> 100$\,TeV neutrinos, we find about 36\,\% of the TANAMI
sources inside these fields. The numbers fluctuate within the range
expected for Poisson statistics among the different AGN classes (see
Table~\ref{tab:count_statistics}). 
It is noteworthy, however,
 that the sample of the 
22 $\gamma$-ray brightest AGN south of $-30^\circ$ declination from the 3LAC catalog
shows no particularly strong agreement with
high-energy neutrino events; 
 only 6 out of the 22 sources fall in the
 $R_{50}$ regions of $>100$\,TeV neutrino events. 
Vice versa, only 5 out of the 13~highest energy
IceCube HESE events in the TANAMI part of the sky (namely HESE-2, HESE-4,
HESE-14, HESE-35, and HESE-52) can possibly be associated with one of
these 22~$\gamma$-ray brightest AGN (see Table~\ref{table:all}). We conclude that the IceCube
signal (above 100\,TeV) cannot simply be attributed to such a small
GeV flux-limited sample of the $\gamma$-ray brightest AGN, but that substantial contributions of a larger population of sources are needed. This finding is
consistent with our previous results \citep{Krauss2014a,Kadler2016} and other groups \citep[e.g.,][]{Ice17,Feyereisen2017}.

%%%%%%%%%%%%%%%%%%%%%%%%

%______________________________________________________________

\section{Summary and conclusions}\label{sec:conclusion}

In this work, we presented first-epoch VLBI images of 39 additional
TANAMI sources, which were included in the program after the first
year of VLBI observations. Most new AGN were added to the sample,
due to their association with a newly detected $\gamma$ ray source by
\lat. For this reason, and the paucity of VLBI observations of southern 
hemisphere sources, this work presents the first VLBI images for
many $\gamma$-bright blazars south of $-30^\circ$ declination.

Complementary to the $\gamma$-ray analysis of a one-year period of \lat data
presented in \citet{Boeck2016}, we discussed the quasi-simultaneous
mas-scale properties of 67\,sources with respect to their $\gamma$-ray
loudness.

Confirming the results by other authors for similar source samples
\citep[e.g.,][]{Kovalev2009}, the $\gamma$-ray bright TANAMI sources
have higher brightness temperatures and are more compact than the
$\gamma$-ray faint ones, indicating higher Doppler beaming factors
\citep{Boeck2016}.  This is also consistent with the findings of
\citet{Lister2015}. 

The group of unclassified sources in the TANAMI sample shows 
mas-scale properties typical for the most compact quasars and BL\,Lac type objects, suggesting
that they also belong to this class.  

There is tentative evidence for a difference in the VLBI morphologies between
$\gamma$-ray faint and $\gamma$-ray bright quasar jets in the sense that
the jets of the $\gamma$-ray brightest quasars are very compact while $\gamma$-ray fainter
quasar jets share morphological similarities with the one-sided jets of radio galaxies.
In contrast to the BL\,Lac objects, whose
broadband spectra can generally be modeled with single-zone SSC models
\citep[e.g.,][]{Tavecchio1998,Cerruti2013}, FSRQs require models
involving external radiation fields to explain their broadband SEDs
\citep[e.g.,][]{Dermer2009}. Hence, the complexity and diversity of
their SEDs and morphologies can be taken as indications of the
existence of different subtypes, including intrinsically $\gamma$-ray
faint, but radio-loud quasars.

We performed a statistical comparison of bright and compact radio-loud
AGN in positional coincidence with high-energy ($>100$\,TeV) IceCube
HESE neutrinos and bright and compact radio-loud AGN far outside the
median-positional-accuracy regions of such neutrinos.  We find no
morphological differences between these two samples and no
clear difference of any characteristic VLBI properties. If
anything, the $\gamma$-dominance values and brightness temperatures of
TANAMI AGN inside the IceCube $>100$\,TeV HESE fields are slightly
lower than for sources outside of these regions.

We also did a simple statistic of the distribution of different subclasses of TANAMI AGN
inside and outside of $>100$\,TeV HESE fields.  The $\gamma$-ray
brightest TANAMI sources show no good positional agreement with high-energy
neutrinos, indicating that the $> 100$\,TeV
IceCube signal is not simply dominated by a small number of the $\gamma$-ray
brightest blazars.  
However, it is
possible that only a subset of variable, radio-loud and $\gamma$-bright AGN
produce high-energy neutrinos \citep{Padovani2016,Kadler2016} and
that a considerable fraction of the detected neutrino events are
associated with faint remote sources below the sensitivity threshold
of currently considered AGN samples. We note that this is also in full agreement with the result
found by the IceCube collaboration \citep{Ice17}.

High-angular resolution radio observations with VLBI techniques remain 
a unique tool to study the relativistic jet flows in high-energy emitting AGN. 
Continued broadband monitoring and correlation studies are necessary
in the endeavor to identify the high-energy emission processes at work in 
AGN and, in the process, to test AGN as possible sources of high-energy 
neutrinos and cosmic rays.

\begin{acknowledgements}
We thank the anonymous referee and B.~Boccardi for helpful comments that improved the manuscript.

The Long Baseline Array is part of the Australia
Telescope National Facility which is funded by the Commonwealth of
Australia for operation as a National Facility managed by CSIRO. 

This study made use of data collected through the AuScope
initiative. AuScope Ltd is funded under the National Collaborative
Research Infrastructure Strategy (NCRIS), an Australian Commonwealth
Government Programme.

This work made use of the Swinburne University of Technology software 
correlator, developed as part of the Australian Major National Research 
Facilities Programme.

 This work was supported by resources provided by the Pawsey Supercomputing 
 Centre with funding from the Australian Government and the Government of 
 Western Australia.

Hartebeesthoek Radio Astronomy Observatory (HartRAO) is a facility of the
     National Research Foundation (NRF) of South Africa.

C.M. acknowledges support from the Studienstiftung des Deutschen Volkes
and the ERC Synergy Grant “BlackHoleCam:
Imaging the Event Horizon of Black Holes” (Grant 610058).

F.K. acknowledges funding from the European Union’s Horizon 2020
research and innovation program under grant agreement No. 653477. 

E.R. acknowledges support from the Spanish MINECO through grants
AYA2012-38491-C02-01 and AYA2015-63939-C2-2-P and from the Generalitat
Valenciana grant PROMETEOII/2014/057. 

This research was funded in part by NASA through Fermi Guest
Investigator grants NNH10ZDA001N, NNH12ZDA001N, and NNH13ZDA001N-FERMI 
(proposal numbers 41213, 61089, and 71326, respectively).

This research was supported by an appointment to the NASA
Postdoctoral Program at the Goddard Space Flight Center, administered
by Oak Ridge Associated Universities through a contract with NASA.

This research has made use of the United States Naval Observatory
(USNO) Radio Reference Frame Image Database (RRFID).

This research has made use of NASA's Astrophysics Data System.

This research has made use of the NASA/IPAC Extragalactic Database
(NED) which is operated by the Jet Propulsion Laboratory, California
Institute of Technology, under contract with the National Aeronautics
and Space Administration.

This research has made use of the SIMBAD database (operated at CDS,
Strasbourg, France).

This research has made use of a collection of ISIS scripts provided by
the Dr. Karl Remeis-Observatory, Bamberg, Germany at
\url{http://www.sternwarte.uni- erlangen.de/isis/} 
and of Astropy, a community-developed core Python
package for Astronomy (Astropy Collaboration, 2013).

\end{acknowledgements}

%-------------------------------------------------------------------
\bibliographystyle{aa}
\bibliography{mnemonic,aaabbrv,tanami}

%%%-----------------------------------------------------------------------%%%

\begin{table*}
\caption{List of TANAMI VLBI observed sources}\label{table:all}
 \centering
\resizebox{.5\textheight}{!}{
\begin{tabular}{llccccccc}
 \hline
 \hline
 Name\tablefootmark{a} & Common Name\tablefootmark{a} & Class\tablefootmark{b} & z\tablefootmark{c} & First epoch & B16\tablefootmark{d} & Radio & $\gamma$-ray   & Inside HESE($>100$\,TeV) \\
                       &                              &
                                                                               &
                                                                                                    & image       &                      & Sample\tablefootmark{e} & Sample\tablefootmark{f} & $R_{50}$   \\
 \hline
0047$-$579 & [HB89] 0047$-$579 & Q & 1.797 &  Paper I & Y & Y &   & \\
0055$-$328 & PKS 0055$-$328    & B & 1.370 & This Work& Y &   &   & \\
0144$-$522 & PKS 0144$-$522    & G & 0.098 &      This Work    &   &   &   & \\
0208$-$512 & [HB89] 0208$-$512 & Q & 0.999 &  Paper I & Y & Y & Y & \\
0227$-$369 & PKS 0227$-$369    & Q & 2.115 & This Work& Y &   &   & \\
0235$-$618 & PKS 0235$-$618    & Q & 0.465 & This Work\tablefootmark{h}     &   &   &   & HESE-20\\
0244$-$470 & PKS 0244$-$470    & Q & 1.385 & This Work& Y &   & Y & \\
0302$-$623 & PKS 0302$-$623    & Q & 1.351 &  This Work\tablefootmark{h}     & Y &   &   & HESE-20\\
0308$-$611 & PKS 0308$-$611    & Q & 1.480 &  This Work\tablefootmark{h}     & Y &   &   & HESE-20\\
0332$-$376 & PMN J0334$-$3725  & B & >0.39 & This Work& Y &   &   & \\
0332$-$403 & [HB89] 0332$-$403 & B & 1.351 &  Paper I & Y & Y & Y & \\
0402$-$362 & PKS 0402$-$362    & Q & 1.417 & This Work& Y &   & Y & \\
0405$-$385 & [HB89] 0405$-$385 & Q & 1.285 &  Paper I & Y &   &   & \\
0412$-$536 & PMNJ0413$-$5332   & Q & 1.024 & This Work& Y &   &   & \\
0426$-$380 & PKS 0426$-$380    & Q & 1.110 & This Work& Y &   & Y & \\
0438$-$436 & [HB89] 0438$-$436 & Q & 2.863 &  Paper I & Y & Y &   & \\
0447$-$439 & PKS 0447$-$439    & B & --    & This Work& Y &   & Y & \\
0454$-$463 & [HB89] 0454$-$463 & Q & 0.853 &  Paper I & Y & Y &   & \\
0506$-$612 & [HB89] 0506$-$612 & Q & 1.093 &  Paper I & Y &   &   & \\
0516$-$621 & PKS 0516$-$621    & Q & 1.300 & This Work& Y &   &   & \\
0518$-$458 & Pic A          & G & 0.035 &  Paper I & Y &   &   & \\
0521$-$365 & ESO 362$-$ G 021  & G & 0.055 &  Paper I & Y & Y & Y & \\
0524$-$485 & PKS 0524$-$485    & Q & 1.300 & This Work& Y &   &   & \\
0527$-$359 & PMNJ0529$-$3555   & U & 0.323 & This Work& Y &   &   & \\
0530$-$485 & PMNJ0531$-$4827   & Q & --    & This Work&   &   & Y & \\
0537$-$441 & [HB89] 0537$-$441 & Q & 0.894 &  Paper I & Y & Y & Y & \\
0625$-$354 & PKS 0625$-$35     & B & 0.055 &  Paper I & Y & Y &   & \\
0637$-$752 & [HB89] 0637$-$752 & Q & 0.653 &  Paper I & Y & Y &   & \textit{HESE-30} \\
0646$-$306 & PKS 0646$-$306    & Q & 0.455 & This Work&   &   &   & HESE-39\\
0700$-$661 & PKS 0700$-$661    & B & >0.39 & This Work& Y &   &   & \\
0717$-$432 & PMN J0718$-$4319  & U & --    & This Work& Y &   &   & \\
0736$-$770 & PKS0736$-$770     & Q & --    & This Work& Y &   &   & \textit{HESE-30} \\
0745$-$330 & PKS0745$-$330     & U & --    & This Work& Y &   &   & \\
0812$-$736 & PMN J0810$-$7530  & U & --    & This Work& Y &   &   & \textit{HESE-30} \\
0902$-$350 & 1FGLJ0904.7$-$3514& U & --    & This Work&   &   &   & \\
0943$-$761 & PKS 0943$-$76     & G & 0.270 &   -- \tablefootmark{j}   &   &   &   & \\ 
1057$-$797 & PKS 1057$-$79     & B & 0.581 & This Work& Y &   &   & \\
1101$-$536 & PKS 1101$-$536    & B & >0.15 & \citet{Krauss2015}& Y &   &   & \textit{HESE-4} \\
1104$-$445 & [HB89] 1104$-$445 & Q & 1.598 &  Paper I & Y & Y &   & \textit{HESE-4} \\ 
1144$-$379 & [HB89] 1144$-$379 & Q & 1.048 &  Paper I & Y &   &   & \\ 
1257$-$326 & PKS 1257$-$326    & Q & 1.256 &  Paper I & Y &   &   & \\ 
1258$-$321 & ESO443$-$G024     & G & 0.017 & This Work& Y &   &   & \\
1313$-$333 & [HB89] 1313$-$333 & Q & 1.210 &  Paper I & Y &   &   & \\
1322$-$428 & Cen A             & G & 0.002 &  Paper I & Y & Y &   & HESE-35\\
1323$-$526 & PMN J1326$-$5256  & B & >0.24 &  Paper I & Y &   &   & HESE-35\\
1325$-$558 & PMN J1329$-$5608  & B & >0.13 & This Work& Y &   & Y & HESE-35\\
1333$-$337 & IC 4296           & G & 0.012 & Paper I& Y &   &   & \textit{HESE-48} \\
1343$-$601 & Cen B       & G & 0.013 & This Work&   &   &   & HESE-35\\
1344$-$376 & PMN J1347$-$3750  & Q & 1.300 & This Work& Y &   &   & \textit{HESE-48} \\
1409$-$651 & Circinus Galaxy   & G & 0.001 & -- \tablefootmark{j} &   &   &   & HESE-35\\
1424$-$418 & [HB89] 1424$-$418 & Q & 1.522 &  Paper I & Y & Y & Y & HESE-35\\
1440$-$389 & PKS 1440$-$389    & B & >0.14 & This Work& Y &   &    & \\
1454$-$354 & PKS 1454$-$354    & Q & 1.424 &  Paper I & Y & Y & Y & \\
1501$-$343 & PMN J1505$-$3432  & B & 1.554 &  Paper I & Y &   &   & \\
1505$-$496 & PMN J1508$-$4953  & Q & 0.776 & This Work& Y &   &   & HESE-35\\
1549$-$790 & PKS 1549$-$79     & G & 0.150 &  Paper I & Y & Y &   & \\
1600$-$445 & PMN J1604$-$4441  & U & >0.01 & This Work& Y &   &   & \\
1600$-$489 & PMN J1603$-$4904  & G & 0.230\tablefootmark{g}&\citet{Mueller2014a}&Y&   & Y & \\
1606$-$667 & PMN J1610$-$6649  & U & --    & This Work& Y &   &   & \\
1610$-$771 & [HB89] 1610$-$771 & Q & 1.710 &  Paper I & Y & Y &   & \\
1613$-$586 & PMN J1617$-$5848  & Q & --    & This Work& Y &   &   & \textit{HESE-52} \\
1646$-$506 & PMNJ1650$-$5044   & U & 0.090 & This Work& Y &   & Y & \textit{HESE-52} \\
1653$-$329 &SwiftJ1656.3$-$3302& Q & 2.400 &  \citet{Krauss2014a}    &   &   &   & HESE-2, HESE-14   \\
1714$-$336 & PMN J1717$-$3342  & B & --    &  Paper I & Y &   & Y & HESE-2, HESE-14\\
1716$-$771 & PKS 1716$-$771    & U & --    &  Paper I & Y &   &   & \\
1718$-$649 & NGC 6328          & G & 0.010 &  Paper I & Y & Y &   & \\ 
1733$-$565 & PKS 1733$-$56     & G & 0.098 &  Paper I & Y & Y &   & \textit{HESE-52} \\
1759$-$396 & PMN J1802$-$3940  & Q & 1.320 &  Paper I & Y &   & Y & HESE-2, HESE-14\\
1804$-$502 & PMN J1808$-$5011  & Q & 1.606 &  Paper I & Y &   &   &         HESE-2   \\
1814$-$637 & PKS 1814$-$63     & G & 0.063 &  Paper I & Y &   &   & \\
1915$-$458 & PKS 1915$-$458    & Q & 2.470 & This Work&   &   &   & HESE-2, \textit{HESE-12} \\
1933$-$400 & PKS 1933$-$400    & Q & 0.965 &  Paper I & Y &   &   &         HESE-2   \\
1954$-$388 & [HB89] 1954$-$388 & Q & 0.630 &  Paper I & Y & Y &   &         HESE-2   \\
2004$-$447 & PKS 2004$-$447    & G & 0.240 &  \citet{Schulz2016}  &   &   & & HESE-2,\textit{HESE-12} \\
2005$-$489 & [HB89] 2005$-$489 & B & 0.071 &  Paper I & Y &   &    & \textit{HESE-12} \\
2027$-$308 & ESO 462$-$ G 027  & G & 0.539 &  Paper I & Y &   &   &         HESE-2   \\
2052$-$474 & [HB89] 2052$-$474 & Q & 1.489 &  Paper I & Y & Y & Y & \\
2106$-$413 & [HB89] 2106$-$413 & Q & 1.058 &  Paper I & Y & Y &   & \\
2123$-$463 & PKS 2123$-$463    & Q & 1.46\tablefootmark{i}  & This Work&   &   &   & \\
2136$-$428 & PMNJ2139$-$4235   & B & --    & This Work& Y &   & Y & \\
2142$-$758 & PKS 2142$-$75     & Q & 1.139 & This Work&   &   & Y & \\
2149$-$306 & PKS 2149$-$306    & Q & 2.345 &  Paper I & Y &   &   & \\
2152$-$699 & ESO 075$-$ G 041  & G & 0.028 &  Paper I & Y &   &   & \\
2154$-$838 & PKS 2155$-$83     & Q & --    & This Work&   &   &   & \\ 
2155$-$304 & [HB89] 2155$-$304 & B & 0.116 &  Paper I & Y &   & Y & \\
2204$-$540 & [HB89] 2204$-$540 & Q & 1.206 &  Paper I & Y & Y &   & \\
2326$-$477 & [HB89] 2326$-$477 & Q & 1.299 &  Paper I & Y & Y &   & \\
2355$-$534 & [HB89] 2355$-$534 & Q & 1.006 &  Paper I & Y &   &   & \\
 \hline
\end{tabular}
}

 \tablefoot{\footnotesize 
\tablefoottext{a}{IAU B1950 and common
     name. } 
 \tablefoottext{b}{Optical classification after \citet[][]{Veron2006}
     and \citet{Shaw2012,Shaw2013a} with
   Q:\,quasar, B:\,BL\,Lac, G:\,galaxy, U:\,unidentified object.}
\tablefoottext{c}{Redshift}
 \tablefoottext{d}{Source included in the radio-$\gamma$ correlation
   study of the first year of \textsl{Fermi}/LAT data by
   \citet[][]{Boeck2016} (B16).}
\tablefoottext{e}{The
radio-selected subsample includes all sources south of declination $-30^\circ$ from the catalog of
\citet{Stickel1994} above a limiting radio flux density of S$_{\rm 5\,GHz}>$2\,Jy with
a flat radio spectrum ($\alpha>-0.5$, S$\sim\nu^{+\alpha}$) between 2.7\,GHz and 5\,GHz (22 sources).} 
\tablefoottext{f}{The  
$\gamma$ ray  sample includes the $\gamma$-ray
brightest AGN
south of declination $-30^\circ$ in the 3LAC catalog (22 sources).}
\tablefoottext{g}{Redshift measurement by \citet{Goldoni2016}.}
\tablefoottext{h}{Results of TANAMI VLBI observations quasi-simultaneous
to the IceCube events were presented in \citet{Krauss2014a}.}
\tablefoottext{i}{Redshift measurement by \citet{DAmmando2012}.}
\tablefoottext{j}{see Fig.~\ref{fig:radplot}}

}
\end{table*}

%%%%%%%%%%                                                                                                                       
\begin{table*}
        \caption{The TANAMI array}\label{table:TANAMIarray}
\centering
%\resizebox{\textwidth}{!}{                                                                                                      
        \begin{tabular}{l c l} \hline\hline
Telescope & Diameter & Location\\
Name (Abbreviation) & [meters] & \\
\hline
Parkes (PA) & 64 & Parkes, New South Wales, Australia \\
ATCA (AT) & 5$\times$22 & Narrabri, New South Wales, Australia  \\
Mopra (MP) & 22 &
Coonabarabran, New South Wales, Australia  \\
Hobart (HO) & 26 & Mt. Pleasant, Tasmania, Australia \\
Ceduna (CD) & 30 & Ceduna, South Australia \\
DSS43$^\mathrm{a}$ & 70 &Tidbinbilla, Australia \\
DSS45$^\mathrm{a}$ & 34 & Tidbinbilla, Australia \\
DSS34$^\mathrm{a}$ & 34 & Tidbinbilla, Australia \\
Hartebeesthoek$^\mathrm{c}$ (HH)  & 26 & Hartebeesthoek, South Africa \\
O'Higgins$^{b}$ (OH) & 9 & O'Higgins, Antarctica \\ 
TIGO$^{b}$ (TC)& 6     &Conc\' epcion, Chile\\
Warkworth (WW)& 12 & Auckland, New Zealand \\
Katherine (KE) & 12 & Northern Territory, Australia \\
Yarragadee (YG) &12 & Western Australia \\
\hline
        \end{tabular}
\tablefoot{
\footnotesize{$^{a}$Operated by the Deep
Space Network of the National Aeronautics and Space %                                                                            
Administration. DSS45 was decommissioned in November 2016
(\url{https://www.cdscc.nasa.gov/Pages/antennas.html}.
}
\footnotesize{$^{b}$Operated by Bundesamt f\"ur Kartographie und
  Geod\"asie (BKG). The telescope was decommissioned in 2014 in Chile
  and moved to in La Plata, Argentina, in 2015 (now: AGCO).}                                         
\footnotesize{$^{c}$Due to a major failure, not available between
  Sept. 2008 and Sept. 2010. }\\
}
\end{table*}
%%%%  
%__________________________________________________________________
%%%%%%%%%%                                                                                                                       
\begin{table*}
        \caption{Array configurations of observations}\label{table:config}
\centering
        \begin{tabular}{l l l} 
\hline\hline
Epoch & Participating Telescopes \\
\hline
2009 Feb. 23 & AT, CD, DSS34, DSS45, HO, MP, OH, PA, TC  \\
2009 Feb. 27 &  AT, CD, DSS45, HO, MP, OH, PA, TC \\ 
2009 Sept. 05 & AT, CD, DSS43, HO, MP, OH, PA, TC      \\ 
2009 Dec. 13 & AT, CD, HO, MP, PA, TC  \\ 
2010 May 07 & AT, CD, DSS43, HO, MP, PA, TC  \\ 
2010 Oct. 28 & AT, CD, DSS34, DSS45, HH, HO, MP, OH, PA, TC \\ 
2011 Apr. 01 & AT, CD, DSS34, HH, HO, MP, PA, TC, WW \\
2011 July 21 & AT, CD, DSS34, DSS43,  HH, HO, MP, PA, TC, WW \\
2011 Aug. 13 & AT, CD, DSS43, HH, HO, KE, MP, PA, TC,
               YG \\
2012 Apr. 27 &  AT, CD, HH, HO, MP, PA, TC, WW \\ 
2012 Sept. 15 & AT, CD, DSS43, HH, HO, KE, PA, TC \\ 
2012 Sept. 16 & AT, CD, DSS 34, DSS45,  HH, HO, KE,
                PA, TC \\ 
\hline
        \end{tabular}
\end{table*}
%%%%

\begin{sidewaystable*}
\caption{Observational parameters of additional sources.}\label{table:images}
 %\resizebox{\textwidth}{!}{
\scriptsize
\begin{tabular}{llcccrrrrrrrrrrr}
 \hline
 \hline
  Name\tablefootmark{a} & Common Name\tablefootmark{a} &
  Class\tablefootmark{b} &
 z\tablefootmark{c} & Morph.\tablefootmark{d} &
 Epoch\tablefootmark{e} &
 $b_\mathrm{maj}$\tablefootmark{f} & $b_\mathrm{min}$\tablefootmark{f}
 & P.A.\tablefootmark{f} & $S_\mathrm{total}$\tablefootmark{g}
 & RMS\tablefootmark{g}  & $S_\mathrm{core}$\tablefootmark{h}  &
 $\theta_\mathrm{min,core}$\tablefootmark{h} & $\theta_\mathrm{maj,core}$\tablefootmark{h}&
 $T_B$\tablefootmark{h}  &
 $S_\mathrm{0.1-100\,GeV}$\tablefootmark{i} \\
 &  &  &  &  & [yyyy-mm-dd] & [mas] & [mas] & [$^\circ$] & [Jy] & [mJy/beam]
 &[Jy] & [mas] & [mas]  & [K] &   [$\times
 10^{-8}\mathrm{ph\,cm^{-2}\,s^{-1}}$] \\
 \hline 
\it 0055$-$328 & \it PKS 0055$-$328 &\it  B & \it 1.370 & \it SS &\it  2009-02-27 & \it 2.61 & \it 1.75 &
\it  31.7 & \it 0.096 &\it  0.1 &\it  0.08 & \it 0.18 & \it 1.30 &\it
   $1.4\times 10^{10}$ & $\mathit{1.6\pm0.5}$ \\

0144$-$522 & \object{PKS 0144$-$522} & G & 0.098 & C & 2008-03-28 & 3.49 & 0.64 & 8.1
  & 0.117 & 0.1 & 0.12 & 0.24 & 0.24 & $3.9\times 10^{10}$ & \\

\it  0227$-$369 &\it  \object{PKS 0227$-$369} &\it  Q & \it 2.115 &\it  C &\it  2009-02-27 &\it  2.46 & \it 1.17 &
 \it 37.4 & \it 0.573 & \it 0.2 &\it  0.57 & \it 0.17 & \it 0.34 &\it $5.1\times 10^{11}$  & $\mathit{8.0\pm0.7}$ \\

0235$-$618 & \object{PKS 0235$-$618} & Q & 0.465 & SS & 2011-04-01 & 1.78 & 0.56 &
4.2 & 0.337 & 0.2 & 0.31 & 0.20 & 0.29 & $1.3\times 10^{11}$ & \\

\it 0244$-$470 & \it \object{PKS 0244$-$470} &\it  Q & \it 1.385 &\it  SS &\it  2009-02-27 &\it  2.41 &\it  1.13 &
\it 43.7 &\it  0.299 &\it  0.2 &\it  0.26 & \it 0.14 &\it  0.39 &\it  $1.9\times 10^{11}$ & $\mathit{9.4\pm0.8}$ \\

\it  0302$-$623 &\it \object{ PKS 0302$-$623} &\it  Q & \it 1.351 & \it IRR &\it  2009-02-27 &\it  1.63 & \it 0.82 &
 \it 40.5 &\it  1.620 &\it  0.2 &\it  1.10 &\it  0.25 &\it  0.64 &\it  $2.8\times 10^{11}$ & $\mathit{5.1\pm1.1}$ \\
 
\it  0308$-$611 &\it   \object{PKS 0308$-$611} &\it   Q &\it   1.480 &\it   SS &\it   2009-09-05 &\it   1.93 &\it   1.31 &
 \it  53.8 &\it   0.828 &\it   0.3 &\it   0.81 &\it   0.30 &\it   0.52 &\it   $2.2\times 10^{11}$ & $\mathit{4.2\pm1.0}$ \\
 
\it  0332$-$376 & \it  \object{PMN J0334$-$3725} & \it  B & $\mathit{>0.39}$& \it  C & \it  2009-02-27 &\it   3.05 & \it  1.29 &
 \it  35.1 &\it   0.322 & \it  0.2 &\it   0.32 &\it   0.10 &\it   0.29 & $\mathit{>2.5}\times 10^{11}$ & $\mathit{2.8\pm0.5}$ \\
 
\it  0402$-$362 &\it   \object{PKS 0402$-$362} &\it   Q & \it  1.417 & \it  SS &\it   2009-09-05 & \it  3.57 & \it  0.87 &
 \it  14.2 &\it   0.576 &\it   0.2 &\it   0.54 &\it   0.37 &\it   0.37 & \it  $1.6\times 10^{11}$ & $\mathit{10.9\pm0.8}$ \\
 
0412$-$536 & \object{PMN\,J0413$-$5332} & Q & 1.024 & C & 2009-12-13 & 3.66 & 1.31 &
 25.0 & 0.033 & 0.4 & 0.03 & $<0.11$ & $<0.11$ & $>4.7\times 10^{10}$ & \\
  
\it 0426$-$380 &\it  \object{PKS 0426$-$380} &\it  Q &\it  1.110 &\it  SS &\it  2009-02-27 &\it  2.53 &\it  0.63 &
\it  29.4 &\it  1.810 & \it 0.4 &\it  1.80 &\it  0.13 &\it  0.32 & \it $1.6\times 10^{12}$ & $\mathit{31.5\pm0.9}$ \\
 
\it 0447$-$439 & \it \object{PKS 0447$-$439}&\it  B & -- & \it SS & \it 2009-02-27 &\it  3.10 &\it  1.43 & \it 36.0
 &\it  0.097 &\it  0.1 &\it  0.09 &\it  0.12 & \it 0.28 & \it $>4.7\times 10^{10}$ &$\mathit{9.7\pm0.6}$ \\
 
\it 0516$-$621 & \it \object{PKS 0516$-$621} & \it Q & \it 1.300 &\it  C &\it  2009-02-23 & \it 2.46 & \it 0.62 &
 \it 78.6 &\it  0.946 &\it  0.4 & \it 0.94 & \it 0.32 &\it  0.32 & \it $3.6\times 10^{11}$ & $\mathit{5.8\pm1.0}$ \\

\it  0524$-$485 &\it  \object{PKS 0524$-$485} &\it  Q & \it 1.300 &\it  SS &\it  2009-02-27 &\it  2.47 & \it 0.92 &
 \it 30.0 &\it  0.579 &\it  0.2 & \it 0.57 &\it  0.23 & \it 0.44 &\it  $2.2\times 10^{11}$ & $\mathit{3.6\pm0.7}$ \\

0527$-$359\tablefootmark{j} & \object{PMN\,J0529$-$3555} & U & 0.323 & C & 2009-12-13& 3.79 & 1.85 &
 7.9 & 0.043 & 0.27 & 0.04 & <0.10 & <0.10 & $>7.5\times 10^{10}$ & \\
 
0530$-$485 & \object{PMN\,J053$-$-4827} & Q & -- & SS  & 2010-10-28 & 3.54 & 0.71 & 7.3
 & 0.161 & 0.1 & 0.15 & 0.22 & 0.37 & $>3.3\times 10^{10}$ & \\

0646$-$306 & \object{PKS 0646$-$306} & Q & 0.455 & SS & 2012-09-16 & 2.17 & 0.59 &
 1.1 & 0.674 & 0.2 & 0.63 & 0.22 & 0.22 & $3.3\times 10^{11}$ & \\
 
\it 0700$-$661 & \it \object{PKS 0700$-$661} &\it  B & \it >0.39 & \it SS & \it 2009-02-27 & \it 1.73 & \it 1.24 &
 \it 55.8 &\it  0.153 &\it  0.1 &\it  0.14 & \it 0.33 & \it 0.35 &\it  $>2.9\times 10^{10}$ & $\mathit{6.0\pm0.7}$ \\

\it  0717$-$432 &\it  \object{PMN J0718$-$4319} &\it  U& -- & \it SS & \it 2009-02-27 & \it 1.87 &\it  0.91 &
 \it 26.5 &\it  0.034 &\it  0.1 &\it  0.03 &\it  <0.04 &\it  <0.04 & \it $>2.6\times 10^{11}$ & $\mathit{1.9\pm0.4}$ \\

0736$-$770 & \object{PKS 0736$-$770} & Q & -- & SS & 2009-12-13 & 1.45 & 1.10 & 1.6 &
 0.315 & 0.3 & 0.31 & 0.22 & 0.30 & $>7.8\times 10^{10}$ & \\

0745$-$330 & \object{PKS 0745$-$330} & U & -- & SS & 2009-12-13 & 3.26 & 1.26 & 16.3
 & 0.671 & 0.4 & 0.60 & 0.32 & 0.64 & $>5.0\times 10^{10}$ & \\

\it  0812$-$736 & \it \object{PMN J0810$-$7530} & \it U& -- &\it  C &\it  2009-02-27 & \it 1.47 &\it  1.13 &
 \it 19.1 &\it  0.037 &\it  0.2&\it  0.04 &\it  <0.04 &\it  <0.04 &\it $ >5.1\times 10^{11}$ & $\mathit{1.4\pm0.4}$ \\

0902$-$350 & \object{PMN J0904$-$3514}  & U& -- & C & 2012-09-15 & 9.78 & 2.69 &
 $-$56.0 & 0.246 & 0.3 & 0.25 & 0.21 & 0.21 & $>9.7\times 10^{10}$ & \\

0943$-$761\tablefootmark{k} & \object{PKS 0943$-$76}    & G & 0.270 &---&2009-12-13  & 52.6 & 3.37 &
      $-$52.5& 0.021 & 1.0 & ---& ---& ---&--- &  \\
 
\it 1057$-$797 & \it \object{PKS 1057$-$79} &\it  B &\it  0.581 & \it C &\it  2009-02-27 & \it 1.05 &\it  0.55 &
 \it $-$31.5 &\it  2.022 &\it  0.5 &\it  1.90 & \it 0.31 & \it 0.36 &\it  $4.7\times 10^{11}$ & $\mathit{6.3\pm1.1}$ \\

1258$-$321 & \object{ESO443$-$G024} & G & 0.017 & SS & 2009-12-13 & 4.27 & 1.78 &
  25.0 & 0.120 & 0.2 & 0.12 & <0.05 & <0.05 & $>7.3\times 10^{11}$ & \\
 
\it 1325$-$558 &\it  \object{PMN J1329$-$5608} & \it B &\it  >0.13 & \it SS & \it 2009-02-23 & \it 2.04 & \it 1.41 &
 \it 85.2 &\it  0.480 & \it 0.3 &\it  0.47 &\it  0.37 & \it 0.44 &\it  $>5.6\times 10^{10}$ & $\mathit{13.9\pm1.6}$ \\

1343$-$601 & \object{Centaurus B} & G & 0.013 & SS & 2011-07-21 & 3.44 & 2.14 &
 21.2 & 2.646 & 0.2 & 0.90 & 1.43 & 1.43 & $0.8\times 10^{10}$ & \\
 
\it 1344$-$376 &\it  \object{PMN J1347$-$3750} & \it Q &\it  1.300 & \it SS &\it  2009-02-27 &\it  3.53 & \it 1.27 &
 \it 14.3 &\it  0.206 & \it 0.1 &\it  0.20 & \it 0.22 &\it  0.30 & \it $1.2\times 10^{11}$& $\mathit{5.0\pm1.0}$ \\

1409$-$651\tablefootmark{k} & \object{Circinus Galaxy} & G & 0.001 & --- & 2010-10-28 & 4.30 & 2.82 &
 64.6 & 0.031 &0.8& --- & --- & --- & --- & \\
 
\it 1440$-$389 &\it  \object{PKS 1440$-$389} &\it  B &\it  >0.14 & \it SS & \it 2009-02-27 &\it  3.17 & \it 1.27 &
 \it 25.0 &\it  0.096 & \it 0.1 &\it  0.09 &\it  0.33 & \it 0.55 &\it  $1.0\times 10^{10}$ & $\mathit{2.3\pm0.4}$ \\

\it  1505$-$496 & \it \object{PMN J1508$-$4953} & \it Q &\it  0.776 & \it SS &\it  2009-02-27 &\it  4.94 &\it  3.38 &
 \it $-$67.4 &\it  0.508 & \it 0.3 & \it 0.50 & \it 0.08 & \it 0.08 & \it $2.2\times 10^{12}$ & $\mathit{4.0\pm1.0}$ \\
 
\it 1600$-$445 & \it \object{PMN J1604$-$4441} & \it U &\it  >0.01 & \it SS &\it  2009-09-05 & \it 2.36 & \it 0.63 &
 \it 17.8 &\it  1.041 &\it  0.4 & \it 0.95 &\it  0.41 &\it  0.56 & \it $>7.2\times 10^{10}$ & $\mathit{21.9\pm1.7}$ \\

\it 1606$-$667 &\it  \object{PMN J1610$-$6649} & \it U& -- & \it C &\it  2009-02-27 &\it  3.04 & \it 1.11 &
\it  $-$52.8 & \it 0.041 &\it  0.2 & \it 0.04 &\it  <0.06 & \it <0.06 & \it $>2.1\times 10^{11}$ & $\mathit{1.2\pm0.3}$ \\
 
1646$-$506 & \object{PMNJ1650$-$5044} & U & 0.090 & SS & 2010-05-07 & 3.07 & 1.53 &
 19.8 & 0.958 & 0.2 & 0.91 & 0.22 & 0.50 & $1.5\times 10^{11}$ & \\

\it 1613$-$586 & \it \object{PMN J1617$-$5848} & \it Q & -- &\it  SS & \it 2009-09-05 &\it  1.02 &\it  0.59 &
 \it 21.6 &\it  2.337 & \it 0.6 & \it 1.59 & \it 0.44 &\it  0.83 & \it $>7.5\times 10^{10}$ & $\mathit{11.2\pm1.7}$ \\

1915$-$458 & \object{PKS 1915$-$458} & Q & 2.470 & SS & 2011-04-01 & 2.03 & 0.46 &
 6.1 & 0.286 & 0.4 & 0.25 & 0.30 & 0.61 & $8.3\times 10^{10}$ & \\

2123$-$463 & \object{PKS 2123$-$463} & Q & 1.46 & SS & 2012-09-16 & 3.44 & 0.78 &
 10.9 & 0.471 & 0.1 & 0.37 & 0.50 & 0.50 & $6.3\times 10^{10}$ & \\

2136$-$428 & \object{PMN\,J2139$-$4235} & B & -- & SS & 2009-12-13 & 4.13 & 2.23 &
 33.8 & 0.096 & 0.2 & 0.09 & <0.06 & <0.06 & $>3.8\times 10^{11}$& \\
 
2142$-$758 & \object{PKS 2142$-$75} & Q & 1.139 & SS & 2011-08-13 & 1.46 & 0.96 &
 $-$51.4 & 1.596 & 0.3 & 1.31 & 0.27 & 0.44 & $4.1\times 10^{11}$ & \\
 
2155-839 & \object{PKS 2155-83} & Q & -- & C & 2012-09-15 & 7.91 & 3.67 & 76.9 &
 0.442 & 0.32 & 0.44 & 1.62 & 2.16 & $>0.2\times 10^{10}$ & \\
  \hline
\end{tabular}
%}

 \tablefoot{\footnotesize 
\textit{All sources marked in italics are included in the analysis
  discussed in Sect.~\ref{sec:Boeck}.}
\tablefoottext{a}{IAU B1950 and common
     name. } 
 \tablefoottext{b}{Optical classification after \citet[][]{Veron2006}
     and \citet{Shaw2012,Shaw2013a} with
   Q:\,quasar, B:\,BL\,Lac, G:\,galaxy, U:\,unidentified object.}
\tablefoottext{c}{Redshift}
 \tablefoottext{d}{Morphology scheme according to
   \citet{Kellermann1998} and Paper~I.}
    \tablefoottext{e}{Date of first TANAMI \X observation shown in Figs.~\ref{fig:newimages1}
to~\ref{fig:newimages4}.}
   \tablefoottext{f}{Major, minor axis, and position angle of
     synthesized beam.}  
\tablefoottext{g}{Total flux density and RMS in the \X
  image.}  
\tablefoottext{h}{Core component parameters: total flux, FWHM of the
  major and minor axis, and brightness temperature.} 
\tablefoottext{i}{0.1-100\,GeV fluxes from \citet[][]{Boeck2016}. All
  other sources have not been analyzed using (quasi-)simultaneous data in \citet[][]{Boeck2016}.
\tablefoottext{j}{Not detected at $\gamma$ rays.}
\tablefoottext{k}{Only observed with one scan and not imaged (see Fig.~\ref{fig:radplot}).}

}
}
\end{sidewaystable*}

\begin{table*}
        \caption{Parameters of tapered images.}\label{table:tapered}
\begin{center}
        \begin{tabular}{ccccccccc} 
\hline\hline
Source & Epoch &  $b_\mathrm{maj}$ & $b_\mathrm{min}$ & P.A. &
                                                                   $S_\mathrm{total}$ & $S_\mathrm{peak}$ 
                                                                 
          & RMS& Taper$^\mathrm{a}$\\
Name & [yyyy-mm-dd] & [mas] & [mas] & $[^\circ]$ &  [Jy] &  [Jy/beam] &  [mJy/beam] & \\ 
\hline
0055$-$328 & 2012-04-27 & 4.89 & 3.91 & $-$69.0 & 0.17 & 0.14&0.05 &100\\
0447$-$439 & 2009-02-27 & 5.26 & 4.12 & $-$89.6 & 0.11 & 0.10 & 0.06  & 100\\
0717$-$432 & 2009-02-27 & 5.97 & 4.90 & $-$79.3 & 0.03 & 0.03 & 0.06 &70\\
\hline
        \end{tabular}
\tablefoot{\footnotesize 
\tablefoottext{a}{Baseline length in M$\lambda$ at which the
  visibility data were downweighted to 10\%.} }
\end{center}
\end{table*}

\begin{table}
        \caption{Count statistics of TANAMI sources within the $R_{50}$ 
of high-energy ($> 100$\,TeV) neutrino events from the IceCube HESE analysis.}\label{tab:count_statistics}
\centering
        \begin{tabular}{@{}lcccc@{}} 
\hline\hline
Class & Number & Inside & Poisson & Fraction\\
 & & $R_{50}(100\text{\,TeV})$ & Error & [\%]\\ 
\hline
Quasars & 46 & 17 & 4 & $37\pm 9$\\
BL\,Lacs & 16 & 5 & 2 & $31^{+13}_{-12}$\\
Galaxies & 17 & 7 & 3 & $41^{+18}_{-17}$\\
Unclassifieds & 9 & 2& 1& $22\pm 11$\\
\hline
All & 88 & 31 & 6 & $35\pm7$\\
\hline
Radio Sample & 22 & 6 & 2 & $27\pm9$ \\
Gamma Sample & 22 & 6 & 2 & $27\pm9$ \\
\hline
\hline
        \end{tabular}
\end{table}

%%%% 

\end{document}